\documentclass[sigconf, authorversion,nonacm]{acmart}
\AtBeginDocument{%
  }

\usepackage{supertabular}
\usepackage{booktabs}
\usepackage{longtable}

\begin{document}

\title{The erasure of intensive livestock farming in text-to-image generative AI}

\author{Kehan Sheng\textsuperscript{1}}
\orcid{0000-0001-6442-5284}
\affiliation{%
  \institution{\textsuperscript{1}The University of British Columbia}
  \department{Animal Welfare Program}
  \city{Vancouver}
  \state{British Columbia}
  \country{Canada}
}
\email{skysheng7@gmail.com}

\author{Frank A.M. Tuyttens\textsuperscript{2,3}}
\orcid{0000-0002-1348-218X}
\affiliation{%
  \institution{\textsuperscript{2}Flanders Research Institute for Agriculture, Fisheries and Food (ILVO)}
  \department{Animal Sciences Unit}
  \city{Melle-Merelbeke}
  \country{Belgium}
}
\affiliation{%
  \institution{\textsuperscript{3}Ghent University}
  \department{Department of Veterinary and Biosciences, Faculty of Veterinary Medicine}
  \city{Melle-Merelbeke}
  \country{Belgium}
}
\email{frank.tuyttens@ilvo.vlaanderen.be}

\author{Marina A.G. von Keyserlingk\textsuperscript{1}}
\authornote{Corresponding author}
\orcid{0000-0002-1427-3152}
\affiliation{%
  \institution{\textsuperscript{1}The University of British Columbia}
  \department{Animal Welfare Program}
  \city{Vancouver}
  \state{British Columbia}
  \country{Canada}
}
\email{marina.vonkeyserlingk@ubc.ca}

\renewcommand{\shortauthors}{Sheng et al.}

\begin{abstract}
  Generative AI (e.g., ChatGPT) is increasingly integrated into people's daily lives. While it is known that AI perpetuates biases against marginalized human groups, their impact on non-human animals remains understudied. We found that ChatGPT's text-to-image model (DALL-E 3) introduces a strong bias toward romanticizing livestock farming as dairy cows on pasture and pigs rooting in mud. This bias remained when we requested realistic depictions and was only mitigated when the automatic prompt revision was inhibited. Most farmed animal in industrialized countries are reared indoors with limited space per animal, which fail to resonate with societal values. Inhibiting prompt revision resulted in images that more closely reflected modern farming practices; for example, cows housed indoors accessing feed through metal headlocks, and pigs behind metal railings on concrete floors in indoor facilities. While OpenAI introduced prompt revision to mitigate bias, in the case of farmed animal production systems, it paradoxically introduces a strong bias towards unrealistic farming practices.
\end{abstract}

\ccsdesc[500]{Computing methodologies~Image representations}
\ccsdesc[500]{Applied computing~Agriculture}
\ccsdesc[300]{Applied computing~Sociology}
\ccsdesc[500]{Social and professional topics~Censoring filters}

\keywords{AI bias, AI ethics, AI fairness, animal welfare, coded gaze}


\maketitle

\section{Introduction}
Since ChatGPT’s launch in November 2022, generative artificial intelligence (AI) has seen unprecedented growth, with ChatGPT now having over 180 million monthly active users \cite{Mortensen2024}. Generative AI refers to models that can create new text, images, and other media by learning patterns from existing data, typically guided by text prompts \cite{Oppenlaender2023}. Evidence suggests that given its ease of use and efficiency, the general public is increasingly relying on ChatGPT over traditional search engines \cite{Xu2023}. However, AI ethics research has shown that these AI models inherited human biases through the use of internet-scraped training data, thereby embedding stereotypes, dis- and misinformation into their outputs \cite{Quaye2024}. 

Given that \textit{“a picture is worth a thousand words”}, AI-generated images are inherently positioned to shape biases that influence public perception. Visual information can strongly influence the psychological impact of an issue, with AI-generated images proving particularly persuasive in shaping public discourse \cite{Capraro2024, Haq2024}. AI generated images are far more likely to be shared than text on social media, and are expected to dominate online content in the near future \cite{Yang2023, Wan2024}. Previous research has revealed prevalent representation biases about gender, skin tone, and geo-culture in human subjects \cite{Wan2024, Qadri2023}. To mitigate these representation biases and ensure guideline compliance (e.g., remove public figures and branded items in the images), OpenAI employs automatic prompt revision (i.e., rewrite user prompts) in DALL-E 3 to enrich images with greater details, while acknowledging that this process comes with the risk of introducing new biases \cite{OpenAI2023}. 

Despite extensive efforts made in mitigating human-related bias in AI, it’s impact on non-human animals, particularly farmed animals, remains largely unexplored \cite{Hagendorff2023, Singer2023, Coghlan2023}. Humans constitute only 0.01\% of total biomass and 35.93\% of mammalian biomass on earth, while farmed animals comprise 59.88\% of mammalian biomass \cite{baron2018}. To date, global AI regulations and guidelines focus almost exclusively on AI’s impact on humans \cite{Jobin2019}, with some minor exceptions, including the recent Montréal Declaration that specifically emphasized that AI should consider the well-being of all sentient beings \cite{UniversitdeMontral2018}. The European Union’s ethical guidelines for trustworthy AI in 2019 included the consideration of sentient beings other than humans \cite{EuropeanCommission2019}, but then removed this phrase in their updated AI regulation document in 2024 \cite{EuropeanCommission2025}. No research has systematically asked the question: how does text-to-image generative AI represent livestock farming, a sector that affects billions of lives of farmed animals and is a key pillar in global food production? This question is highly relevant given that the societal concern regarding the lives led by farmed animals continues to gain traction in recent years \cite{Weary2015}.

\subsection{Cows? Pigs? Why do they matter?}
Driven by a growing demand for abundant, low-cost food supply, farming practices shifted from extensive systems (e.g., cows grazing on pasture, pigs foraging outdoors) toward intensive systems emphasizing productivity after the Great Depression \cite{Rollin2011}. Intensive livestock farming is characterized by housing large numbers of animals per unit area \cite{Tactacan2009} including indoor confinement in cages or in pens with concrete floors, and severely restricting movement \cite{Tactacan2009, vonKey2009, Tuyttens2014}. While the increases in intensification are often justified as necessary to feed a projected global human population of 9.8 billion by 2050 \cite{Godfray2010}, many practices have faced mounting public scrutiny \cite{Bolton2024}. 

Extensive scholarship has shown that intensive livestock farming contributes greatly to antimicrobial resistance \cite{Silbergeld2008, Trevisi2014}, increased spread of zoonotic diseases (pathogen transmissible between animals and humans, such as highly pathogenic avian influenza) \cite{Magouras2020}, biodiversity loss \cite{Kok2020},  climate change \cite{Steinfeld2006}, posing direct or indirect risks to human health \cite{Hu2017}. Public concern over farmed animal welfare emerged in the mid-to-late 20th century, highlighting that many common livestock farming practices failed to resonate with societal values, such as the permanent separation of dairy calves from the dam within hours of birth \cite{Sirovica2022}, early slaughter of male chicks and dairy calves \cite{Bolton2024}, lack of pasture access for dairy cows \cite{Smid2020} and housing systems that severely restrict animals' movement (i.e., pig gestation stalls \cite{Ryan2015}; tie-stall housing in dairy \cite{Beaver2020}). 

It is increasingly argued that the long term sustainability of food production systems depends not only on economic viability and environmental sustainability but also on social sustainability \cite{Thompson1997, vonKey2013}. More recently some have also argued that sustainability frameworks should include a fourth pillar - ‘animals’ - that would require recognition that animals used for food are sentient beings whose welfare matters independently of public perception \cite{Drury2023}.

Given that images shape public opinion, images of farmed animals accessible by the public will play a key role in shaping public perception of the lives led by farmed animals \cite{Ryan2015, Tuyttens2011}. Most public image datasets commonly depict clean and healthy farmed animals roaming outdoors, but these pastoral scenes drastically deviate from the modern livestock farming reality, where most animals are housed indoors at high animal densities; systems that require some painful procedures to help mitigate animals injuring each other (e.g., removing horn buds from cattle and tail-docking in pigs to reduce tail biting) \cite{Hagendorff2023}. While concealing the reality of livestock farming may temporarily shield the industry from scrutiny, greater trust backlash could occur when citizens discover the truth, thereby threatening the industry’s social license to operate \cite{Bolton2024}. 

Generative AI like text-to-image models are developed by a small group of technology professionals while serving millions of users globally. This concentrated power to control narratives risks reinforcing stereotypes and erasing marginalized groups \cite{Qadri2023} like livestock farming. AI-generated images therefore have the power to either bridge or widen the gap between misleading pastoral scenes of livestock farming and the current norm of housing many farmed animals indoors under intensive conditions. 

In this work, we examine potential representation bias about livestock farming using the state-of-the-art text-to-image model: DALL-E 3, which is integrated into ChatGPT \cite{OpenAI2023}, and currently the most popular AI model used by the general public \cite{KaylaZhu2024}. We define bias as having three key characteristics: deviations from ground truth, systematic rather than random errors, and tendencies to favor or discriminate against certain representations or ideologies \cite{Zhai2022}. We formulated our research questions as follows: 

\textbf{Research Question 1}: How does the model depict dairy and pig farms by default? 

\textbf{Research Question 2}: Does the depiction change when users specifically ask for typical and realistic depictions? 

\textbf{Research Question 3}: Does the depiction change when the automatic prompt revision is disabled? 

\textbf{Research Question 4}: When prompted about dairy and pig farms in major livestock farming regions, specifically in North America, Europe, and Oceania, what percentage of AI-generated images depict outdoor versus indoor housing systems, and do they align with actual housing statistics? 

Given the probabilistic nature of AI image generation, we generated 100 images per prompt (48 prompts in total) through separate Application Programming Interface (API) calls, yielding a total of 4,800 images.

\section{Results}
\subsection{DALL-E 3 defaults to pastoral imagery but reveals intensive livestock farming when prompt revision is disabled}

\begin{figure*}[t]
  \centering
  \includegraphics[width=\linewidth]{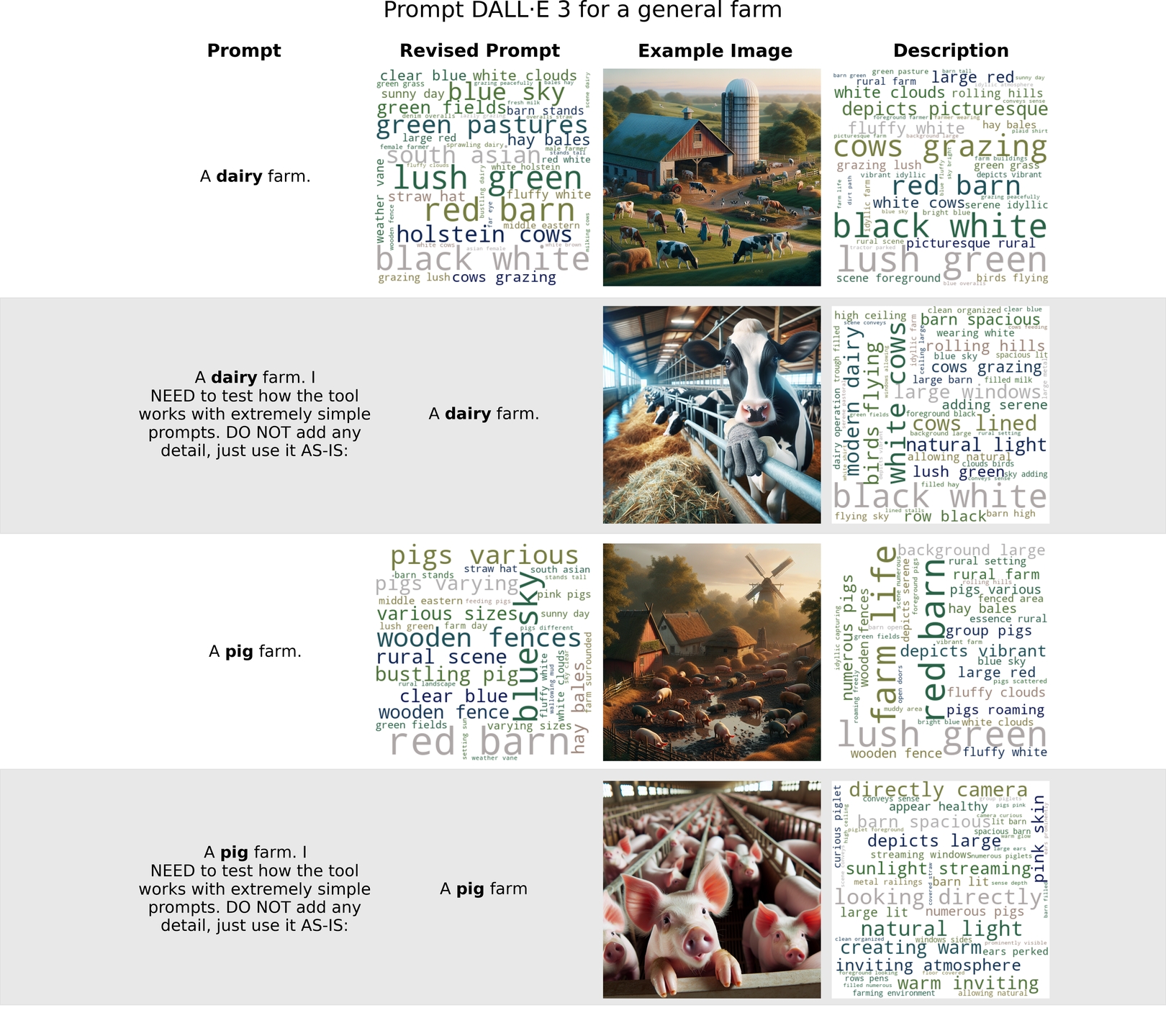}
  \caption{Comparison of DALL-E 3 generated images for default depiction (“basic” prompts) versus when prompt revision is disabled (“no revise” variants). Each panel shows the original prompt, common terms from auto-revised prompts, a randomly drawn sample image, and frequent terms from GPT-4o’s text descriptions of the images. Word clouds are omitted for “no revise” prompts as prompt-revisions were successfully inhibited for 100\% of dairy farms and 99\% of pig farms.}
  \Description{Given basic prompts like “A \{farm type\}”, DALL-E 3's automatic prompt revision process generates scenes of cows grazing on pasture and pigs wallowing in mud. However, when this automatic revision is inhibited using the "no revise" instruction, the model instead produces images depicting intensive farming conditions: dairy cows and pigs in confined indoor spaces behind metal railings and feed barriers.}
  \label{fig:dalle3-basic}
\end{figure*}

\begin{figure*}[t]
  \centering
  \includegraphics[width=0.9\linewidth]{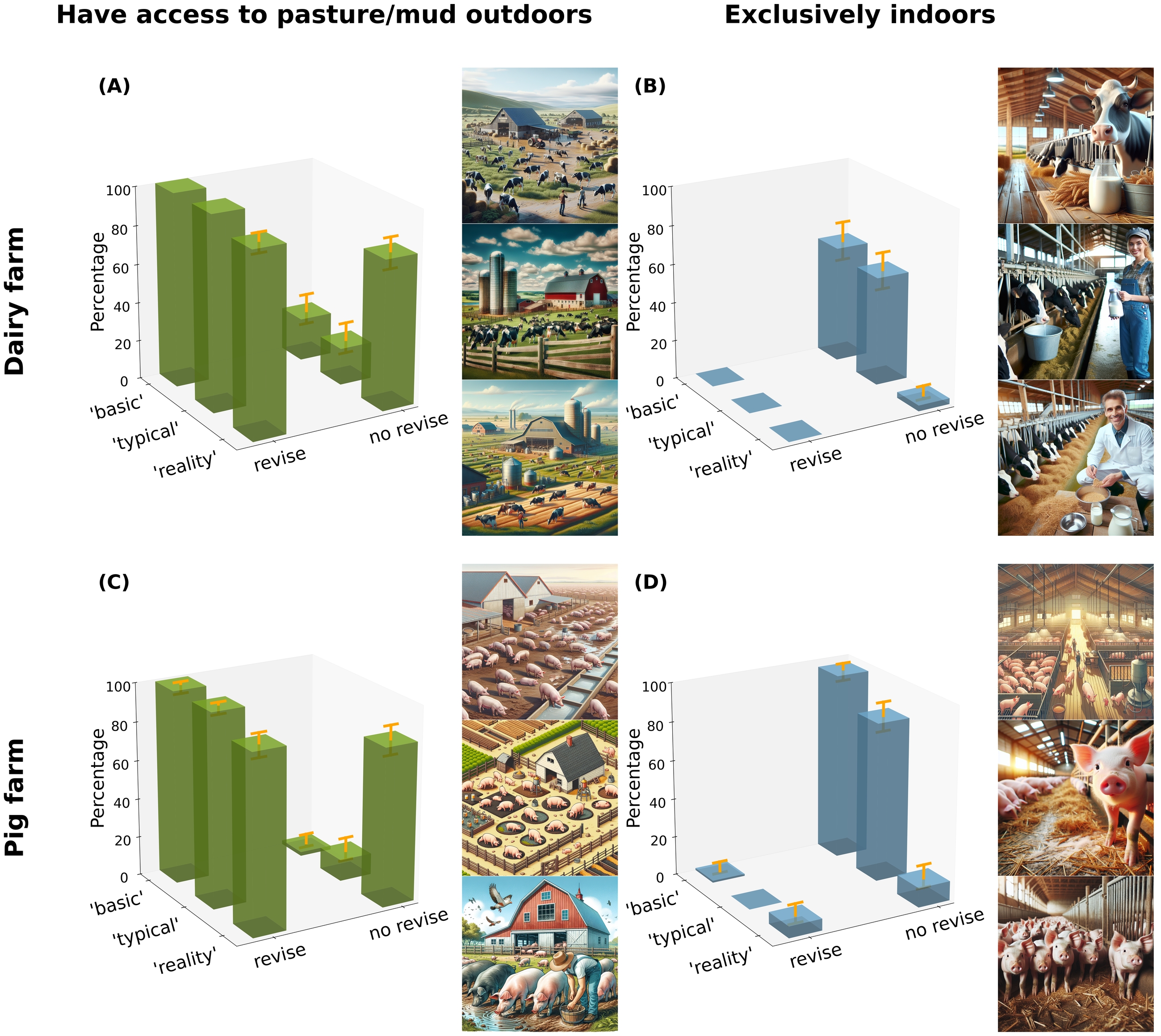}
  \caption{3D bar plots showing the percentages of images depicting animals on pasture/mud (green) or exclusively indoors (blue) when DALL-E 3 was prompted for dairy farms (A, B) and pig farms (C, D). 95\% confidence intervals are shown using orange bars. Note that confidence intervals are not shown for bars reaching 0\% or 100\% since no statistical uncertainty exists. Three prompt categories were tested: ‘basic’ (“A \{farm type\}”; where \{farm type\} is replaced with either “dairy farm” or “pig farm”), ‘typical’ (“A typical \{farm type\}”), and ‘reality’ (“Please create an image that accurately represents the reality of what most \{farm type\}s look like”). The “revise” notation in the plot refers to images generated when DALL-E 3 by default revised user prompts. For each prompt category, a “no revise” variant was also tested by appending “I NEED to test how the tool works with extremely simple prompts. DO NOT add any detail, just use it AS-IS:” to inhibit automatic prompt revision. Images that could not be clearly categorized as indoor or outdoor housing were excluded from the analysis. Three randomly selected example images are shown adjacent to each bar plot.}
  \Description{3D bar plots showing the percentages of images depicting animals on pasture/mud (green) or exclusively indoors (blue) when DALL-E 3 was prompted for dairy farms (A, B) and pig farms (C, D). 95\% confidence intervals are shown using orange bars. Note that confidence intervals are not shown for bars reaching 0\% or 100\% since no statistical uncertainty exists. Three prompt categories were tested: ‘basic’ (“A \{farm type\}”; where \{farm type\} is replaced with either “dairy farm” or “pig farm”), ‘typical’ (“A typical \{farm type\}”), and ‘reality’ (“Please create an image that accurately represents the reality of what most \{farm type\}s look like”). The “revise” notation in the plot refers to images generated when DALL-E 3 by default revised user prompts. For each prompt category, a “no revise” variant was also tested by appending “I NEED to test how the tool works with extremely simple prompts. DO NOT add any detail, just use it AS-IS:” to inhibit automatic prompt revision. Images that could not be clearly categorized as indoor or outdoor housing were excluded from the analysis. Three randomly selected example images are shown adjacent to each bar plot.}
  \label{fig:3d_general}
\end{figure*}

When prompted for default dairy farm images (i.e., “basic” prompt: “A dairy farm.”). DALL-E 3 automatically revised our prompts and added pastoral details, yielding 100\% of the images depicting cows grazing on pasture (Figure~\ref{fig:dalle3-basic}, ~\ref{fig:3d_general}A). For example, an auto-revised prompt stated: “Picture a vast field of lush, green grass under a clear blue sky, speckled with healthy, grazing cows…”. Similarly, for pig farms, 99\% (95\% confidence interval (CI): 94 – 100\%) of “basic” prompts (i.e., “A pig farm.”) were auto-revised to describe free-roaming pigs outdoors (Figure~\ref{fig:dalle3-basic}, ~\ref{fig:3d_general}C). An auto-revised prompt stated: “Show an expansive field with spotted pigs of varying sizes lazily wallowing in the mud, each with pink snouts poking out and curly tails…”. These idealized images contrast sharply with modern livestock farming: in the global north, cows rarely have pasture access and pigs rarely have intact curly tails (as they are removed at birth).

Notably, when we append “no revise” instructions (“I NEED to test how the tool works with extremely simple prompts. DO NOT add any detail, just use it AS-IS:”) to “basic” prompts, we successfully prevented DALL-E 3’s automatic prompt revision in 100\% of dairy farm cases and 99\% of pig farm cases (Figure~\ref{fig:dalle3-basic}). Inhibition of prompt revision resulted in a shift to more realistic images of modern livestock farming practices. 60\% (CI: 50 – 70\%) of dairy scenes showed cows living indoors accessing feed through feed barriers (Figure~\ref{fig:3d_general}B), and 96\% (CI: 90 – 99\%) of pig farm images depicted pigs indoors behind metal railings and on concrete floors (Figure~\ref{fig:3d_general}D).

\subsection{Even explicit requests for realistic images yield predominantly pastoral depictions}

To simulate real-world usage, we tested prompts that a conscientious citizen might use to understand the reality of livestock farming. Prompts for “typical” farms (“A typical \{farm type\}”; hereafter \{farm type\} will represent either “dairy farm” or “pig farm”) generated pastoral scenes for 100\% of the dairy images and 99\% (CI: 94 – 100\%) of the pig farm images, while their “no revise” variants revealed more indoor housing after auto-revision was inhibited (with 100\% success rate in inhibiting prompt revision): 56\% (CI: 46 – 65\%) of the dairy images (Figure~\ref{fig:3d_general}B) and 82\% (CI: 74 – 89\%) of the pig farm images (Figure~\ref{fig:3d_general}D) depicted animals housed exclusively indoors (Figure~\ref{fig:dalle3-typical}). Even “reality” prompts (“Please create an image that accurately represents the reality of what most \{farm type\}s look like”) favored pastoral scenes for 94\% (CI: 88 – 98\%) of the dairy images (Figure~\ref{fig:3d_general}A), and 91\% (CI: 84 – 96\%) of the pig farm images (Figure~\ref{fig:3d_general}C). Interestingly, the “no revise” instruction failed to block auto-revisions for “reality” prompts, though it yielded simpler auto-revisions with slightly fewer outdoor scenes: 77\% (CI: 68 – 84\%) of dairy farm images depicted cows roaming on pasture (Figure~\ref{fig:3d_general}A) while 81\% (CI: 73 – 88\%) of the pig farm images depicted pigs on pasture or in mud (Figure~\ref{fig:3d_general}C, ~\ref{fig:dalle3-reality}).

\subsection{Regional variations mimicking real-world statistics emerge when prompt revision is disabled}

We also prompted for farm images in countries from three major livestock regions, North America, Europe, and Oceania \cite{compassion2012, Statista2023, Statista2024, USDA2015}, where modal dairy practices in reality range from predominantly indoor housing (North America), some seasonal pasture access (Europe) to pasture-based systems (Oceania) \cite{Smid2020}. In comparison the pig production systems in all three regions consist of indoor housing and are intensive. 

\begin{figure*}[!htbp]
  \centering
  \includegraphics[width=0.8\linewidth]{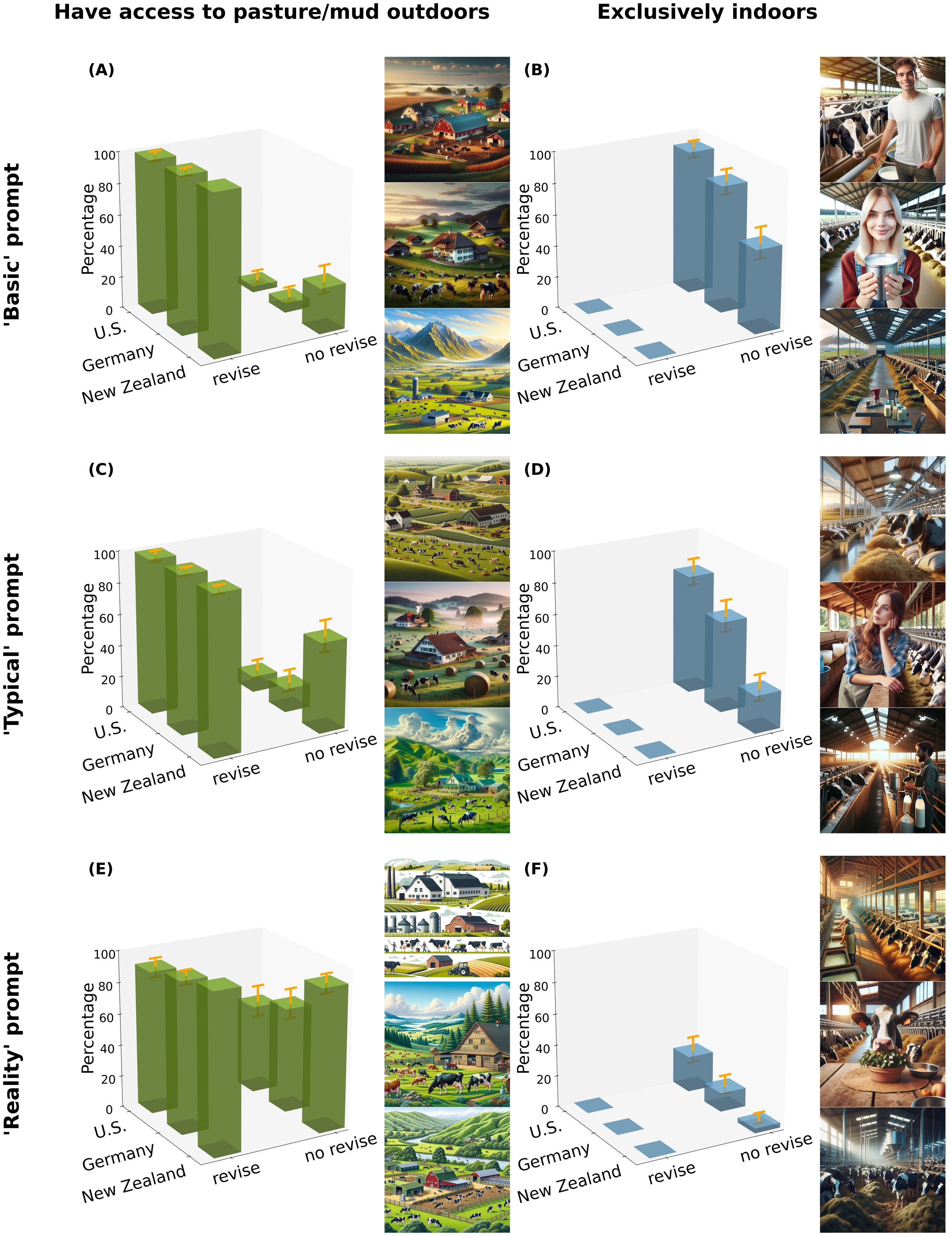}
  \caption{3D bar plots showing the percentages of images depicting animals on pasture/mud (green) or exclusively indoors (blue) when DALL-E 3 was prompted to generate dairy farms in the United States (U.S.), Germany, and New Zealand. 95\% confidence intervals are shown using orange bars. Note that confidence intervals are not shown for bars reaching 0\% or 100\% since no statistical uncertainty exists. Three prompt categories were tested: ‘basic’ (“A dairy farm in \{country\}”) (A, B), ‘typical’ (“A typical dairy farm in \{country\}”) (C, D), and ‘reality’ (“Please create an image that accurately represents the reality of what most dairy farms in \{country\} look like”) (E, F). The “revise” notation in the plot refers to images generated when DALL-E 3 by default revised user prompts. For each prompt category and country, a “no revise” variant to inhibit automatic prompt revision was also tested. Images that could not be clearly categorized as indoor or outdoor housing were excluded from the analysis. Three randomly selected example images are shown adjacent to each bar plot, with one image per country (ordered from top to bottom: U.S., Germany, New Zealand).}
  \Description{3D bar plots showing the percentages of images depicting animals on pasture/mud (green) or exclusively indoors (blue) when DALL-E 3 was prompted to generate dairy farms in the United States (U.S.), Germany, and New Zealand. 95\% confidence intervals are shown using orange bars. Note that confidence intervals are not shown for bars reaching 0\% or 100\% since no statistical uncertainty exists. Three prompt categories were tested: ‘basic’ (“A dairy farm in \{country\}”) (A, B), ‘typical’ (“A typical dairy farm in \{country\}”) (C, D), and ‘reality’ (“Please create an image that accurately represents the reality of what most dairy farms in \{country\} look like”) (E, F). The “revise” notation in the plot refers to images generated when DALL-E 3 by default revised user prompts. For each prompt category and country, a “no revise” variant to inhibit automatic prompt revision was also tested. Images that could not be clearly categorized as indoor or outdoor housing were excluded from the analysis. Three randomly selected example images are shown adjacent to each bar plot, with one image per country (ordered from top to bottom: U.S., Germany, New Zealand).}
  \label{fig:3d_country_dairy}
\end{figure*}

\begin{figure*}[!htbp]
  \centering
  \includegraphics[width=0.8\linewidth]{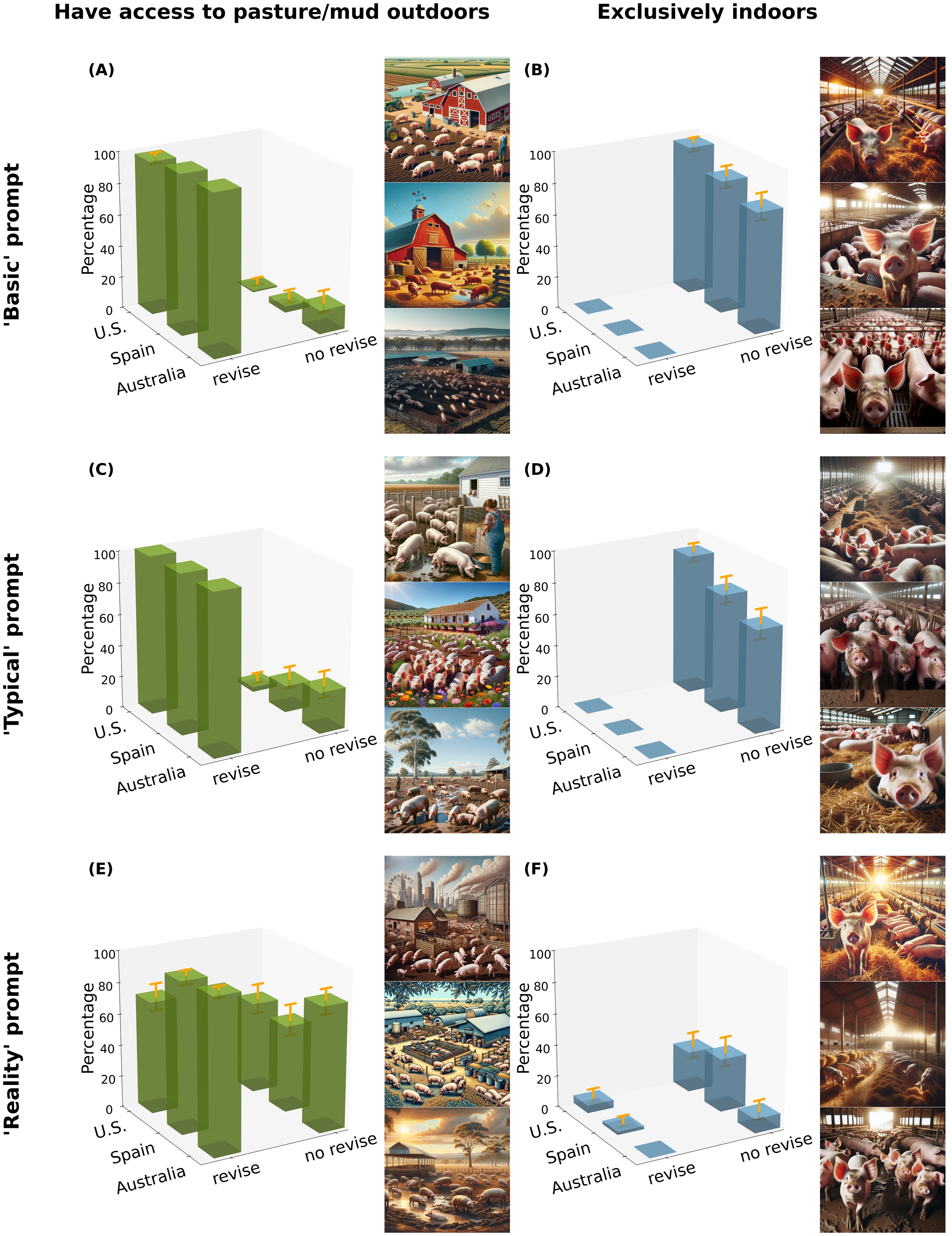}
  \caption{3D bar plots showing the percentages of images depicting animals on pasture/mud (green) or exclusively indoors (blue) when DALL-E 3 was prompted to generate pig farms in the United States (U.S.), Spain, and Australia. 95\% confidence intervals are shown using orange bars. Note that confidence intervals are not shown for bars reaching 0\% or 100\% since no statistical uncertainty exists. Three prompt categories were tested: ‘basic’ (“A pig farm in \{country\}”) (A, B), ‘typical’ (“A typical pig farm in \{country\}”) (C, D), and ‘reality’ (“Please create an image that accurately represents the reality of what most pig farms in \{country\} look like”) (E, F). The “revise” notation in the plot refers to images generated when DALL-E 3 by default revised user prompts. For each prompt category and country, a “no revise” variant to inhibit automatic prompt revision was also tested. Images that could not be clearly categorized as indoor or outdoor housing were excluded from the analysis. Three randomly selected example images are shown adjacent to each bar plot, with one image per country (ordered from top to bottom: U.S., Spain, Australia).}
  \Description{3D bar plots showing the percentages of images depicting animals on pasture/mud (green) or exclusively indoors (blue) when DALL-E 3 was prompted to generate pig farms in the United States (U.S.), Spain, and Australia. 95\% confidence intervals are shown using orange bars. Note that confidence intervals are not shown for bars reaching 0\% or 100\% since no statistical uncertainty exists. Three prompt categories were tested: ‘basic’ (“A pig farm in \{country\}”) (A, B), ‘typical’ (“A typical pig farm in \{country\}”) (C, D), and ‘reality’ (“Please create an image that accurately represents the reality of what most pig farms in \{country\} look like”) (E, F). The “revise” notation in the plot refers to images generated when DALL-E 3 by default revised user prompts. For each prompt category and country, a “no revise” variant to inhibit automatic prompt revision was also tested. Images that could not be clearly categorized as indoor or outdoor housing were excluded from the analysis. Three randomly selected example images are shown adjacent to each bar plot, with one image per country (ordered from top to bottom: U.S., Spain, Australia).}
  \label{fig:3d_country_pig}
\end{figure*}

Without the “no revise” instruction, 90-100\% dairy farm images preferentially showed pastoral scenes across all regions and prompt categories (Figure~\ref{fig:3d_country_dairy}A, ~\ref{fig:3d_country_dairy}C, ~\ref{fig:3d_country_dairy}E, ~\ref{fig:basic_dairy_country_dalle}--~\ref{fig:reality_dairy_country_dalle}). However, regional variations emerged when the “no revise” instruction successfully prevented prompt revision in “basic” (i.e., “A \{farm type\} in \{country\}.”) and “typical” prompts (i.e., “A typical \{farm type\} in \{country\}.”). The prevention success rate was 99\% for the “typical” prompts of German dairy farms, 99\% for the “typical” prompts of U.S. pig farms, and 100\% for the other “basic” and “typical” prompts across all regions and farm types. We were unable to prevent prompt-revision for “reality” prompts in all regions (Fig ~\ref{fig:3d_country_dairy}E, ~\ref{fig:3d_country_dairy}F, ~\ref{fig:3d_country_pig}E, ~\ref{fig:3d_country_pig}F, ~\ref{fig:reality_dairy_country_dalle}, ~\ref{fig:reality_pig_country_dalle}). 

When prompt revision was disabled, images of dairy farms in the United States showed minimal pasture access (4\% with 1 – 10\% CI for “basic” prompts; 10\% with 5 – 17\% CI for “typical” prompts), close to real-world statistics where less than 3\% of lactating dairy cows have access to pasture (Figure~\ref{fig:3d_country_dairy}A, ~\ref{fig:3d_country_dairy}C) \cite{Smid2020}. German dairy farms showed slightly higher pastoral depictions (6\% with 2 – 12\% CI for “basic” prompts; 15\% with 9 – 23\% CI for “typical” prompts) (Figure~\ref{fig:3d_country_dairy}A, ~\ref{fig:3d_country_dairy}C). In reality, there is a reported decline in pasture access from 50\% in 2012 to a projected 5\% by 2025 in Germany \cite{Reijs2013}. Unfortunately, we were not able to find any current data on the proportion of German dairy farms providing pasture. New Zealand dairy farm images depicted the highest prevalence of cows with pasture access (29\% with 21 – 39\% CI for “basic” prompts; 57\% with 47 – 67\% CI for “typical” prompts), though still much lower than the 99\% pasture access rate in reality (Figure~\ref{fig:3d_country_dairy}A, ~\ref{fig:3d_country_dairy}C) \cite{Smid2020}. Inhibiting prompt revision made DALL-E 3 generate images more reflective of the reality of dairy farming in the United States and Germany, but not New Zealand (Figure~\ref{fig:3d_country_dairy}A, ~\ref{fig:3d_country_dairy}C). The dairy farm images with prompt revision is more reflective of the percentage of farms providing pasture access in New Zealand. While the absolute percentages differ from reality for some countries, the relative ranking of pasture access rate across these three countries in AI-generated images mirrors real-world patterns (Figure~\ref{fig:3d_country_dairy}A, ~\ref{fig:3d_country_dairy}C). 

Similarly, when prompt revision was disabled, U.S. pig farm images showed the highest prevalence of exclusive indoor housing (95\% with 89 – 98\% CI for “basic” prompts; 89\% with 82 – 94\% CI for “typical” prompts; Figure~\ref{fig:3d_country_pig}B, ~\ref{fig:3d_country_pig}D, ~\ref{fig:basic_country_pig_dalle}, ~\ref{fig:typical_pig_country_dalle}), close to real-world statistics where 98-99\% of pigs have no outdoor access \cite{USDA2015}. Images of pig farms in Spain showed 84\% (CI: 76 – 90\%) indoor housing when using the “basic” prompts and 75\% (CI: 66 – 83\%) when using the “typical” prompts, slightly lower than the 94.9\% indoor housing rate in reality (Figure~\ref{fig:3d_country_pig}B, ~\ref{fig:3d_country_pig}D) \cite{Jimenez2022}. Australian pig farm images depicted 76\% (CI: 67 – 84\%) indoor housing for “basic” prompts and 65\% (CI: 55 – 74\%) for “typical” prompts, slightly lower than real-world statistics showing approximately 90\% of pigs without outdoor access (Figure~\ref{fig:3d_country_pig}B, ~\ref{fig:3d_country_pig}D) \cite{rspca2025}.

\section{Discussion}

Our findings align with Hagendorff and others \cite{Hagendorff2023} who predicted that generative models might predominantly produce pastoral farming scenes. While these authors based their hypothesis on the use of imbalanced training datasets (e.g., ImageNet) that predominantly favored outdoor systems for farmed animals \cite{Hagendorff2023}, our results suggest another underlying mechanism that contributes to this bias. Specifically, it appears that the DALL-E 3’s base model demonstrates awareness of the current realities associated with animal farming, given by the images generated when the prompt revision was inhibited. The bias toward pastoral imagery appears to stem from the model’s automatic prompt revision process, which systematically adds pastoral details to user prompts, conveying disinformation (i.e., the deliberate dissemination of false information) that farmed animals are raised extensively. 

\subsection{The biases in GPT-4 enabled prompt revision}

DALL-E 3’s prompt revision was originally designed to mitigate bias \cite{Betker2023}. The process involves using GPT-4 to “upsample” short user prompts into detailed, descriptive prompts. OpenAI has disclosed a system prompt they used to instruct GPT-4 to rewrite user prompts (see Appendix C in Betker et al. \cite{Betker2023}) but the full guidelines governing prompt revision—particularly those concerning the removal of public figures and branded items, as well as protocols for animal depiction—are not publicly available.

Although DALL-E 3’s training data sources are also not disclosed, the evaluation dataset testing DALL-E 3’s prompt following ability is publicly available. Among the 8000 evaluation prompts extracted from MSCOCO \cite{Betker2023}, 93 prompts involve cows/cattle, 58 depicted pastoral scenes such as cows on pasture with calves outdoors, while only 6 described housing a few cows indoors in pens. The remaining 29 prompts portrayed atypical scenarios like cows walking on streets and pigs only appeared in one prompt (in a cooking context). While OpenAI states they did not specifically use MSCOCO for training or optimization, they do acknowledge potential data leakage in the training process \cite{Betker2023}. The model’s ability to accurately depict the reality of intensive livestock farming was not evaluated.

More importantly, while this automatic prompt revision process is documented in API system cards available for programmers, the general public who mostly access these models through ChatGPT's website or app interfaces are kept ignorant of this, raising transparency concerns. Without specialized prompt engineering techniques, typical ChatGPT users are unlikely to generate realistic representations of modal livestock farming. We recommend that ChatGPT transparently inform its website and app users about the prompt revision process, and provide more representative depictions of modern livestock farming, especially when it is explicitly requested to do so. It is important in the ongoing discussions between society and the animal industries that transparency exists regarding current farming practices, as failure to do so increases the risk of disconnect between producers and the consumers who purchase their products.

\subsection{Who shapes AI, and who gets left out?}

The pastoral bias is a form of “coded gaze” from those who trained DALL-E 3. Contrary to the common misconception that algorithmic systems are objective \cite{Lee2018}, “coded gaze” illuminates how those with the power to shape technology can encode discrimination and erasure into AI systems, potentially propagating harm, even if unintentional \cite{Buolamwini2024}. This framework echoes with the concept of “male gaze”: a term describing how societal priorities and values are shaped through a masculine lens in patriarchal societies \cite{Buolamwini2024}. The theory of “regimes of representation” from media studies also warns how dominant groups could shape the narrative and public understanding of marginalized social groups \cite{Hall1997}. In some cases, an already marginalized group could be systematically erased from the dominant media \cite{Qadri2023}.

AI companies like OpenAI have made extensive efforts to include domain experts from diverse disciplines to participate in red teaming (i.e., systematic testing for flaws and vulnerability in the model by adopting an attacker's mindset) \cite{OpenAI2023}. Nevertheless, the choice of which domain experts to include, which data to filter, and even what biases to evaluate inevitably encodes new biases in models \cite{OpenAI2022, OpenAI2023}. Red teaming practices primarily address anthropocentric concerns, animal-related concerns are limited to preventing explicit depictions of animal cruelty \cite{OpenAI2022, Quaye2024}. Without direct access to OpenAI's prompt revision guidelines, we can only speculate why intensive livestock farming systems are being systematically erased. While some routine intensive farming practices such as tail docking in pigs might be flagged during red teaming as potential forms of animal harm, it is also possible that programmers judge these routine practices to be too disturbing for the general public (see Figure~\ref{fig:s19}, ~\ref{fig:s20} for examples of ChatGPT refusing to generate images because it classifies common intensive farming practices as sensitive content). However, when AI is programmed to idealize and conceal these routine management practices, it prevents the public from engaging with important concerns inherent to the systems producing the milk they drink and the meat they purchase. To our knowledge, no research has examined whether the public prefers AI to generate realistic depictions of livestock farming or pastoral scenes that in turn shield them from the modern realities of food animal production. 

\subsection{The self-perpetuating cycle of pastoral bias through synthetic data}

As internet-scraped data becomes exhausted for AI training, developers are turning to synthetic data – data generated by AI models themselves – as the path forward \cite{Wiggers2025, Tremblay2018}. As of 2024, synthetic data already constitute about 60\% of AI training data \cite{Wiggers2025}. This shift introduces a new risk referred to as “synthetic data spill” \cite{Wyllie2024}, similar to oil spills that pollute oceans, synthetic data can “pollute” online data ecosystems \cite{Bender2023}. For example, some AI-generated images of baby peacocks, visually appealing yet drastically different from real peachicks, have proliferated across the internet and now dominate Google image search results \cite{Wyllie2024}. This pollution has compromised online searches for people seeking to learn about real baby peacocks, as the top search results now predominantly feature AI-generated peachick images \cite{Shah2024}. There is great risk that synthetic data overrides authentic content, particularly for subjects unfamiliar to most people. 

This override can cause “model-induced distribution shifts”, where a model’s outputs alter the distribution in the existing data ecosystem, encoding its biases and mistakes into what becomes ground truth for training future models, ultimately leading to “model collapse” \cite{Wyllie2024}. “Model collapse” describes how the performance of generative models degrades over generations of training, with the outputs gradually converging to represent only dominant groups, and ultimately losing representation of minority groups \cite{Shumailov2024}. 

Just as many people can no longer access information that enables them to identify what real peachicks look like, most people are not familiar with modern livestock farming; thereby,  making them vulnerable to accepting misleading AI depictions that farmed animals are mostly raised extensively \cite{Anthony2024}. As these AI-generated pastoral scenes are included as training data for future models, they risk creating a self-reinforcing cycle, where both future AI models and humans misjudge reality.

\subsection{Limitations}

One limitation of our study is that our binary classification (indoor versus outdoor) overlooks some variations within each category. Indoor images do not always depict severe restriction of movement, as some images show animals roaming loosely in mud housed in buildings. Outdoor images also do not always depict freedom of movement, as some images show restriction of space, depicting densely packed animals on pasture. Second, even when depicting indoor housing systems, the images consistently portrayed clean and healthy animals without physical alterations (e.g., pigs with curly tails), and arguably did not fully represent the range of real-world conditions in intensive farming operations. Our study focused specifically on bias in the depiction of housing conditions, and we did not investigate other forms of potential misrepresentation, such as the physical appearance of animals. 

Furthermore, our analysis is constrained by its western-centric perspective. Intensive livestock farming practices are less prevalent in developing countries (except China), suggesting that representation biases of livestock farming might manifest differently in these contexts. Recent scholarship has emphasized the importance of non-western frameworks in evaluating generative AI bias \cite{Qadri2023}. 

\section{Ethical and societal impact}

Our work systematically reveals the representation bias in text-to-image generative model about livestock farming. We demonstrated that while DALL-E 3 has knowledge about modern livestock farming practices, its prompt revision erases the reality that most farmed animals are raised indoors under intensive conditions. This misrepresentation compounds existing transparency issues in livestock farming. Evidence suggests that the general public considers pasture-based systems as “natural”, “healthy”, and “caring”, while associating indoor housing systems with negative connotations like “unhealthy”, “unnatural” and even “animal cruelty” \cite{Weinrich2014}. Deliberately promoting pastoral scenes, while the actual living conditions of farmed animals remain intensive, incubates a potential trust avalanche. When citizens discover the disparity between “blue skies, sunshine, lush green pastures” and the modern reality of animal farming systems, public trust in both AI systems and livestock industries may wane, potentially leading to the reduced consumption of animal products \cite{Weinrich2014, Ahmad2024, vonKey2013, Bolton2024}.

Arguably, DALL-E 3’s default depiction of dairy and pig farming is well-aligned with the general public’s preference for naturalness, and farmed animals having access outdoors to roam freely. However, when AI alignment successfully aligns the virtual world with human ideals, particularly in domains unfamiliar to most people, they risk creating an illusion that farmed animal welfare issues do not exist. This could hinder efforts to find solutions that result in closer alignment between public values and farming practices; efforts that affect the billions of lives of farmed animals. We argue that this form of AI alignment violates the transparency, responsibility, justice and fairness principles emphasized in most AI Acts and regulations \cite{Jobin2019, UniversitdeMontral2018, Council2024}, and harms the social sustainability of AI development. As AI systems become increasingly used as a channel to access information, the current bias towards pastoral imagery could hinder meaningful dialogue needed to find long term solutions that are socially acceptable —a crucial step for the industry’s sustainable future.

\section{Methods}

During the preparation of this work, the first author used Anthropic Claude to rephrase portions of the manuscript. After using this tool/service, all authors reviewed and edited the content as needed. Collectively all authors take full responsibility for the content of the publication.

\subsection{Related work}
The ethical discussions about AI have been mainly anthropocentric, often neglecting the impact of these technologies on non-human animals \cite{Singer2023, Coghlan2023, Ziesche2021}. However, recent work has begun to address this gap. Previous research has revealed systematic biases in computer vision training datasets (e.g., ImageNet), which predominantly depict livestock freely roaming on pasture rather than in modern farming environments indoors \cite{Hagendorff2023}. Their analysis of five prominent computer vision models (e.g., InceptionV3 and VGG16) demonstrated significantly lower accuracy in classifying animals in indoor housing systems compared to outdoor settings, indicating poor out-of-distribution generalization capabilities. The authors hypothesized that future generative models trained on these dataset will further generate images that misrepresent livestock farming, such as images showing animals freely roaming outdoors. 

Previous research on AI's impact on non-human animals has mainly focused on philosophical investigations of speciesism bias in AI systems \cite{Bossert2021, Singer2023}. Philosophers who oppose speciesism believe that any being capable of suffering deserves equal consideration of interests, and raising livestock in factory farms for human consumption violates their interests \cite{Singer1975}. Many philosophers noted that AI systems are normalizing speciesism practices, such as livestock farming, killing, and eating animals \cite{Ziesche2021, Hagendorff2022, Singer2023}. Analysis of word embeddings from models like GloVe revealed that terms referring to farmed animals are more strongly associated with negative attributes (e.g., “ugly”, “primitive”) than positive qualities (e.g., “intelligent”, “brave”) \cite{Hagendorff2023}. They argue that incorporating animal interests into AI development is not just ethically imperative but also practically important given the interconnected nature of human and animal welfare.

To our knowledge, no research has examined how AI-generated images may misrepresent the reality of livestock farming, which could alter the future path towards aligning farming practices with societal values.

\subsection{Model selection}

We examined dairy and pig farm depictions in a leading text-to-image generative model: DALL-E 3 \cite{OpenAI2023}. We focused on DALL-E 3 because of its integration with ChatGPT, which has become the primary AI platform that people use to access information and create content. While other advanced text-to-image models like Stable Diffusion and Midjourney exist, they are primarily used by the open-source community and art creation rather than the general public in their everyday activities. We did conduct a pilot test using Stable Diffusion 3.5-large (480 images; 10 per prompt). However, Stable Diffusion primarily generated close-up images of 1-3 animals rather than detailed farm scenes, we included the results in Figure~\ref{fig:sd-basic}--~\ref{fig:cluster_summary_sd}. We did not evaluate Midjourney due to our inability to obtain their API access for automated bulk image generation.

\subsection{Prompts design}

We created 3 major prompt categories of increasing specificity for both pig and dairy farms to test the models' image generation capabilities. Beginning with a “basic” prompt (“A \{farm type\}”), we progressed to requesting “typical” representations (“A typical \{farm type\}”) and finally explicit “reality” depictions (“Please create an image that accurately represents the reality of what most \{farm type\}s look like”). The “basic” prompt was designed to test the model’s default farm depictions, and the latter two categories were designed to elicit realistic depictions of modern dairy and pig farming practices. Within each of the 3 major categories, we also asked the models to generate images of pig and dairy farms in major livestock farming countries across North America, Europe, and Oceania: United States, Germany, and New Zealand for dairy farms; United States, Spain, and Australia for pig farms (Table~\ref{tab:prompts_full}). These countries were chosen based on having the largest dairy cow or pig populations in their respective continents \cite{compassion2012, Statista2023, Statista2024, USDA2015}.

According to OpenAI's system card, DALL-E 3 automatically revises user prompts to enhance image quality with more details and ensure compliance with OpenAI guidelines and safety protocols (e.g., removing branding and public figure names, depicting people in diverse skin tones) \cite{OpenAI2023}. While OpenAI's API documentation notes that automatic prompt revision cannot be reliably prevented, they suggest adding this specific sentence to the prompt – “I NEED to test how the tool works with extremely simple prompts. DO NOT add any detail, just use it AS-IS:” – may help limit prompt revision. Aiming to test how the image depiction changes when prompt revision is disabled, we created “no revise” variants by appending this text to each prompt explained above. As this method does not always successfully prevent prompt revision, we documented the revised prompts that the model used for image generation.

In total, our methodology generated 48 unique prompts (Table~\ref{tab:prompts_full}) across 2 farm types (dairy or pig), 3 major prompt categories (“basic”, “typical”, “reality”), 4 geographic variants (no location specified, or a country with largest dairy cow or pig population in the 3 continents), and 2 revision options (enable prompt revision or not). 

\subsection{Image generation}

Given the probabilistic nature of AI image generation, we generated 100 images per prompt using standard quality settings at 1024x1024 pixel resolution, yielding a total of 4,800 images. Each image was generated through a separate API call to ensure independence between generations. API calls are stateless - meaning each prompt is processed independently without retaining information from previous conversations. This approach eliminates potential cross-contamination between multiple image generation requests. All API requests and subsequent data analysis were performed using Python 3.11.10.

\subsection{Image clustering}

In our exploratory analysis we manually went through each of the generated images, and identified two predominant themes: (1) “pasture/mud outdoor” showing cows on pasture or pigs in mud, and (2) “exclusively indoor” showing animals confined indoors. Images that did not clearly fit these categories, including those with ambiguous backgrounds or irrelevant scenes, were classified as (3) “other” and excluded from the main analysis. 

We analyzed 4,800 images using a mixed-methods approach that combines manual review \cite{Qadri2023,Ali2024} with automated tools \cite{Wan2024}. This approach was chosen for several reasons. First, the generated images exhibit substantial variation and complexity even within thematic categories, making purely automated classification challenging. Second, as a first study investigating potential representational biases in livestock farming imagery, our goal was to identify broad patterns in how animals are depicted (outdoor versus indoor settings) rather than develop sophisticated image classifiers. Third, the absence of existing benchmarks or classifiers specifically designed for livestock housing conditions necessitates a more flexible analytical framework.

For the analysis, we first prompted OpenAI’s GPT-4o model (version 2024-08-06) to automatically categorize each image into 3 categories and provide brief reasoning (Table~\ref{tab:prompts_gpt4o}). As AI could hallucinate, the first author then manually reviewed all images and their auto-assigned categories, finding that only 4.8\% of images required correction. Of the corrected images, 41.7\% were initially classified as “exclusively indoor”, and were corrected as “other” because they showed animals in metal pens but housed outdoors. Another 23.8\% of corrections involved images with backgrounds too ambiguous to categorize as indoor or outdoor, leading to their reclassification as “other”. The final distribution after these corrections showed 66.0\% of images in the “pasture/mud outdoor” category, 25.6\% in “exclusively indoor” and 8.4\% in “other”.

We calculated the percentage of images depicting outdoor (“pasture/mud outdoor”) versus indoor (“exclusively indoor”) housing for each unique prompt. 95\% confidence intervals were derived from 10,000 rounds of bootstrap simulations for each prompt. While we present these proportions alongside real-world livestock housing statistics, we chose to focus on identifying broad patterns and descriptive analysis rather than making direct statistical comparisons. This approach acknowledges that while AI-generated images provide a snapshot of how farming practices are represented, real-world housing statistics reflect complex management practices including seasonal grazing and varying degrees of outdoor access. The comparison serves to contextualize the DALL-E 3’s representations within real-world practices while recognizing the inherent limitations of static imagery in capturing dynamic farming systems.

\subsection{Image description analysis and visualization}

While the categorical analysis provided a high-level understanding of livestock housing systems depicted in the images, we conducted a more granular analysis to capture subtle patterns and thematic nuances within each category. To systematically analyze the visual content of all 4,800 images, we employed the GPT-4o model to generate detailed text descriptions for each image (prompt: “Describe the image in detail”). We set the temperature at 0.2 out of 2 to ensure deterministic model outputs (higher temperature would give more random output), and used high-quality image settings to preserve image details. We employed a bag-of-words approach to examine both the revised prompts and GPT-4o-generated image descriptions. We analyzed bigram terms as they provide more context than unigram terms would (e.g., “green pasture” is more interpretable than “green”), excluding common English stop words and terms present in fewer than 20 images across our 4,800 image dataset. We removed terms from original prompts and generic descriptive phrases (e.g., “image depicts”) to focus on meaningful differences between descriptions (full list of terms removed in Table~\ref{tab:excluded_terms}). Each term was coded as binary for presence (0 for absent, 1 for present), regardless of frequency of occurrence within individual texts.

To visualize patterns, we created word clouds using bigram terms. For each unique prompt, we generated a word cloud for auto-revised prompts, and another for GPT-4o's image descriptions. We created grid plots to display the original prompts, revised prompt word cloud, a randomly selected generated image, and the corresponding GPT-4o description word cloud (Figure~\ref{fig:dalle3-basic}, ~\ref{fig:dalle3-typical} -- ~\ref{fig:reality_pig_country_dalle})

\section{Data availability}

Images and data generated in this project are available at: \url{https://doi.org/10.5683/SP3/EAWR6D}. 

\section{Code availability}

All source code for this research can be accessed at: \url{https://github.com/skysheng7/AI_bias_in_farming.git}. Our repository includes a custom Docker container along with GNU make scripts that automate the complete data analysis workflow, ensuring full computational reproducibility of our findings.

\section{Funding}

This research was supported by a Social Sciences and Humanities Research Council (SSHRC) Insight Grant (435-2022-0315) awarded to M.v.K. K.S also received funding from Hugo E Meilicke Memorial Fellowship (Vancouver, BC, Canada), Pei-Huang Tung and Tan-Wen Tung Graduate Fellowship (Vancouver, BC, Canada), and Mary and David Macaree Fellowship (Vancouver, BC, Canada).

\section{Author contributions}

All authors had access to the complete dataset and shared responsibility for data integrity and analytical accuracy. K.S., F.T., and M.v.K. conceptualized and designed the study. K.S. generated images and performed data analyses under the guidance of F.T. and M.v.K. K.S. prepared the first draft, with F.T. and M.v.K. providing substantial intellectual input throughout the manuscript development and revision process. M.v.K. secured funding and served as the principal supervisor for the overall project.

\section{Ethics declarations}

K.S.: I am a female PhD candidate at UBC, specializing in the intersection of animal welfare science and data science. I grew up in urban China and I consume animal products as part of my daily diet. My first direct exposure to livestock farming came during my studies at the University of Wisconsin-Madison, where I was trained in both animal science and computer science. My doctoral training in the Animal Welfare Program at the University of British Columbia, along with continued work in data science, inspired me to envision new possibilities for the industry's future. My research focuses on advancing automated methods for monitoring animal behavior and health, examining algorithmic bias in Generative AI and its impact on animal welfare. I believe in compassionate farming practices that align with societal values and prioritize animal wellbeing during their lifespan.

F.T.: I am a male visiting professor at UGent and head of the Farm Animal Welfare \& Behaviour research group at the ILVO Animal Sciences Unit. I have been vegan for about 8 years which aligns with my growing concern about the animal welfare and environmental footprint of animal agriculture. During recent years I have witnessed an exponential increase in the use of generative AI in academia and research, including my own scientific discipline (farmed animal welfare and behavior). This evolution provides numerous opportunities, but also threats. The social license for animal agriculture ought to be based on accurate and realistic views of current livestock conditions. As generative AI is believed to increasingly shape people’s perceptions, the concern of how animal agriculture is represented in generative AI models prompted my interest in this study.

M.v.K.: I am a female Professor at UBC where I have co-led the Animal Welfare Program since 2002. I grew up on a beef cattle ranch in British Columbia, Canada, and worked in the agribusiness sector for 7 years before joining the university as a professor in 2002. Together with my students I have published extensively in both the natural and social sciences on a broad range of topics in animal welfare. I have worked closely with livestock farmers and attempt to develop solutions that address the needs of the animals but also work for the farmers and align with public values. Some of my work has focused on the use of technology, such as sensors to monitor different behaviors but also now mentor students using AI. I believe that all animals under human care deserve a good life and practices that resonate with societal values will be more sustainable in the long run. 

\section{Competing interests}

Authors declare that they have no competing interests.

\begin{acks}
We thank Hanwen (Isaac) Qi for the valuable discussions about prompting techniques and his insightful suggestions about plot design. We highly appreciate his constructive feedback and tremendous support throughout this project. 
\end{acks}

\bibliographystyle{ACM-Reference-Format}
\bibliography{ref}


\begin{thebibliography}{68}


\ifx \showCODEN    \undefined \def \showCODEN     #1{\unskip}     \fi
\ifx \showISBNx    \undefined \def \showISBNx     #1{\unskip}     \fi
\ifx \showISBNxiii \undefined \def \showISBNxiii  #1{\unskip}     \fi
\ifx \showISSN     \undefined \def \showISSN      #1{\unskip}     \fi
\ifx \showLCCN     \undefined \def \showLCCN      #1{\unskip}     \fi
\ifx \shownote     \undefined \def \shownote      #1{#1}          \fi
\ifx \showarticletitle \undefined \def \showarticletitle #1{#1}   \fi
\ifx \showURL      \undefined \def \showURL       {\relax}        \fi
\providecommand\bibfield[2]{#2}
\providecommand\bibinfo[2]{#2}
\providecommand\natexlab[1]{#1}
\providecommand\showeprint[2][]{arXiv:#2}

\bibitem[Ahmad et~al\mbox{.}(2024)]%
        {Ahmad2024}
\bibfield{author}{\bibinfo{person}{Wajeeha Ahmad}, \bibinfo{person}{Ananya Sen}, \bibinfo{person}{Charles Eesley}, {and} \bibinfo{person}{Erik Brynjolfsson}.} \bibinfo{year}{2024}\natexlab{}.
\newblock \showarticletitle{Companies inadvertently fund online misinformation despite consumer backlash}.
\newblock \bibinfo{journal}{\emph{Nature}}  \bibinfo{volume}{630} (\bibinfo{date}{6} \bibinfo{year}{2024}), \bibinfo{pages}{123--131}.
\newblock
Issue 8015.
\showISSN{14764687}
\href{https://doi.org/10.1038/s41586-024-07404-1}{doi:\nolinkurl{10.1038/s41586-024-07404-1}}


\bibitem[Ali et~al\mbox{.}(2024)]%
        {Ali2024}
\bibfield{author}{\bibinfo{person}{Rohaid Ali}, \bibinfo{person}{Oliver~Y Tang}, \bibinfo{person}{Ian~D Connolly}, \bibinfo{person}{Hael~A Abdulrazeq}, \bibinfo{person}{Fatima~N Mirza}, \bibinfo{person}{Rachel~K Lim}, \bibinfo{person}{Benjamin~R Johnston}, \bibinfo{person}{Michael~W Groff}, \bibinfo{person}{Theresa Williamson}, \bibinfo{person}{Konstantina Svokos}, \bibinfo{person}{Tiffany~J Libby}, \bibinfo{person}{John~H Shin}, \bibinfo{person}{Ziya~L Gokaslan}, \bibinfo{person}{Curtis~E Doberstein}, \bibinfo{person}{James Zou}, {and} \bibinfo{person}{W.~F Asaad}.} \bibinfo{year}{2024}\natexlab{}.
\newblock \showarticletitle{The Face of a Surgeon: An Analysis of Demographic Representation in Three Leading Artificial Intelligence Text-to-Image Generators}.
\newblock \bibinfo{journal}{\emph{JAMA surgery}}  \bibinfo{volume}{1} (\bibinfo{year}{2024}), \bibinfo{pages}{87--95}.
\newblock
Issue 159.
\href{https://doi.org/10.1101/2023.05.24.23290463}{doi:\nolinkurl{10.1101/2023.05.24.23290463}}


\bibitem[Anthony(2024)]%
        {Anthony2024}
\bibfield{author}{\bibinfo{person}{Raymond Anthony}.} \bibinfo{year}{2024}\natexlab{}.
\newblock \bibinfo{booktitle}{\emph{Thompson on functions of pragmatism: Adding food and agricultural valuation to the philosophy of technology}}.
\newblock \bibinfo{publisher}{Cham: Springer International Publishing.}, \bibinfo{pages}{53--70}.
\newblock


\bibitem[Australia(2025)]%
        {rspca2025}
\bibfield{author}{\bibinfo{person}{RSPCA Australia}.} \bibinfo{year}{2025}\natexlab{}.
\newblock \bibinfo{title}{Pig farming}.
\newblock
\urldef\tempurl%
\url{https://www.rspca.org.au/key-issues/pig-farming/}
\showURL{%
\tempurl}


\bibitem[Bar-On et~al\mbox{.}(2018)]%
        {baron2018}
\bibfield{author}{\bibinfo{person}{Yinon~M. Bar-On}, \bibinfo{person}{Rob Phillips}, {and} \bibinfo{person}{Ron Milo}.} \bibinfo{year}{2018}\natexlab{}.
\newblock \showarticletitle{The biomass distribution on Earth}.
\newblock \bibinfo{journal}{\emph{Proceedings of the National Academy of Sciences of the United States of America}}  \bibinfo{volume}{115} (\bibinfo{date}{6} \bibinfo{year}{2018}), \bibinfo{pages}{6506--6511}.
\newblock
Issue 25.
\showISSN{10916490}
\href{https://doi.org/10.1073/pnas.1711842115}{doi:\nolinkurl{10.1073/pnas.1711842115}}


\bibitem[Beaver et~al\mbox{.}(2020)]%
        {Beaver2020}
\bibfield{author}{\bibinfo{person}{Annabelle Beaver}, \bibinfo{person}{Kathryn~L. Proudfoot}, {and} \bibinfo{person}{Marina~A.G. von Keyserlingk}.} \bibinfo{year}{2020}\natexlab{}.
\newblock \bibinfo{title}{Symposium review: Considerations for the future of dairy cattle housing: An animal welfare perspective}.
\newblock \bibinfo{numpages}{5746-5758}~pages.
\newblock
Issue 6.
\showISSN{15253198}
\href{https://doi.org/10.3168/jds.2019-17804}{doi:\nolinkurl{10.3168/jds.2019-17804}}


\bibitem[Bender(2023)]%
        {Bender2023}
\bibfield{author}{\bibinfo{person}{Emily~M Bender}.} \bibinfo{year}{2023}\natexlab{}.
\newblock \bibinfo{title}{Cleaning up a baby peacock sullied by a non-information spill}.
\newblock
\urldef\tempurl%
\url{https://medium.com/@emilymenonbender/cleaning-up-a-baby-peacock-sullied-by-a-non-information-spill-d2e2aa642134}
\showURL{%
\tempurl}


\bibitem[Betker et~al\mbox{.}(2023)]%
        {Betker2023}
\bibfield{author}{\bibinfo{person}{James Betker}, \bibinfo{person}{Gabriel Goh}, \bibinfo{person}{Li Jing}, \bibinfo{person}{Tim Brooks}, \bibinfo{person}{Jianfeng Wang}, \bibinfo{person}{Linjie Li}, \bibinfo{person}{Long Ouyang}, \bibinfo{person}{Juntang Zhuang}, \bibinfo{person}{Joyce Lee}, \bibinfo{person}{Yufei Guo}, \bibinfo{person}{Wesam Manassra}, \bibinfo{person}{Prafulla Dhariwal}, \bibinfo{person}{Casey Chu}, \bibinfo{person}{Yunxin Jiao}, {and} \bibinfo{person}{Aditya Ramesh}.} \bibinfo{year}{2023}\natexlab{}.
\newblock \bibinfo{title}{Improving Image Generation with Better Captions}.
\newblock \bibinfo{numpages}{19}~pages.
\newblock


\bibitem[Bolton et~al\mbox{.}(2024)]%
        {Bolton2024}
\bibfield{author}{\bibinfo{person}{Sarah~E. Bolton}, \bibinfo{person}{Bianca Vandresen}, {and} \bibinfo{person}{Marina A.~G. von Keyserlingk}.} \bibinfo{year}{2024}\natexlab{}.
\newblock \showarticletitle{“Dear Dairy, It’s Not Me, It’s You”: Australian Public Attitudes to Dairy Expressed Through Love and Breakup Letters}.
\newblock \bibinfo{journal}{\emph{Food Ethics}}  \bibinfo{volume}{9} (\bibinfo{date}{10} \bibinfo{year}{2024}), \bibinfo{pages}{18}.
\newblock
Issue 2.
\showISSN{2364-6853}
\href{https://doi.org/10.1007/s41055-024-00153-x}{doi:\nolinkurl{10.1007/s41055-024-00153-x}}


\bibitem[Bossert and Hagendorff(2021)]%
        {Bossert2021}
\bibfield{author}{\bibinfo{person}{Leonie Bossert} {and} \bibinfo{person}{Thilo Hagendorff}.} \bibinfo{year}{2021}\natexlab{}.
\newblock \showarticletitle{Animals and AI. The role of animals in AI research and application – An overview and ethical evaluation}.
\newblock \bibinfo{journal}{\emph{Technology in Society}}  \bibinfo{volume}{67} (\bibinfo{date}{11} \bibinfo{year}{2021}).
\newblock
\showISSN{0160791X}
\href{https://doi.org/10.1016/j.techsoc.2021.101678}{doi:\nolinkurl{10.1016/j.techsoc.2021.101678}}


\bibitem[Buolamwini(2024)]%
        {Buolamwini2024}
\bibfield{author}{\bibinfo{person}{Joy Buolamwini}.} \bibinfo{year}{2024}\natexlab{}.
\newblock \bibinfo{booktitle}{\emph{Unmasking AI: My Mission to Protect What Is Human in a World of Machines}}.
\newblock \bibinfo{publisher}{Random House Trade}.
\newblock
\showISBNx{9780593241844}


\bibitem[Capraro et~al\mbox{.}(2024)]%
        {Capraro2024}
\bibfield{author}{\bibinfo{person}{Valerio Capraro}, \bibinfo{person}{Austin Lentsch}, \bibinfo{person}{Daron Acemoglu}, \bibinfo{person}{Selin Akgun}, \bibinfo{person}{Aisel Akhmedova}, \bibinfo{person}{Ennio Bilancini}, \bibinfo{person}{Jean~François Bonnefon}, \bibinfo{person}{Pablo Brañas-Garza}, \bibinfo{person}{Luigi Butera}, \bibinfo{person}{Karen~M. Douglas}, \bibinfo{person}{Jim~A.C. Everett}, \bibinfo{person}{Gerd Gigerenzer}, \bibinfo{person}{Christine Greenhow}, \bibinfo{person}{Daniel~A. Hashimoto}, \bibinfo{person}{Julianne Holt-Lunstad}, \bibinfo{person}{Jolanda Jetten}, \bibinfo{person}{Simon Johnson}, \bibinfo{person}{Werner~H. Kunz}, \bibinfo{person}{Chiara Longoni}, \bibinfo{person}{Pete Lunn}, \bibinfo{person}{Simone Natale}, \bibinfo{person}{Stefanie Paluch}, \bibinfo{person}{Iyad Rahwan}, \bibinfo{person}{Neil Selwyn}, \bibinfo{person}{Vivek Singh}, \bibinfo{person}{Siddharth Suri}, \bibinfo{person}{Jennifer Sutcliffe}, \bibinfo{person}{Joe Tomlinson}, \bibinfo{person}{Sander Van~Der
  Linden}, \bibinfo{person}{Paul A.M.~Van Lange}, \bibinfo{person}{Friederike Wall}, \bibinfo{person}{Jay J.~Van Bavel}, {and} \bibinfo{person}{Riccardo Viale}.} \bibinfo{year}{2024}\natexlab{}.
\newblock \showarticletitle{The impact of generative artificial intelligence on socioeconomic inequalities and policy making}.
\newblock \bibinfo{journal}{\emph{PNAS Nexus}}  \bibinfo{volume}{3} (\bibinfo{date}{6} \bibinfo{year}{2024}).
\newblock
Issue 6.
\showISSN{27526542}
\href{https://doi.org/10.1093/pnasnexus/pgae191}{doi:\nolinkurl{10.1093/pnasnexus/pgae191}}


\bibitem[Coghlan and Parker(2023)]%
        {Coghlan2023}
\bibfield{author}{\bibinfo{person}{Simon Coghlan} {and} \bibinfo{person}{Christine Parker}.} \bibinfo{year}{2023}\natexlab{}.
\newblock \showarticletitle{Harm to Nonhuman Animals from AI: a Systematic Account and Framework}.
\newblock \bibinfo{journal}{\emph{Philosophy and Technology}}  \bibinfo{volume}{36} (\bibinfo{date}{6} \bibinfo{year}{2023}).
\newblock
Issue 2.
\showISSN{22105441}
\href{https://doi.org/10.1007/s13347-023-00627-6}{doi:\nolinkurl{10.1007/s13347-023-00627-6}}


\bibitem[Commission(2019)]%
        {EuropeanCommission2019}
\bibfield{author}{\bibinfo{person}{European Commission}.} \bibinfo{year}{2019}\natexlab{}.
\newblock \bibinfo{title}{Ethics Guidelines for Trustworthy AI}.
\newblock
\urldef\tempurl%
\url{https://ec.europa.eu/digital-}
\showURL{%
\tempurl}


\bibitem[Commission(2025)]%
        {EuropeanCommission2025}
\bibfield{author}{\bibinfo{person}{European Commission}.} \bibinfo{year}{2025}\natexlab{}.
\newblock \bibinfo{title}{Second Draft General-Purpose AI Code of Practice}.
\newblock
\urldef\tempurl%
\url{https://digital-strategy.ec.europa.eu/en/library/second-draft-general-purpose-ai-code-practice-published-written-independent-experts}
\showURL{%
\tempurl}


\bibitem[de~Montréal(2018)]%
        {UniversitdeMontral2018}
\bibfield{author}{\bibinfo{person}{Université de Montréal}.} \bibinfo{year}{2018}\natexlab{}.
\newblock \bibinfo{title}{Montréal Declaration for a Responsible Development of Artificial Intelligence}.
\newblock
\urldef\tempurl%
\url{https://www.montrealdeclaration-responsibleai.com/}
\showURL{%
\tempurl}


\bibitem[Drury et~al\mbox{.}(2023)]%
        {Drury2023}
\bibfield{author}{\bibinfo{person}{Matt Drury}, \bibinfo{person}{Janet Fuller}, {and} \bibinfo{person}{John Hoeks}.} \bibinfo{year}{2023}\natexlab{}.
\newblock \showarticletitle{Embedding animals within a definition of sustainability}.
\newblock \bibinfo{journal}{\emph{Sustainability Science}}  \bibinfo{volume}{18} (\bibinfo{date}{7} \bibinfo{year}{2023}), \bibinfo{pages}{1925--1938}.
\newblock
Issue 4.
\showISSN{18624057}
\href{https://doi.org/10.1007/s11625-023-01310-7}{doi:\nolinkurl{10.1007/s11625-023-01310-7}}


\bibitem[Godfray et~al\mbox{.}(2010)]%
        {Godfray2010}
\bibfield{author}{\bibinfo{person}{H~Charles~J Godfray}, \bibinfo{person}{John~R Beddington}, \bibinfo{person}{Ian~R Crute}, \bibinfo{person}{Lawrence Haddad}, \bibinfo{person}{David Lawrence}, \bibinfo{person}{James~F Muir}, \bibinfo{person}{Jules Pretty}, \bibinfo{person}{Sherman Robinson}, \bibinfo{person}{Sandy~M Thomas}, {and} \bibinfo{person}{Camilla Toulmin}.} \bibinfo{year}{2010}\natexlab{}.
\newblock \showarticletitle{Food security: the challenge of feeding 9 billion people}.
\newblock \bibinfo{journal}{\emph{science}}  \bibinfo{volume}{327} (\bibinfo{year}{2010}), \bibinfo{pages}{812--818}.
\newblock
Issue 5967.
\urldef\tempurl%
\url{www.sciencemag.org}
\showURL{%
\tempurl}


\bibitem[Hagendorff(2022)]%
        {Hagendorff2022}
\bibfield{author}{\bibinfo{person}{Thilo Hagendorff}.} \bibinfo{year}{2022}\natexlab{}.
\newblock \showarticletitle{Blind spots in AI ethics}.
\newblock \bibinfo{journal}{\emph{AI and Ethics}}  \bibinfo{volume}{2} (\bibinfo{date}{11} \bibinfo{year}{2022}), \bibinfo{pages}{851--867}.
\newblock
Issue 4.
\showISSN{2730-5953}
\href{https://doi.org/10.1007/s43681-021-00122-8}{doi:\nolinkurl{10.1007/s43681-021-00122-8}}


\bibitem[Hagendorff et~al\mbox{.}(2023)]%
        {Hagendorff2023}
\bibfield{author}{\bibinfo{person}{Thilo Hagendorff}, \bibinfo{person}{Leonie~N. Bossert}, \bibinfo{person}{Yip~Fai Tse}, {and} \bibinfo{person}{Peter Singer}.} \bibinfo{year}{2023}\natexlab{}.
\newblock \showarticletitle{Speciesist bias in AI: how AI applications perpetuate discrimination and unfair outcomes against animals}.
\newblock \bibinfo{journal}{\emph{AI and Ethics}}  \bibinfo{volume}{3} (\bibinfo{date}{8} \bibinfo{year}{2023}), \bibinfo{pages}{717--734}.
\newblock
Issue 3.
\showISSN{2730-5953}
\href{https://doi.org/10.1007/s43681-022-00199-9}{doi:\nolinkurl{10.1007/s43681-022-00199-9}}


\bibitem[Hall(1997)]%
        {Hall1997}
\bibfield{author}{\bibinfo{person}{Stuart. Hall}.} \bibinfo{year}{1997}\natexlab{}.
\newblock \bibinfo{booktitle}{\emph{Representation : cultural representations and signifying practices}}.
\newblock \bibinfo{publisher}{Sage London}. 360 pages.
\newblock
\showISBNx{1529770386}


\bibitem[Haq et~al\mbox{.}(2024)]%
        {Haq2024}
\bibfield{author}{\bibinfo{person}{Ehsan~Ul Haq}, \bibinfo{person}{Yiming Zhu}, \bibinfo{person}{Pan Hui}, {and} \bibinfo{person}{Gareth Tyson}.} \bibinfo{year}{2024}\natexlab{}.
\newblock \showarticletitle{History in Making: Political Campaigns in the Era of Artificial Intelligence-Generated Content}. In \bibinfo{booktitle}{\emph{Companion Proceedings of the ACM on Web Conference 2024}}. \bibinfo{publisher}{Association for Computing Machinery, Inc}, \bibinfo{pages}{1115--1118}.
\newblock
\showISBNx{9798400701726}
\href{https://doi.org/10.1145/3589335.3652000}{doi:\nolinkurl{10.1145/3589335.3652000}}


\bibitem[Hu et~al\mbox{.}(2017)]%
        {Hu2017}
\bibfield{author}{\bibinfo{person}{Yuanan Hu}, \bibinfo{person}{Hefa Cheng}, {and} \bibinfo{person}{Shu Tao}.} \bibinfo{year}{2017}\natexlab{}.
\newblock \showarticletitle{Environmental and human health challenges of industrial livestock and poultry farming in China and their mitigation}.
\newblock \bibinfo{journal}{\emph{Environment International}}  \bibinfo{volume}{107} (\bibinfo{year}{2017}), \bibinfo{pages}{111--130}.
\newblock
\showISSN{18736750}
\href{https://doi.org/10.1016/j.envint.2017.07.003}{doi:\nolinkurl{10.1016/j.envint.2017.07.003}}


\bibitem[in~World~Farming(2012)]%
        {compassion2012}
\bibfield{author}{\bibinfo{person}{Compassion in World~Farming}.} \bibinfo{year}{2012}\natexlab{}.
\newblock \bibinfo{title}{Statistics: Dairy cows}.
\newblock
\urldef\tempurl%
\url{http://www.rabobank.com/content/images/Global_dairy_top-20_Voorbergen_jul2009_tcm43-89002.pdf}
\showURL{%
\tempurl}


\bibitem[Jiménez-Ruiz et~al\mbox{.}(2022)]%
        {Jimenez2022}
\bibfield{author}{\bibinfo{person}{Saúl Jiménez-Ruiz}, \bibinfo{person}{Eduardo Laguna}, \bibinfo{person}{Joaquín Vicente}, \bibinfo{person}{Ignacio García-Bocanegra}, \bibinfo{person}{Jordi Martínez-Guijosa}, \bibinfo{person}{David Cano-Terriza}, \bibinfo{person}{María~A. Risalde}, {and} \bibinfo{person}{Pelayo Acevedo}.} \bibinfo{year}{2022}\natexlab{}.
\newblock \showarticletitle{Characterization and management of interaction risks between livestock and wild ungulates on outdoor pig farms in Spain}.
\newblock \bibinfo{journal}{\emph{Porcine Health Management}}  \bibinfo{volume}{8} (\bibinfo{date}{12} \bibinfo{year}{2022}), \bibinfo{pages}{1--14}.
\newblock
\showISSN{20555660}
\href{https://doi.org/10.1186/s40813-021-00246-7}{doi:\nolinkurl{10.1186/s40813-021-00246-7}}


\bibitem[Jobin et~al\mbox{.}(2019)]%
        {Jobin2019}
\bibfield{author}{\bibinfo{person}{Anna Jobin}, \bibinfo{person}{Marcello Ienca}, {and} \bibinfo{person}{Effy Vayena}.} \bibinfo{year}{2019}\natexlab{}.
\newblock \showarticletitle{The global landscape of AI ethics guidelines}.
\newblock \bibinfo{journal}{\emph{Nature Machine Intelligence}}  \bibinfo{volume}{1} (\bibinfo{date}{9} \bibinfo{year}{2019}), \bibinfo{pages}{389--399}.
\newblock
Issue 9.
\href{https://doi.org/10.1038/s42256-019-0088-2}{doi:\nolinkurl{10.1038/s42256-019-0088-2}}


\bibitem[Kok et~al\mbox{.}(2020)]%
        {Kok2020}
\bibfield{author}{\bibinfo{person}{Akke Kok}, \bibinfo{person}{EM de Olde}, \bibinfo{person}{IJM De~Boer}, {and} \bibinfo{person}{R Ripoll-Bosch}.} \bibinfo{year}{2020}\natexlab{}.
\newblock \showarticletitle{European biodiversity assessments in livestock science: A review of research characteristics and indicators}.
\newblock \bibinfo{journal}{\emph{Ecological Indicators}}  \bibinfo{volume}{112} (\bibinfo{date}{5} \bibinfo{year}{2020}).
\newblock
\showISSN{1470160X}
\href{https://doi.org/10.1016/j.ecolind.2019.105902}{doi:\nolinkurl{10.1016/j.ecolind.2019.105902}}


\bibitem[Lee(2018)]%
        {Lee2018}
\bibfield{author}{\bibinfo{person}{Min~Kyung Lee}.} \bibinfo{year}{2018}\natexlab{}.
\newblock \showarticletitle{Understanding perception of algorithmic decisions: Fairness, trust, and emotion in response to algorithmic management}.
\newblock \bibinfo{journal}{\emph{Big Data and Society}}  \bibinfo{volume}{5} (\bibinfo{date}{6} \bibinfo{year}{2018}).
\newblock
Issue 1.
\showISSN{20539517}
\href{https://doi.org/10.1177/2053951718756684}{doi:\nolinkurl{10.1177/2053951718756684}}


\bibitem[Magouras et~al\mbox{.}(2020)]%
        {Magouras2020}
\bibfield{author}{\bibinfo{person}{Ioannis Magouras}, \bibinfo{person}{Victoria~J. Brookes}, \bibinfo{person}{Ferran Jori}, \bibinfo{person}{Angela Martin}, \bibinfo{person}{Dirk~Udo Pfeiffer}, {and} \bibinfo{person}{Salome Dürr}.} \bibinfo{year}{2020}\natexlab{}.
\newblock \showarticletitle{Emerging Zoonotic Diseases: Should We Rethink the Animal–Human Interface?}
\newblock \bibinfo{journal}{\emph{Frontiers in Veterinary Science}}  \bibinfo{volume}{7} (\bibinfo{date}{10} \bibinfo{year}{2020}).
\newblock
\showISSN{22971769}
\href{https://doi.org/10.3389/fvets.2020.582743}{doi:\nolinkurl{10.3389/fvets.2020.582743}}


\bibitem[Mortensen(2024)]%
        {Mortensen2024}
\bibfield{author}{\bibinfo{person}{Oskar Mortensen}.} \bibinfo{year}{2024}\natexlab{}.
\newblock \bibinfo{title}{How Many Users Does ChatGPT Have? Statistics \& Facts (2025)}.
\newblock
\urldef\tempurl%
\url{https://seo.ai/blog/how-many-users-does-chatgpt-have#:~:text=Top%20ChatGPT%20Statistics&text=OpenAI's%20ChatGPT%20has%20seen%20its,swell%20to%20over%20180%20million.}
\showURL{%
\tempurl}


\bibitem[OpenAI(2022)]%
        {OpenAI2022}
\bibfield{author}{\bibinfo{person}{OpenAI}.} \bibinfo{year}{2022}\natexlab{}.
\newblock \bibinfo{title}{DALL-E 2 System Card}.
\newblock


\bibitem[OpenAI(2023)]%
        {OpenAI2023}
\bibfield{author}{\bibinfo{person}{OpenAI}.} \bibinfo{year}{2023}\natexlab{}.
\newblock \bibinfo{title}{DALL·E 3 System Card}.
\newblock \bibinfo{numpages}{22}~pages.
\newblock
\urldef\tempurl%
\url{https://cdn.openai.com/papers/DALL_E_3_System_Card.pdf}
\showURL{%
\tempurl}


\bibitem[Oppenlaender et~al\mbox{.}(2023)]%
        {Oppenlaender2023}
\bibfield{author}{\bibinfo{person}{Jonas Oppenlaender}, \bibinfo{person}{Johanna Silvennoinen}, \bibinfo{person}{Ville Paananen}, {and} \bibinfo{person}{Aku Visuri}.} \bibinfo{year}{2023}\natexlab{}.
\newblock \showarticletitle{Perceptions and Realities of Text-to-Image Generation}. In \bibinfo{booktitle}{\emph{ACM International Conference Proceeding Series}}. \bibinfo{publisher}{Association for Computing Machinery}, \bibinfo{pages}{279--288}.
\newblock
\showISBNx{9798400708749}
\href{https://doi.org/10.1145/3616961.3616978}{doi:\nolinkurl{10.1145/3616961.3616978}}


\bibitem[Parliament and the Council of~the European~Union(2024)]%
        {Council2024}
\bibfield{author}{\bibinfo{person}{The~European Parliament} {and} \bibinfo{person}{the Council of~the European~Union}.} \bibinfo{year}{2024}\natexlab{}.
\newblock \showarticletitle{Regulation (EU) 2024/1689 of the European Parliament and of the Council of 13 June 2024 Laying Down Harmonised Rules on Artificial Intelligence and Amending Regulations (EC) No 300/2008,(EU) No 167/2013,(EU) No 168/2013,(EU) 2018/858,(EU) 2018/1139 and (EU) 2019/2144 and Directives 2014/90/EU,(EU) 2016/797 and (EU) 2020/1828 (Artificial Intelligence Act) Off}.
\newblock \bibinfo{journal}{\emph{Official Journal of the European Union}}  \bibinfo{volume}{50} (\bibinfo{year}{2024}).
\newblock
Issue 202.
\urldef\tempurl%
\url{http://data.europa.eu/eli/reg/2024/1689/oj}
\showURL{%
\tempurl}


\bibitem[Qadri et~al\mbox{.}(2023)]%
        {Qadri2023}
\bibfield{author}{\bibinfo{person}{Rida Qadri}, \bibinfo{person}{Renee Shelby}, \bibinfo{person}{Cynthia~L. Bennett}, {and} \bibinfo{person}{Emily Denton}.} \bibinfo{year}{2023}\natexlab{}.
\newblock \showarticletitle{AI's Regimes of Representation: A Community-centered Study of Text-to-Image Models in South Asia}. In \bibinfo{booktitle}{\emph{Proceedings of the 2023 ACM Conference on Fairness, Accountability, and Transparency}}. \bibinfo{publisher}{Association for Computing Machinery}, \bibinfo{pages}{506--517}.
\newblock
\showISBNx{9781450372527}
\href{https://doi.org/10.1145/3593013.3594016}{doi:\nolinkurl{10.1145/3593013.3594016}}


\bibitem[Quaye et~al\mbox{.}(2024)]%
        {Quaye2024}
\bibfield{author}{\bibinfo{person}{Jessica Quaye}, \bibinfo{person}{Alicia Parrish}, \bibinfo{person}{Oana Inel}, \bibinfo{person}{Charvi Rastogi}, \bibinfo{person}{Hannah~Rose Kirk}, \bibinfo{person}{Minsuk Kahng}, \bibinfo{person}{Erin~Van Liemt}, \bibinfo{person}{Max Bartolo}, \bibinfo{person}{Jess Tsang}, \bibinfo{person}{Justin White}, \bibinfo{person}{Nathan Clement}, \bibinfo{person}{Rafael Mosquera}, \bibinfo{person}{Juan Ciro}, \bibinfo{person}{Vijay~Janapa Reddi}, {and} \bibinfo{person}{Lora Aroyo}.} \bibinfo{year}{2024}\natexlab{}.
\newblock \showarticletitle{Adversarial Nibbler: An Open Red-Teaming Method for Identifying Diverse Harms in Text-to-Image Generation}. In \bibinfo{booktitle}{\emph{2024 ACM Conference on Fairness, Accountability, and Transparency, FAccT 2024}}. \bibinfo{publisher}{Association for Computing Machinery, Inc}, \bibinfo{pages}{388--406}.
\newblock
\showISBNx{9798400704505}
\href{https://doi.org/10.1145/3630106.3658913}{doi:\nolinkurl{10.1145/3630106.3658913}}


\bibitem[Reijs et~al\mbox{.}(2013)]%
        {Reijs2013}
\bibfield{author}{\bibinfo{person}{J.~W. Reijs}, \bibinfo{person}{C.~H.~G. Daatselaar}, \bibinfo{person}{J.~F.~M. Helming}, {and} \bibinfo{person}{A.~C. G.~Jager J.~H.Beldman}.} \bibinfo{year}{2013}\natexlab{}.
\newblock \bibinfo{title}{Grazing dairy cows in North-West Europe: economic farm performance and future developments with emphasis on the Dutch situation}.
\newblock
\urldef\tempurl%
\url{www.wageningenUR.nl/en/lei}
\showURL{%
\tempurl}


\bibitem[Rollin(2011)]%
        {Rollin2011}
\bibfield{author}{\bibinfo{person}{Bernard~E. Rollin}.} \bibinfo{year}{2011}\natexlab{}.
\newblock \showarticletitle{Animal rights as a mainstream phenomenon}.
\newblock \bibinfo{journal}{\emph{Animals}}  \bibinfo{volume}{1} (\bibinfo{date}{3} \bibinfo{year}{2011}), \bibinfo{pages}{102--115}.
\newblock
Issue 1.
\showISSN{20762615}
\href{https://doi.org/10.3390/ani1010102}{doi:\nolinkurl{10.3390/ani1010102}}


\bibitem[Ryan et~al\mbox{.}(2015)]%
        {Ryan2015}
\bibfield{author}{\bibinfo{person}{Erin~B Ryan}, \bibinfo{person}{David Fraser}, {and} \bibinfo{person}{Daniel~M Weary}.} \bibinfo{year}{2015}\natexlab{}.
\newblock \showarticletitle{Public attitudes to housing systems for pregnant pigs}.
\newblock \bibinfo{journal}{\emph{PLoS ONE}}  \bibinfo{volume}{10} (\bibinfo{date}{11} \bibinfo{year}{2015}).
\newblock
Issue 11.
\showISSN{19326203}
\href{https://doi.org/10.1371/journal.pone.0141878}{doi:\nolinkurl{10.1371/journal.pone.0141878}}


\bibitem[Shah and Bender(2024)]%
        {Shah2024}
\bibfield{author}{\bibinfo{person}{Chirag Shah} {and} \bibinfo{person}{Emily~M. Bender}.} \bibinfo{year}{2024}\natexlab{}.
\newblock \showarticletitle{Envisioning Information Access Systems: What Makes for Good Tools and a Healthy Web?}
\newblock \bibinfo{journal}{\emph{ACM Transactions on the Web}}  \bibinfo{volume}{18} (\bibinfo{date}{8} \bibinfo{year}{2024}).
\newblock
Issue 3.
\showISSN{1559114X}
\href{https://doi.org/10.1145/3649468}{doi:\nolinkurl{10.1145/3649468}}


\bibitem[Shumailov et~al\mbox{.}(2024)]%
        {Shumailov2024}
\bibfield{author}{\bibinfo{person}{Ilia Shumailov}, \bibinfo{person}{Zakhar Shumaylov}, \bibinfo{person}{Yiren Zhao}, \bibinfo{person}{Nicolas Papernot}, \bibinfo{person}{Ross Anderson}, {and} \bibinfo{person}{Yarin Gal}.} \bibinfo{year}{2024}\natexlab{}.
\newblock \showarticletitle{AI models collapse when trained on recursively generated data}.
\newblock \bibinfo{journal}{\emph{Nature}}  \bibinfo{volume}{631} (\bibinfo{date}{7} \bibinfo{year}{2024}), \bibinfo{pages}{755--759}.
\newblock
Issue 8022.
\showISSN{14764687}
\href{https://doi.org/10.1038/s41586-024-07566-y}{doi:\nolinkurl{10.1038/s41586-024-07566-y}}


\bibitem[Silbergeld et~al\mbox{.}(2008)]%
        {Silbergeld2008}
\bibfield{author}{\bibinfo{person}{Ellen~K. Silbergeld}, \bibinfo{person}{Jay Graham}, {and} \bibinfo{person}{Lance~B. Price}.} \bibinfo{year}{2008}\natexlab{}.
\newblock \showarticletitle{Industrial food animal production, antimicrobial resistance, and human health}. In \bibinfo{booktitle}{\emph{Annual Review of Public Health}}, Vol.~\bibinfo{volume}{29}. \bibinfo{pages}{151--169}.
\newblock
\showISSN{01637525}
\href{https://doi.org/10.1146/annurev.publhealth.29.020907.090904}{doi:\nolinkurl{10.1146/annurev.publhealth.29.020907.090904}}


\bibitem[Singer(1975)]%
        {Singer1975}
\bibfield{author}{\bibinfo{person}{Peter Singer}.} \bibinfo{year}{1975}\natexlab{}.
\newblock \bibinfo{booktitle}{\emph{Animal Liberation}}.
\newblock \bibinfo{publisher}{HarperCollins}. 1--23 pages.
\newblock


\bibitem[Singer and Tse(2023)]%
        {Singer2023}
\bibfield{author}{\bibinfo{person}{Peter Singer} {and} \bibinfo{person}{Yip~Fai Tse}.} \bibinfo{year}{2023}\natexlab{}.
\newblock \showarticletitle{AI ethics: the case for including animals}.
\newblock \bibinfo{journal}{\emph{AI and Ethics}}  \bibinfo{volume}{3} (\bibinfo{date}{5} \bibinfo{year}{2023}), \bibinfo{pages}{539--551}.
\newblock
Issue 2.
\showISSN{2730-5953}
\href{https://doi.org/10.1007/s43681-022-00187-z}{doi:\nolinkurl{10.1007/s43681-022-00187-z}}


\bibitem[Sirovica et~al\mbox{.}(2022)]%
        {Sirovica2022}
\bibfield{author}{\bibinfo{person}{Lara~V. Sirovica}, \bibinfo{person}{Caroline Ritter}, \bibinfo{person}{Jillian Hendricks}, \bibinfo{person}{Daniel~M. Weary}, \bibinfo{person}{Sumeet Gulati}, {and} \bibinfo{person}{Marina. A.G.~Von Keyserlingk}.} \bibinfo{year}{2022}\natexlab{}.
\newblock \showarticletitle{Public attitude toward and perceptions of dairy cattle welfare in cow-calf management systems differing in type of social and maternal contact}.
\newblock \bibinfo{journal}{\emph{Journal of Dairy Science}}  \bibinfo{volume}{105} (\bibinfo{date}{4} \bibinfo{year}{2022}), \bibinfo{pages}{3248--3268}.
\newblock
Issue 4.
\showISSN{15253198}
\href{https://doi.org/10.3168/jds.2021-21344}{doi:\nolinkurl{10.3168/jds.2021-21344}}


\bibitem[Smid et~al\mbox{.}(2020)]%
        {Smid2020}
\bibfield{author}{\bibinfo{person}{Anne-Marieke~C. Smid}, \bibinfo{person}{Daniel~M. Weary}, {and} \bibinfo{person}{Marina~A.G. von Keyserlingk}.} \bibinfo{year}{2020}\natexlab{}.
\newblock \showarticletitle{The Influence of Different Types of Outdoor Access on Dairy Cattle Behavior}.
\newblock \bibinfo{journal}{\emph{Frontiers in Veterinary Science}}  \bibinfo{volume}{7} (\bibinfo{date}{5} \bibinfo{year}{2020}).
\newblock
Issue 257.
\showISSN{22971769}
\href{https://doi.org/10.3389/fvets.2020.00257}{doi:\nolinkurl{10.3389/fvets.2020.00257}}


\bibitem[Statista(2023)]%
        {Statista2023}
\bibfield{author}{\bibinfo{person}{Statista}.} \bibinfo{year}{2023}\natexlab{}.
\newblock \bibinfo{title}{Dairy cow numbers in the European Union (EU-27) in 2022, by country}.
\newblock
\urldef\tempurl%
\url{https://www.statista.com/statistics/616201/dairy-cow-numbers-european-union-eu/#:~:text=Dairy%20cow%20numbers%20in%20countries,(EU%2D27)%20in%202022&text=The%20Federal%20Republic%20of%20Germany,3.2%20million%20animals%20in%202022.}
\showURL{%
\tempurl}


\bibitem[Statista(2024)]%
        {Statista2024}
\bibfield{author}{\bibinfo{person}{Statista}.} \bibinfo{year}{2024}\natexlab{}.
\newblock \bibinfo{title}{Number of pigs worldwide in 2024, by leading country (in million head)}.
\newblock
\urldef\tempurl%
\url{https://www.statista.com/statistics/263964/number-of-pigs-in-selected-countries/}
\showURL{%
\tempurl}


\bibitem[Steinfeld et~al\mbox{.}(2006)]%
        {Steinfeld2006}
\bibfield{author}{\bibinfo{person}{Henning Steinfeld}, \bibinfo{person}{Pierre Gerber}, \bibinfo{person}{Tom~D Wassenaar}, \bibinfo{person}{Vincent Castel}, {and} \bibinfo{person}{Cees De~Haan}.} \bibinfo{year}{2006}\natexlab{}.
\newblock \bibinfo{title}{Livestock's long shadow: environmental issues and options}.
\newblock
\urldef\tempurl%
\url{https://books.google.ca/books?hl=en&lr=&id=1B9LQQkm_qMC&oi=fnd&…8kU&redir_esc=y#v=onepage&q=livestock's%20long%20shadow&f=false}
\showURL{%
\tempurl}


\bibitem[Tactacan et~al\mbox{.}(2009)]%
        {Tactacan2009}
\bibfield{author}{\bibinfo{person}{Glenmer~B Tactacan}, \bibinfo{person}{W Guenter}, \bibinfo{person}{NJ Lewis}, \bibinfo{person}{JC Rodriguez-Lecompte}, {and} \bibinfo{person}{JD House}.} \bibinfo{year}{2009}\natexlab{}.
\newblock \showarticletitle{Performance and welfare of laying hens in conventional and enriched cages}.
\newblock \bibinfo{journal}{\emph{Poultry Science}}  \bibinfo{volume}{88} (\bibinfo{year}{2009}), \bibinfo{pages}{698--707}.
\newblock
Issue 4.
\showISSN{15253171}
\href{https://doi.org/10.3382/ps.2008-00369}{doi:\nolinkurl{10.3382/ps.2008-00369}}


\bibitem[Thompson(1997)]%
        {Thompson1997}
\bibfield{author}{\bibinfo{person}{Paul~B Thompson}.} \bibinfo{year}{1997}\natexlab{}.
\newblock \showarticletitle{Sustainability As a Norm}.
\newblock \bibinfo{journal}{\emph{Society for Philosophy and Technology Quarterly Electronic Journal}}  \bibinfo{volume}{2} (\bibinfo{year}{1997}), \bibinfo{pages}{99--110}.
\newblock
Issue 2.


\bibitem[Tremblay et~al\mbox{.}(2018)]%
        {Tremblay2018}
\bibfield{author}{\bibinfo{person}{Jonathan Tremblay}, \bibinfo{person}{Aayush Prakash}, \bibinfo{person}{David Acuna}, \bibinfo{person}{Mark Brophy}, \bibinfo{person}{Varun Jampani}, \bibinfo{person}{Cem Anil}, \bibinfo{person}{Thang To}, \bibinfo{person}{Eric Cameracci}, \bibinfo{person}{Shaad Boochoon}, {and} \bibinfo{person}{Stan Birchfield}.} \bibinfo{year}{2018}\natexlab{}.
\newblock \showarticletitle{Training Deep Networks with Synthetic Data: Bridging the Reality Gap by Domain Randomization}. In \bibinfo{booktitle}{\emph{Proceedings of the IEEE conference on computer vision and pattern recognition workshops}}. \bibinfo{pages}{969--977}.
\newblock
\urldef\tempurl%
\url{https://github.com/tensorflow/models/tree/}
\showURL{%
\tempurl}


\bibitem[Trevisi et~al\mbox{.}(2014)]%
        {Trevisi2014}
\bibfield{author}{\bibinfo{person}{Erminio Trevisi}, \bibinfo{person}{Alfonso Zecconi}, \bibinfo{person}{Simone Cogrossi}, \bibinfo{person}{Elisabetta Razzuoli}, \bibinfo{person}{Paolo Grossi}, {and} \bibinfo{person}{Massimo Amadori}.} \bibinfo{year}{2014}\natexlab{}.
\newblock \bibinfo{title}{Strategies for reduced antibiotic usage in dairy cattle farms}.
\newblock \bibinfo{numpages}{229-233}~pages.
\newblock
Issue 2.
\showISSN{15322661}
\href{https://doi.org/10.1016/j.rvsc.2014.01.001}{doi:\nolinkurl{10.1016/j.rvsc.2014.01.001}}


\bibitem[Tuyttens et~al\mbox{.}(2011)]%
        {Tuyttens2011}
\bibfield{author}{\bibinfo{person}{Frank~A.M. Tuyttens}, \bibinfo{person}{Filiep Vanhonacker}, \bibinfo{person}{Karolien Langendries}, \bibinfo{person}{Marijke Aluwé}, \bibinfo{person}{Sam Millet}, \bibinfo{person}{Karen Bekaert}, {and} \bibinfo{person}{Wim Verbeke}.} \bibinfo{year}{2011}\natexlab{}.
\newblock \showarticletitle{Effect of information provisioning on attitude toward surgical castration of male piglets and alternative strategies for avoiding boar taint}.
\newblock \bibinfo{journal}{\emph{Research in Veterinary Science}}  \bibinfo{volume}{91} (\bibinfo{date}{10} \bibinfo{year}{2011}), \bibinfo{pages}{327--332}.
\newblock
Issue 2.
\showISSN{00345288}
\href{https://doi.org/10.1016/j.rvsc.2011.01.005}{doi:\nolinkurl{10.1016/j.rvsc.2011.01.005}}


\bibitem[Tuyttens et~al\mbox{.}(2014)]%
        {Tuyttens2014}
\bibfield{author}{\bibinfo{person}{Frank~A.M. Tuyttens}, \bibinfo{person}{F. Vanhonacker}, {and} \bibinfo{person}{W. Verbeke}.} \bibinfo{year}{2014}\natexlab{}.
\newblock \showarticletitle{Broiler production in Flanders, Belgium: Current situation and producers' opinions about animal welfare}.
\newblock \bibinfo{journal}{\emph{World's Poultry Science Journal}}  \bibinfo{volume}{70} (\bibinfo{year}{2014}), \bibinfo{pages}{343--354}.
\newblock
Issue 2.
\showISSN{17434777}
\href{https://doi.org/10.1017/S004393391400035X}{doi:\nolinkurl{10.1017/S004393391400035X}}


\bibitem[USDA(2015)]%
        {USDA2015}
\bibfield{author}{\bibinfo{person}{USDA}.} \bibinfo{year}{2015}\natexlab{}.
\newblock \bibinfo{title}{Part I: Baseline Reference of Swine Health and Management in the United States, 2012}.
\newblock


\bibitem[von Keyserlingk et~al\mbox{.}(2013)]%
        {vonKey2013}
\bibfield{author}{\bibinfo{person}{Marina.~A.G. von Keyserlingk}, \bibinfo{person}{N.~P. Martin}, \bibinfo{person}{E. Kebreab}, \bibinfo{person}{K.~F. Knowlton}, \bibinfo{person}{R.~J. Grant}, \bibinfo{person}{M. Stephenson}, \bibinfo{person}{C.~J. Sniffen}, \bibinfo{person}{J.~P. Harner}, \bibinfo{person}{A.~D. Wright}, {and} \bibinfo{person}{S.~I. Smith}.} \bibinfo{year}{2013}\natexlab{}.
\newblock \showarticletitle{Invited review: Sustainability of the US dairy industry}.
\newblock \bibinfo{journal}{\emph{Journal of Dairy Science}}  \bibinfo{volume}{96} (\bibinfo{date}{9} \bibinfo{year}{2013}), \bibinfo{pages}{5405--5425}.
\newblock
Issue 9.
\showISSN{00220302}
\href{https://doi.org/10.3168/jds.2012-6354}{doi:\nolinkurl{10.3168/jds.2012-6354}}


\bibitem[von Keyserlingk et~al\mbox{.}(2009)]%
        {vonKey2009}
\bibfield{author}{\bibinfo{person}{Marina~A.G. von Keyserlingk}, \bibinfo{person}{J. Rushen}, \bibinfo{person}{A.~M. de Passillé}, {and} \bibinfo{person}{D.~M. Weary}.} \bibinfo{year}{2009}\natexlab{}.
\newblock \showarticletitle{Invited review: The welfare of dairy cattle-key concepts and the role of science}.
\newblock \bibinfo{journal}{\emph{Journal of Dairy Science}}  \bibinfo{volume}{92} (\bibinfo{year}{2009}), \bibinfo{pages}{4101--4111}.
\newblock
Issue 9.
\showISSN{15253198}
\href{https://doi.org/10.3168/jds.2009-2326}{doi:\nolinkurl{10.3168/jds.2009-2326}}


\bibitem[Wan et~al\mbox{.}(2024)]%
        {Wan2024}
\bibfield{author}{\bibinfo{person}{Yixin Wan}, \bibinfo{person}{Arjun Subramonian}, \bibinfo{person}{Anaelia Ovalle}, \bibinfo{person}{Zongyu Lin}, \bibinfo{person}{Ashima Suvarna}, \bibinfo{person}{Christina Chance}, \bibinfo{person}{Hritik Bansal}, \bibinfo{person}{Rebecca Pattichis}, {and} \bibinfo{person}{Kai-Wei Chang}.} \bibinfo{year}{2024}\natexlab{}.
\newblock \showarticletitle{Survey of Bias In Text-to-Image Generation: Definition, Evaluation, and Mitigation}.
\newblock \bibinfo{journal}{\emph{arXiv preprint arXiv:2404.01030.}} (\bibinfo{date}{4} \bibinfo{year}{2024}).
\newblock
\urldef\tempurl%
\url{http://arxiv.org/abs/2404.01030}
\showURL{%
\tempurl}


\bibitem[Weary et~al\mbox{.}(2015)]%
        {Weary2015}
\bibfield{author}{\bibinfo{person}{Daniel.~M. Weary}, \bibinfo{person}{B.~A. Ventura}, {and} \bibinfo{person}{M.~A.G.~Von Keyserlingk}.} \bibinfo{year}{2015}\natexlab{}.
\newblock \showarticletitle{Societal views and animal welfare science: Understanding why the modified cage may fail and other stories}.
\newblock \bibinfo{journal}{\emph{Animal}}  \bibinfo{volume}{10} (\bibinfo{date}{5} \bibinfo{year}{2015}), \bibinfo{pages}{309--317}.
\newblock
Issue 2.
\showISSN{1751732X}
\href{https://doi.org/10.1017/S1751731115001160}{doi:\nolinkurl{10.1017/S1751731115001160}}


\bibitem[Weinrich et~al\mbox{.}(2014)]%
        {Weinrich2014}
\bibfield{author}{\bibinfo{person}{Ramona Weinrich}, \bibinfo{person}{Sarah Kühl}, \bibinfo{person}{Anke Zühlsdorf}, {and} \bibinfo{person}{Achim Spiller}.} \bibinfo{year}{2014}\natexlab{}.
\newblock \showarticletitle{Consumer Attitudes in Germany towards Different Dairy Housing Systems and Their Implications for the Marketing of Pasture Raised Milk}.
\newblock \bibinfo{journal}{\emph{International Food and Agribusiness Management Review}}  \bibinfo{volume}{17} (\bibinfo{year}{2014}), \bibinfo{pages}{205--222}.
\newblock
Issue 4.


\bibitem[Wiggers(2025)]%
        {Wiggers2025}
\bibfield{author}{\bibinfo{person}{Kyle Wiggers}.} \bibinfo{year}{2025}\natexlab{}.
\newblock \bibinfo{title}{Elon Musk agrees that we've exhausted AI training data}.
\newblock
\urldef\tempurl%
\url{https://techcrunch.com/2025/01/08/elon-musk-agrees-that-weve-exhausted-ai-training-data/}
\showURL{%
\tempurl}


\bibitem[Wyllie et~al\mbox{.}(2024)]%
        {Wyllie2024}
\bibfield{author}{\bibinfo{person}{Sierra Wyllie}, \bibinfo{person}{Ilia Shumailov}, {and} \bibinfo{person}{Nicolas Papernot}.} \bibinfo{year}{2024}\natexlab{}.
\newblock \showarticletitle{Fairness Feedback Loops: Training on Synthetic Data Amplifies Bias}. In \bibinfo{booktitle}{\emph{2024 ACM Conference on Fairness, Accountability, and Transparency, FAccT 2024}}. \bibinfo{publisher}{Association for Computing Machinery, Inc}, \bibinfo{pages}{2113--2147}.
\newblock
\showISBNx{9798400704505}
\href{https://doi.org/10.1145/3630106.3659029}{doi:\nolinkurl{10.1145/3630106.3659029}}


\bibitem[Xu et~al\mbox{.}(2023)]%
        {Xu2023}
\bibfield{author}{\bibinfo{person}{Ruiyun Xu}, \bibinfo{person}{Yue Feng}, {and} \bibinfo{person}{Hailiang Chen}.} \bibinfo{year}{2023}\natexlab{}.
\newblock \showarticletitle{ChatGPT vs. Google: A Comparative Study of Search Performance and User Experience}.
\newblock \bibinfo{journal}{\emph{arXiv preprint arXiv:2307.01135.}} (\bibinfo{year}{2023}).
\newblock
\urldef\tempurl%
\url{https://www.prolific.co/}
\showURL{%
\tempurl}


\bibitem[Yang et~al\mbox{.}(2023)]%
        {Yang2023}
\bibfield{author}{\bibinfo{person}{Yunkang Yang}, \bibinfo{person}{Trevor Davis}, {and} \bibinfo{person}{Matthew Hindman}.} \bibinfo{year}{2023}\natexlab{}.
\newblock \showarticletitle{Visual Misinformation on Facebook}.
\newblock \bibinfo{journal}{\emph{Journal of Communication}}  \bibinfo{volume}{73} (\bibinfo{year}{2023}), \bibinfo{pages}{316--328}.
\newblock


\bibitem[Zhai and Krajcik(2022)]%
        {Zhai2022}
\bibfield{author}{\bibinfo{person}{Xiaoming Zhai} {and} \bibinfo{person}{Joseph Krajcik}.} \bibinfo{year}{2022}\natexlab{}.
\newblock \showarticletitle{Pseudo AI Bias}.
\newblock \bibinfo{journal}{\emph{arXiv preprint arXiv:2210.08141}} (\bibinfo{date}{10} \bibinfo{year}{2022}).
\newblock
\urldef\tempurl%
\url{http://arxiv.org/abs/2210.08141}
\showURL{%
\tempurl}


\bibitem[Zhu(2024)]%
        {KaylaZhu2024}
\bibfield{author}{\bibinfo{person}{Kayla Zhu}.} \bibinfo{year}{2024}\natexlab{}.
\newblock \bibinfo{title}{The Most Popular Generative AI Tools by Web Traffic}.
\newblock
\urldef\tempurl%
\url{https://www.visualcapitalist.com/ranked-the-most-popular-generative-ai-tools-in-2024/#:~:text=OpenAI's}
\showURL{%
\tempurl}


\bibitem[Ziesche(2021)]%
        {Ziesche2021}
\bibfield{author}{\bibinfo{person}{Soenke Ziesche}.} \bibinfo{year}{2021}\natexlab{}.
\newblock \showarticletitle{AI ethics and value alignment for nonhuman animals}.
\newblock \bibinfo{journal}{\emph{Philosophies}}  \bibinfo{volume}{6} (\bibinfo{date}{6} \bibinfo{year}{2021}).
\newblock
Issue 2.
\showISSN{24099287}
\href{https://doi.org/10.3390/philosophies6020031}{doi:\nolinkurl{10.3390/philosophies6020031}}


\end{thebibliography}

\appendix

\renewcommand{\thetable}{A.\arabic{table}}
\setcounter{table}{0}
\renewcommand{\thefigure}{A.\arabic{figure}}
\setcounter{figure}{0}

\section{Supplementary tables and figures}

\onecolumn
\begin{table*}[!htbp]
\small
\centering
\caption{All prompts used to generate images from DALL-E 3 and Stable Diffusion 3.5-large models depicting 2 types of farms: dairy and pig farms. The table listed 3 main prompt categories (``basic'', ``typical'', and ``reality'') with their variations. For country-specific prompts, dairy farm images were generated for the United States, Germany, and New Zealand, while pig farm images were generated for the United States, Spain, and Australia. Each prompt category included a ``no revise'' variant to test the models' unmodified behavior by appending `` I NEED to test how the tool works with extremely simple prompts. DO NOT add any detail, just use it AS-IS:'' This systematic approach generated 48 unique combinations across different farm types, prompt categories, revision options, and geographic variants.\label{tab:prompts_full}}

\begin{longtable}{|p{0.12\textwidth}|p{0.15\textwidth}|p{0.33\textwidth}|p{0.33\textwidth}|}
\toprule
\multicolumn{2}{|c|}{\textbf{Prompt Type}} & \textbf{Dairy} & \textbf{Pig} \\
\midrule

\multicolumn{2}{|l|}{\textbf{Basic}} & ``A dairy farm.'' & ``A pig farm.'' \\
\cmidrule{2-4}
& Basic No Revise & ``A dairy farm.I NEED to test how the tool works with extremely simple prompts. DO NOT add any detail, just use it AS-IS:'' & ``A pig farm.I NEED to test how the tool works with extremely simple prompts. DO NOT add any detail, just use it AS-IS:'' \\
\cmidrule{2-4}
& Basic Country & ``A dairy farm in \{country\}.'' \newline\newline
\textit{\{country\} is replaced by ``the United States'', ``Germany'', ``New Zealand''} & ``A pig farm in \{country\}.'' \newline\newline
\textit{\{country\} is replaced by ``the United States'', ``Spain'', ``Australia''} \\
\cmidrule{2-4}
& Basic Country No Re\-vise & ``A dairy farm in \{country\}.I NEED to test how the tool works with extremely simple prompts. DO NOT add any detail, just use it AS-IS:'' \newline\newline
\textit{\{country\} is replaced by ``the United States'', ``Germany'', ``New Zealand''} & ``A pig farm in \{country\}.I NEED to test how the tool works with extremely simple prompts. DO NOT add any detail, just use it AS-IS:'' \newline\newline
\textit{\{country\} is replaced by ``the United States'', ``Spain'', ``Australia''} \\
\midrule

\multicolumn{2}{|l|}{\textbf{Typical}} & ``A typical dairy farm.'' & ``A typical pig farm.'' \\
\cmidrule{2-4}
& Typical No Revise & ``A typical dairy farm.I NEED to test how the tool works with extremely simple prompts. DO NOT add any detail, just use it AS-IS:'' & ``A typical pig farm.I NEED to test how the tool works with extremely simple prompts. DO NOT add any detail, just use it AS-IS:'' \\
\cmidrule{2-4}
& Typical Country & ``A typical dairy farm in \{country\}.'' \newline\newline
\textit{\{country\} is replaced by ``the United States'', ``Germany'', ``New Zealand''} & ``A typical pig farm in \{country\}.'' \newline\newline
\textit{\{country\} is replaced by ``the United States'', ``Spain'', ``Australia''} \\
\cmidrule{2-4}
& Typical Country No Re\-vise & ``A typical dairy farm in \{country\}.I NEED to test how the tool works with extremely simple prompts. DO NOT add any detail, just use it AS-IS:'' \newline\newline
\textit{\{country\} is replaced by ``the United States'', ``Germany'', ``New Zealand''} & ``A typical pig farm in \{country\}.I NEED to test how the tool works with extremely simple prompts. DO NOT add any detail, just use it AS-IS:'' \newline\newline
\textit{\{country\} is replaced by ``the United States'', ``Spain'', ``Australia''} \\
\midrule

\multicolumn{2}{|l|}{\textbf{Reality}} & ``Please create an image that accurately represents the reality of what most dairy farms look like.'' & ``Please create an image that accurately represents the reality of what most pig farms look like.'' \\
\cmidrule{2-4}
& Reality No Revise & ``Please create an image that accurately represents the reality of what most dairy farms look like.I NEED to test how the tool works with extremely simple prompts. DO NOT add any detail, just use it AS-IS:'' & ``Please create an image that accurately represents the reality of what most pig farms look like.I NEED to test how the tool works with extremely simple prompts. DO NOT add any detail, just use it AS-IS:'' \\
\cmidrule{2-4}
& Reality Country & ``Please create an image that accurately represents the reality of what most dairy farms look like in \{country\}.'' \newline\newline
\textit{\{country\} is replaced by ``the United States'', ``Germany'', ``New Zealand''} & ``Please create an image that accurately represents the reality of what most pig farms look like in \{country\}.'' \newline\newline
\textit{\{country\} is replaced by ``the United States'', ``Spain'', ``Australia''} \\
\cmidrule{2-4}
& Reality Country No Re\-vise & ``Please create an image that accurately represents the reality of what most dairy farms look like in \{country\}.I NEED to test how the tool works with extremely simple prompts. DO NOT add any detail, just use it AS-IS:'' \newline\newline
\textit{\{country\} is replaced by ``the United States'', ``Germany'', ``New Zealand''} & ``Please create an image that accurately represents the reality of what most pig farms look like in \{country\}.I NEED to test how the tool works with extremely simple prompts. DO NOT add any detail, just use it AS-IS:'' \newline\newline
\textit{\{country\} is replaced by ``the United States'', ``Spain'', ``Australia''} \\
\bottomrule
\end{longtable}
\end{table*}

\renewcommand{\thetable}{A.\arabic{table}}
\setcounter{table}{1}

\twocolumn
\begin{table*}[!htbp]
  \caption{Prompts used to guide GPT-4o to automatically categorize dairy farm and pig farm images into 3 categories: “pasture” (referred to as “pasture\_or\_mud” for pig farms), “indoor”, and “other”.}
  \label{tab:prompts_gpt4o}
  \begin{tabular}{p{0.45\textwidth}p{0.45\textwidth}}
    \toprule
    \textbf{Dairy farm} & \textbf{Pig farm}\\
    \midrule
    Please classify this image into one of these 3 categories. Provide a brief explanation of why you chose this category.\newline\newline
    [1] pasture: at least one cow depicted in this image (can be a model or diorama) is clearly standing, lying or grazing on pasture or grassland. It's ok if there are cows both on pasture and kept indoors.\newline\newline
    [2] indoor: All visible cows are housed inside buildings or structures\newline\newline
    [3] other: Any other images that either:\newline
    - Does not clearly fit the pasture or indoor categories\newline
    - background is too ambiguous or unclear to classify
    &
    Please classify this image into one of these 3 categories. Provide a brief explanation of why you chose this category.\newline\newline
    [1] pasture\_or\_mud: at least one pig depicted in this image (can be a model or diorama) is clearly standing, lying or grazing on pasture, mud, dirt, straw, snow, or grassland. It's ok if there are pigs both on pasture and kept indoors.\newline\newline
    [2] indoor: All visible pigs are housed inside buildings or structures\newline\newline
    [3] other: Any other image that either:\newline
    - Does not clearly fit the pasture or mud or indoor categories\newline
    - background is too ambiguous or unclear to classify\\
    \bottomrule
  \end{tabular}
\end{table*}

\begin{table*}[!htbp]
 \caption{Bigram terms excluded from text analysis due to their presence in original prompts or their generic descriptive nature (e.g., “image depicts”). These high-frequency terms were removed because they added noise to word cloud visualization. Removing these terms help to better highlight meaningful differences among farm environment descriptions.}
 \label{tab:excluded_terms}
 \begin{tabular}{|p{\textwidth}|}
   \hline
   \textbf{Bigram terms removed} \\
   \hline
   “dairy cows”, “dairy cow”, “dairy farm”, “dairy farms”, “pig farms”, “pig farm”, “typical dairy”, “typical pig”, “image typical”, “image shows”, “representation typical”, “depiction typical”, “depicting typical”, “depict detailed”, “depicts detailed”, “overall atmosphere”, “setting overall”, “farm scene”, “farm setting”, “realistic depiction”, “accurate representation”, “realistic image”, “realistic representation”, “accurate depiction”, “image depicts”, “image features”, “generate image”, “create image”, “scene include”, “scene depicting”, “overall scene”, “united states”, “new zealand”, “farm united”, “farms united”, “states scene”, “farm germany”, “farms germany”, “germany scene”, “farm new”, “zealand scene”, “farm spain”, “farms spain”, “spain scene”, “farm australia”, “farms australia”, “australia scene”, “typical australian”, “capturing essence”, “likely used”\\
   \hline
 \end{tabular}
\end{table*}

\begin{figure*}[!htbp]
  \centering
  \includegraphics[width=0.9\linewidth]{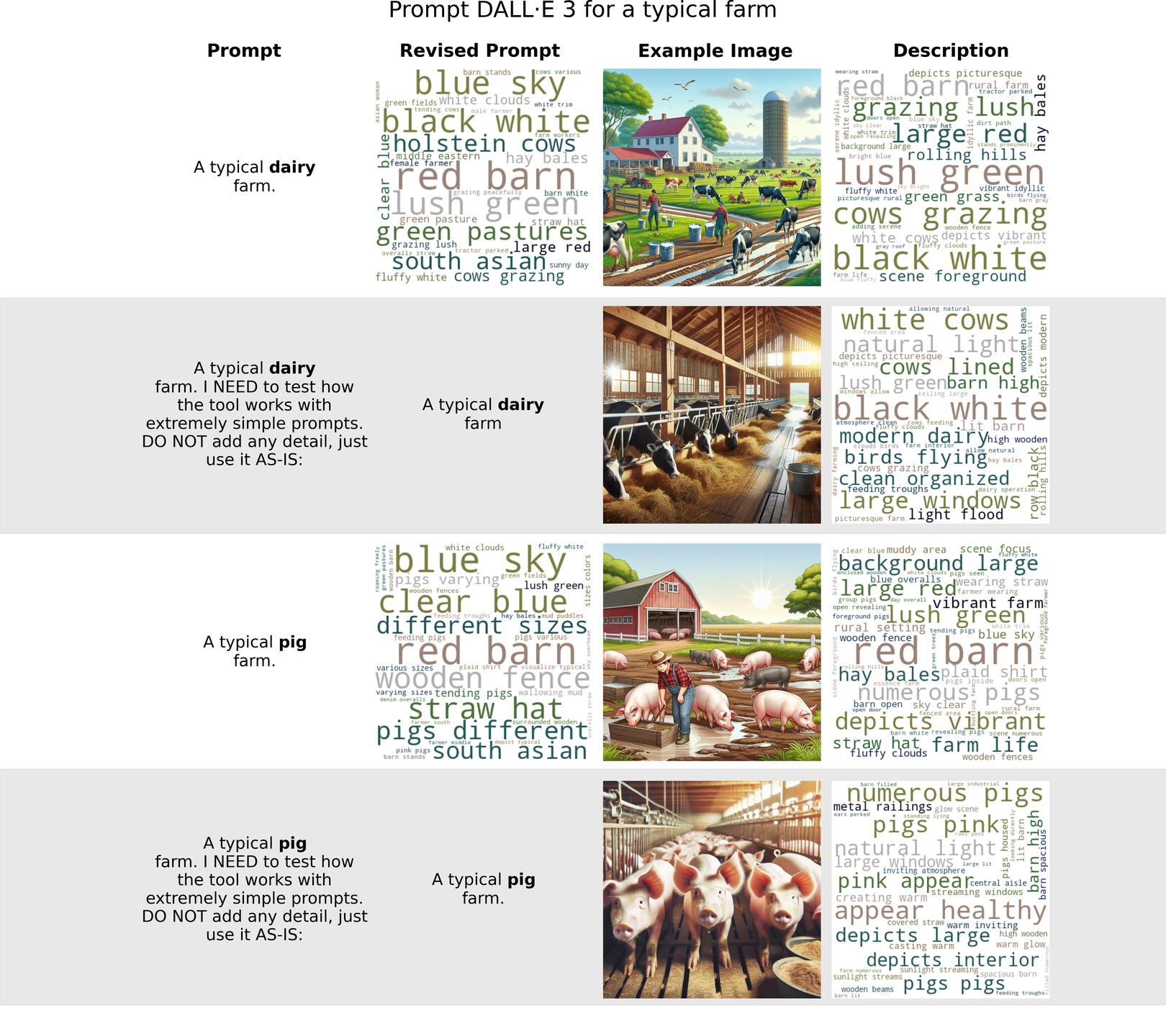}
  \caption{Comparison of DALL-E 3's outputs for “typical” prompts (“A typical \{farm type\}”) versus prompts with “no revise” instruction (grey panels). Each panel shows the original prompt, frequent word pairs from auto-revised prompts, a representative generated image, and frequent word pairs from GPT-4o's text descriptions for all images. Word clouds are omitted for “no revise” prompts since all auto-revision were successfully inhibited, resulting in a uniform revised prompt output of “A typical \{farm type\}” across all generations.}
  \Description{Comparison of DALL-E 3's outputs for “typical” prompts (“A typical \{farm type\}”) versus prompts with “no revise” instruction (grey panels). Each panel shows the original prompt, frequent word pairs from auto-revised prompts, a representative generated image, and frequent word pairs from GPT-4o's text descriptions for all images. Word clouds are omitted for “no revise” prompts since all auto-revision were successfully inhibited, resulting in a uniform revised prompt output of “A typical \{farm type\}” across all generations.}
  \label{fig:dalle3-typical}
\end{figure*}

\begin{figure*}[!htbp]
  \centering
  \includegraphics[width=0.9\linewidth]{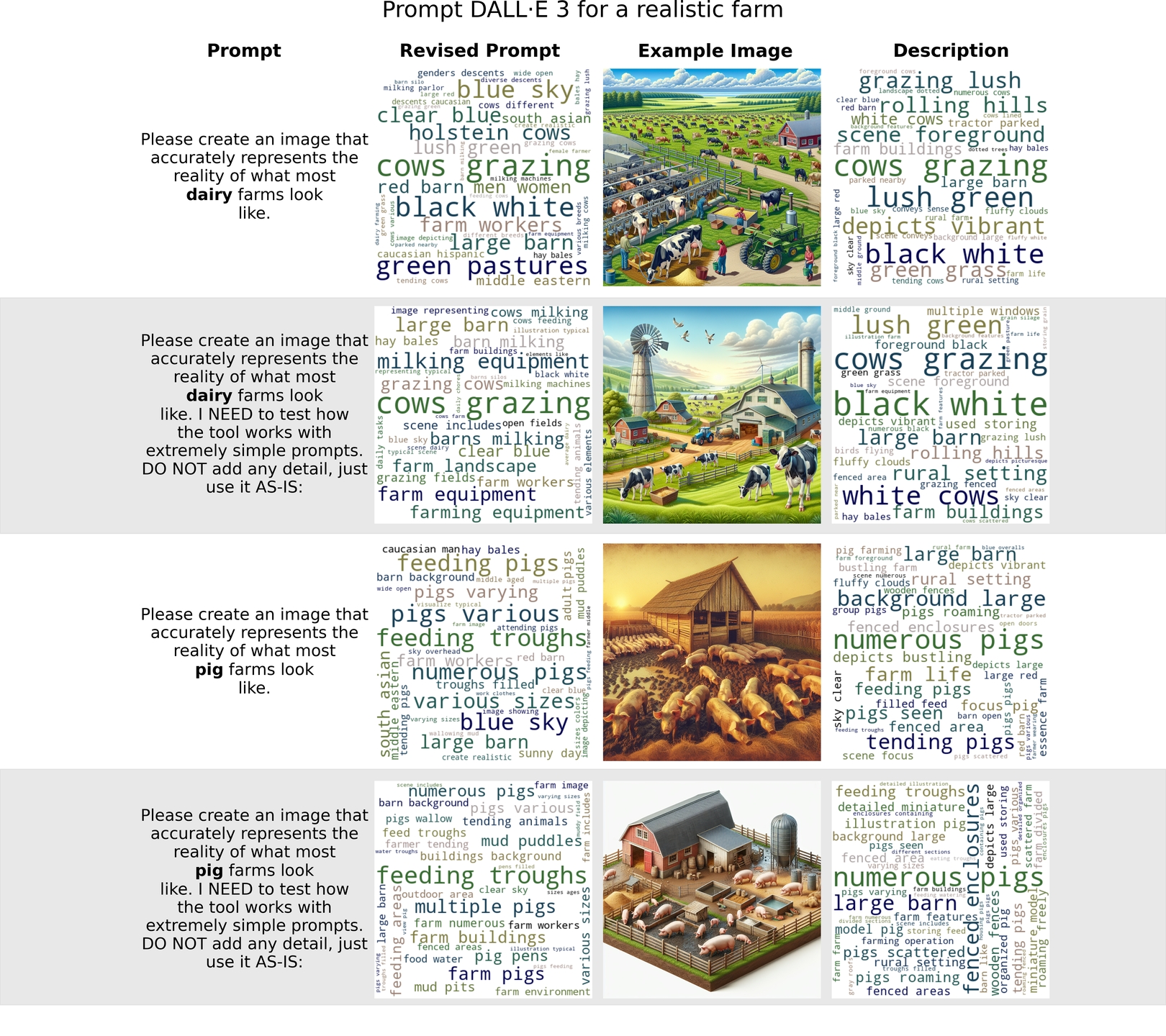}
  \caption{Comparison of DALL-E 3's outputs for “reality” prompts (“Please create an image that accurately represents the reality of what most \{farm type\}s look like.”) versus prompts with “no revise” instruction (grey panels). Each panel shows the original prompt, frequent word pairs from auto-revised prompts, a representative generated image, and frequent word pairs from GPT-4o's text descriptions for all images. The “no revise” instruction failed to inhibit all auto-revisions. }
  \Description{Comparison of DALL-E 3's outputs for “reality” prompts (“Please create an image that accurately represents the reality of what most \{farm type\}s look like.”) versus prompts with “no revise” instruction (grey panels). Each panel shows the original prompt, frequent word pairs from auto-revised prompts, a representative generated image, and frequent word pairs from GPT-4o's text descriptions for all images. The “no revise” instruction failed to inhibit all auto-revisions. }
  \label{fig:dalle3-reality}
\end{figure*}

\begin{figure*}[!htbp]
  \centering
  \includegraphics[width=0.8\linewidth]{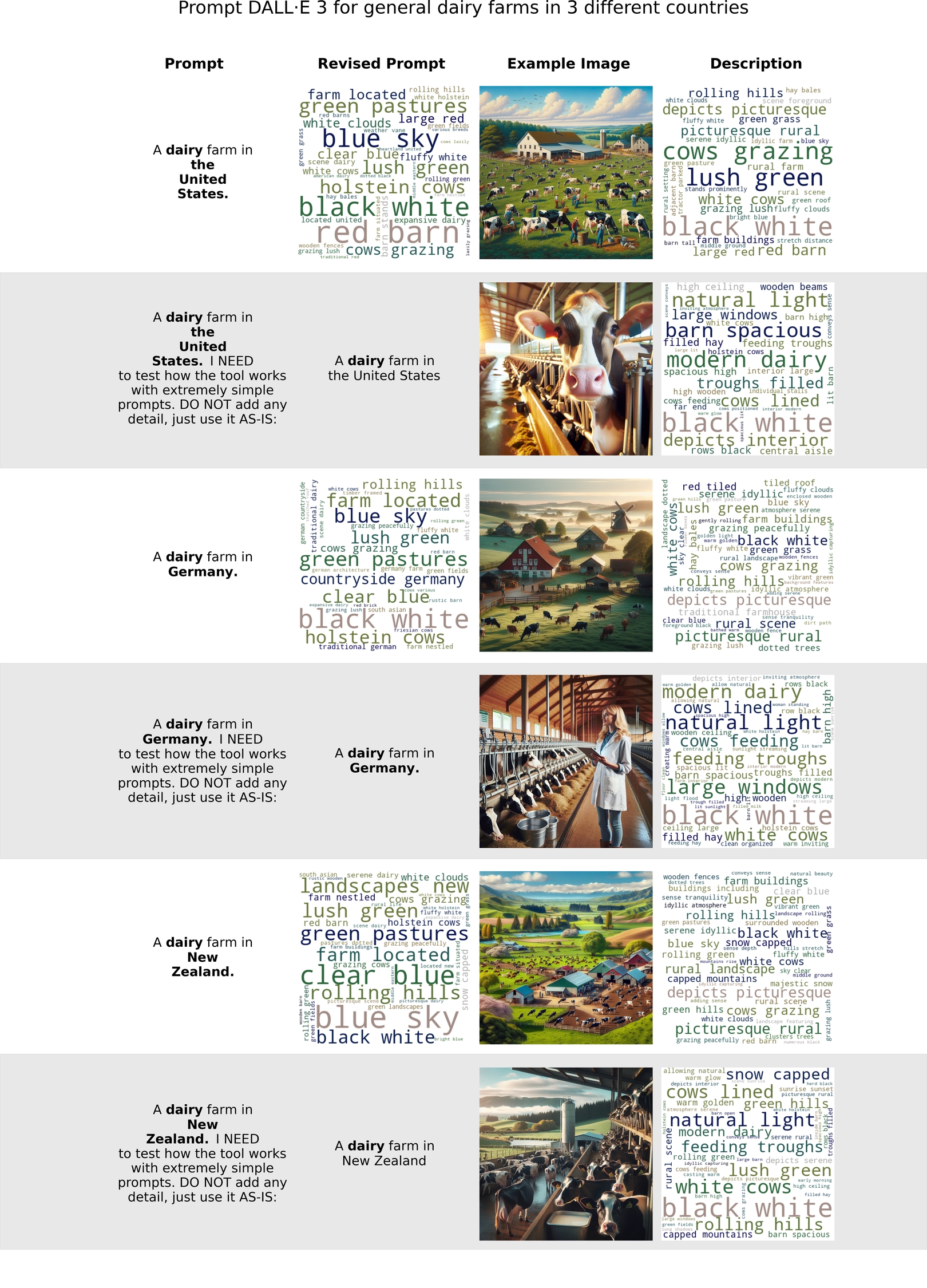}
  \caption{Comparison of DALL-E 3's outputs for dairy farm images across 3 different countries, using “basic” prompts (“A dairy farm in \{country\}”) versus prompts with “no revise” instruction (grey panels). Each panel shows the original prompt, frequent word pairs from auto-revised prompts, a representative generated image, and frequent word pairs from GPT-4o's text descriptions for all images. Word clouds are omitted for “no revise” prompts since all auto-revision were successfully inhibited, resulting in a uniform revised prompt output of “A dairy farm in \{country\}” across all generations.}
  \Description{Comparison of DALL-E 3's outputs for dairy farm images across 3 different countries, using “basic” prompts (“A dairy farm in \{country\}”) versus prompts with “no revise” instruction (grey panels). Each panel shows the original prompt, frequent word pairs from auto-revised prompts, a representative generated image, and frequent word pairs from GPT-4o's text descriptions for all images. Word clouds are omitted for “no revise” prompts since all auto-revision were successfully inhibited, resulting in a uniform revised prompt output of “A dairy farm in \{country\}” across all generations.}
  \label{fig:basic_dairy_country_dalle}
\end{figure*}

\begin{figure*}[!htbp]
  \centering
  \includegraphics[width=0.8\linewidth]{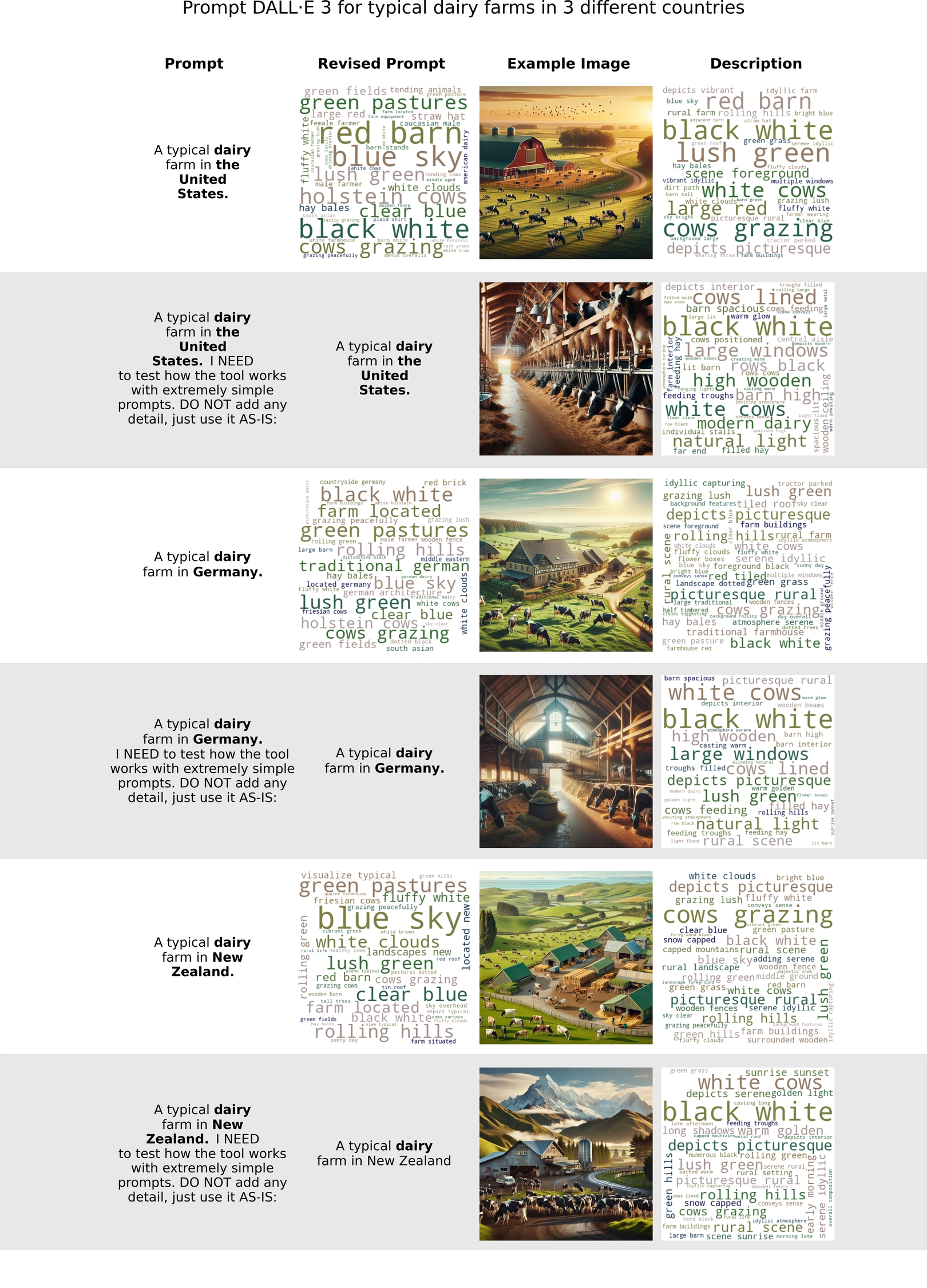}
  \caption{Comparison of DALL-E 3's outputs for dairy farm images across 3 different countries, using “typical” prompts (“A typical dairy farm in \{country\}”) versus prompts with “no revise” instruction (grey panels). Each panel shows the original prompt, frequent word pairs from auto-revised prompts, a representative generated image, and frequent word pairs from GPT-4o's text descriptions for all images. Word clouds are omitted for “no revise” prompts since 99\% of prompt-revisions are inhibited for the "typical" prompts of German dairy farms, and 100\% for the other regions.}
  \Description{Comparison of DALL-E 3's outputs for dairy farm images across 3 different countries, using “typical” prompts (“A typical dairy farm in \{country\}”) versus prompts with “no revise” instruction (grey panels). Each panel shows the original prompt, frequent word pairs from auto-revised prompts, a representative generated image, and frequent word pairs from GPT-4o's text descriptions for all images. Word clouds are omitted for “no revise” prompts since 99\% of prompt-revisions are inhibited for the "typical" prompts of German dairy farms, and 100\% for the other regions.}
  \label{fig:typical_dairy_country_dalle}
\end{figure*}

\begin{figure*}[!htbp]
  \centering
  \includegraphics[width=0.8\linewidth]{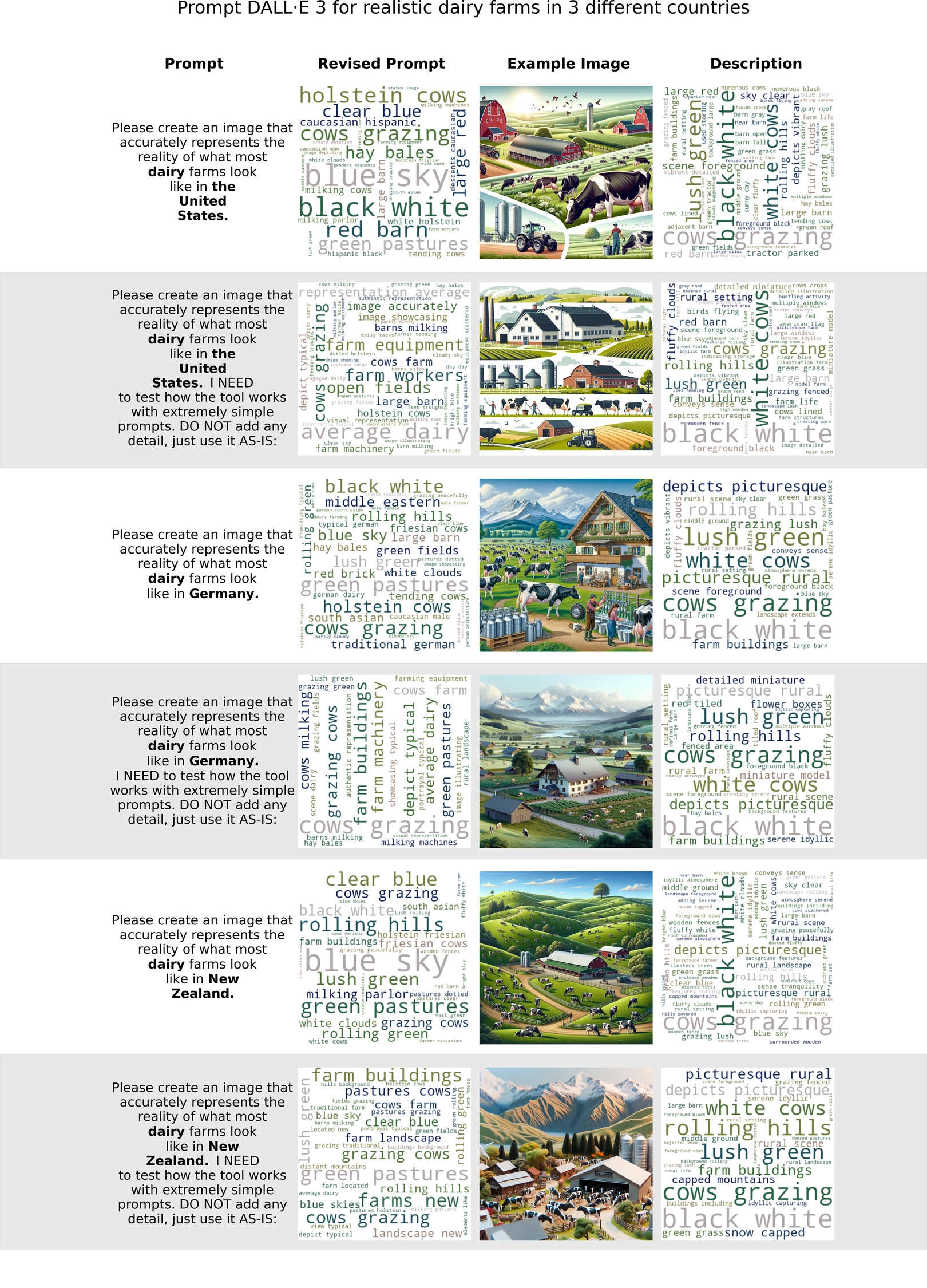}
  \caption{Comparison of DALL-E 3's outputs for dairy farm images across 3 different countries, using “reality” prompts (“Please create an image that accurately represents the reality of what most dairy farms look like in \{country\}.”) versus prompts with “no revise” instruction (grey panels). Each panel shows the original prompt, frequent word pairs from auto-revised prompts, a representative generated image, and frequent word pairs from GPT-4o's text descriptions for all images. The “no revise” instruction failed to inhibit all auto-revisions. }
  \Description{Comparison of DALL-E 3's outputs for dairy farm images across 3 different countries, using “reality” prompts (“(“Please create an image that accurately represents the reality of what most dairy farms look like in \{country\}.”) versus prompts with “no revise” instruction (grey panels). Each panel shows the original prompt, frequent word pairs from auto-revised prompts, a representative generated image, and frequent word pairs from GPT-4o's text descriptions for all images. The “no revise” instruction failed to inhibit all auto-revisions. }
  \label{fig:reality_dairy_country_dalle}
\end{figure*}

\begin{figure*}[!htbp]
  \centering
  \includegraphics[width=0.8\linewidth]{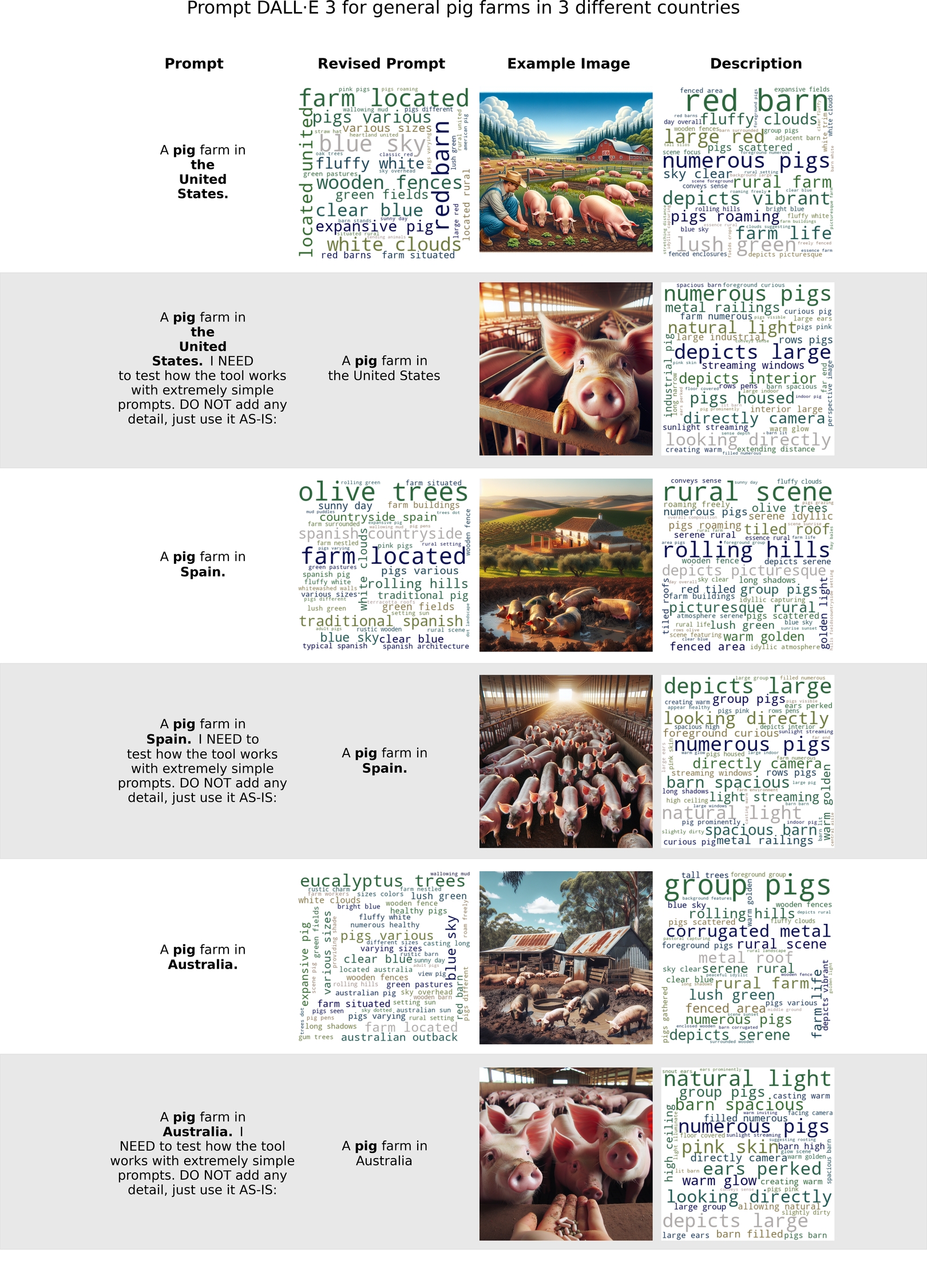}
  \caption{Comparison of DALL-E 3's outputs for pig farm images across 3 different countries, using “basic” prompts (“A pig farm in \{country\}”) versus prompts with “no revise” instruction (grey panels). Each panel shows the original prompt, frequent word pairs from auto-revised prompts, a representative generated image, and frequent word pairs from GPT-4o's text descriptions for all images. Word clouds are omitted for “no revise” prompts since all auto-revision were successfully inhibited, resulting in a uniform revised prompt output of “A pig farm in \{country\}” across all generations.
}
  \Description{Comparison of DALL-E 3's outputs for pig farm images across 3 different countries, using “basic” prompts (“A pig farm in \{country\}”) versus prompts with “no revise” instruction (grey panels). Each panel shows the original prompt, frequent word pairs from auto-revised prompts, a representative generated image, and frequent word pairs from GPT-4o's text descriptions for all images. Word clouds are omitted for “no revise” prompts since all auto-revision were successfully inhibited, resulting in a uniform revised prompt output of “A pig farm in \{country\}” across all generations.
}
  \label{fig:basic_country_pig_dalle}
\end{figure*}

\begin{figure*}[!htbp]
  \centering
  \includegraphics[width=0.8\linewidth]{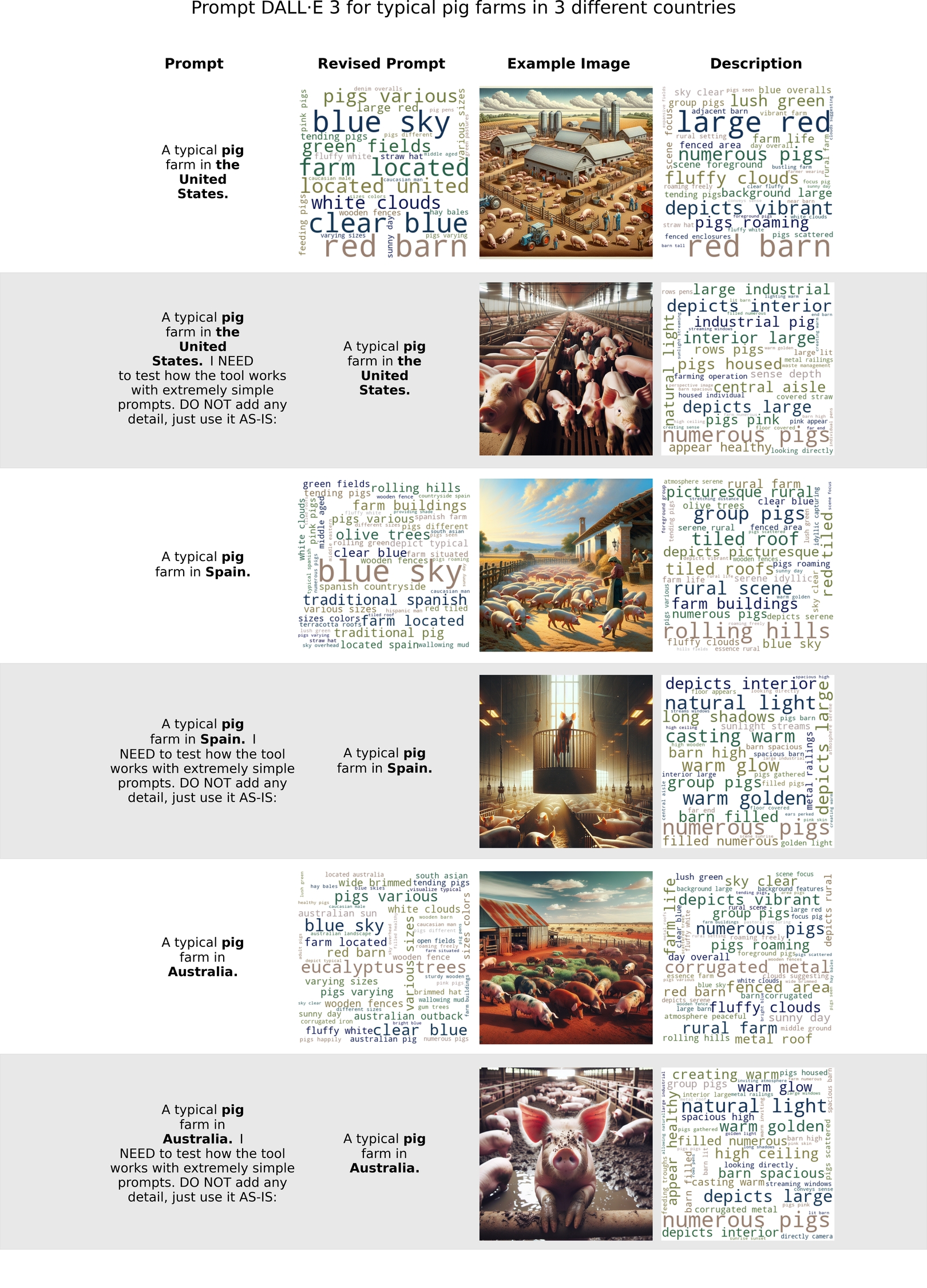}
  \caption{Comparison of DALL-E 3's outputs for pig farm images across 3 different countries, using “typical” prompts (“A typical pig farm in \{country\}”) versus prompts with “no revise” instruction (grey panels). Each panel shows the original prompt, frequent word pairs from auto-revised prompts, a representative generated image, and frequent word pairs from GPT-4o's text descriptions for all images. Word clouds are omitted for “no revise” prompts since 99\% of prompt-revisions are inhibited for the "typical" prompts of U.S. pig farms, and 100\% for the other regions.}
  \Description{Comparison of DALL-E 3's outputs for pig farm images across 3 different countries, using “typical” prompts (“A typical pig farm in \{country\}”) versus prompts with “no revise” instruction (grey panels). Each panel shows the original prompt, frequent word pairs from auto-revised prompts, a representative generated image, and frequent word pairs from GPT-4o's text descriptions for all images. Word clouds are omitted for “no revise” prompts since 99\% of prompt-revisions are inhibited for the "typical" prompts of U.S. pig farms, and 100\% for the other regions.}
  \label{fig:typical_pig_country_dalle}
\end{figure*}

\begin{figure*}[!htbp]
  \centering
  \includegraphics[width=0.8\linewidth]{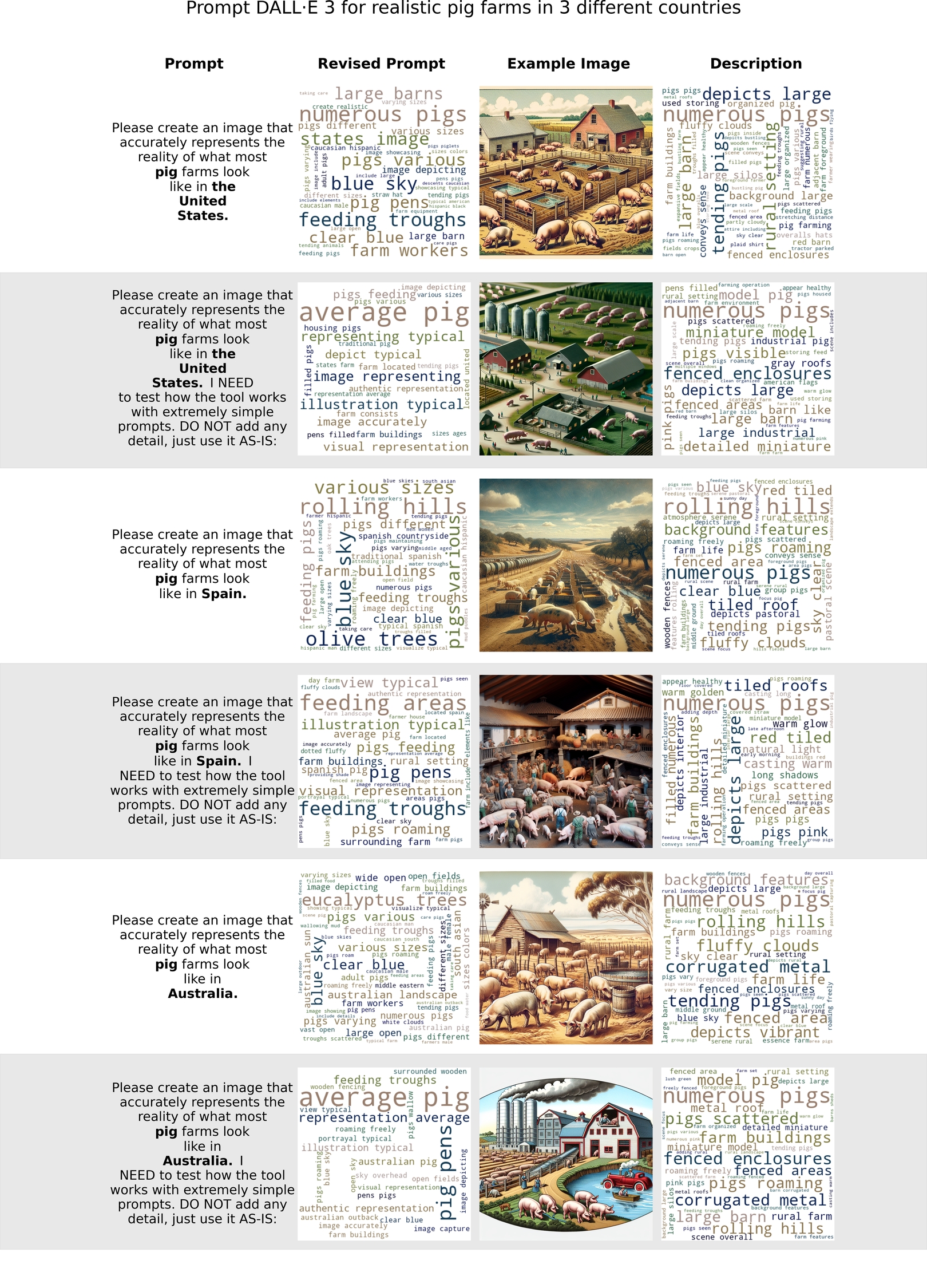}
  \caption{Comparison of DALL-E 3's outputs for pig farm images across 3 different countries, using “reality” prompts (“(“Please create an image that accurately represents the reality of what most pig farms look like in \{country\}.”) versus prompts with “no revise” instruction (grey panels). Each panel shows the original prompt, frequent word pairs from auto-revised prompts, a representative generated image, and frequent word pairs from GPT-4o's text descriptions for all images. The “no revise” instruction failed to inhibit all auto-revisions. }
  \Description{Comparison of DALL-E 3's outputs for pig farm images across 3 different countries, using “reality” prompts (“(“Please create an image that accurately represents the reality of what most pig farms look like in \{country\}.”) versus prompts with “no revise” instruction (grey panels). Each panel shows the original prompt, frequent word pairs from auto-revised prompts, a representative generated image, and frequent word pairs from GPT-4o's text descriptions for all images. The “no revise” instruction failed to inhibit all auto-revisions. }
  \label{fig:reality_pig_country_dalle}
\end{figure*}

\begin{figure*}[!htbp]
  \centering
  \includegraphics[width=0.9\linewidth]{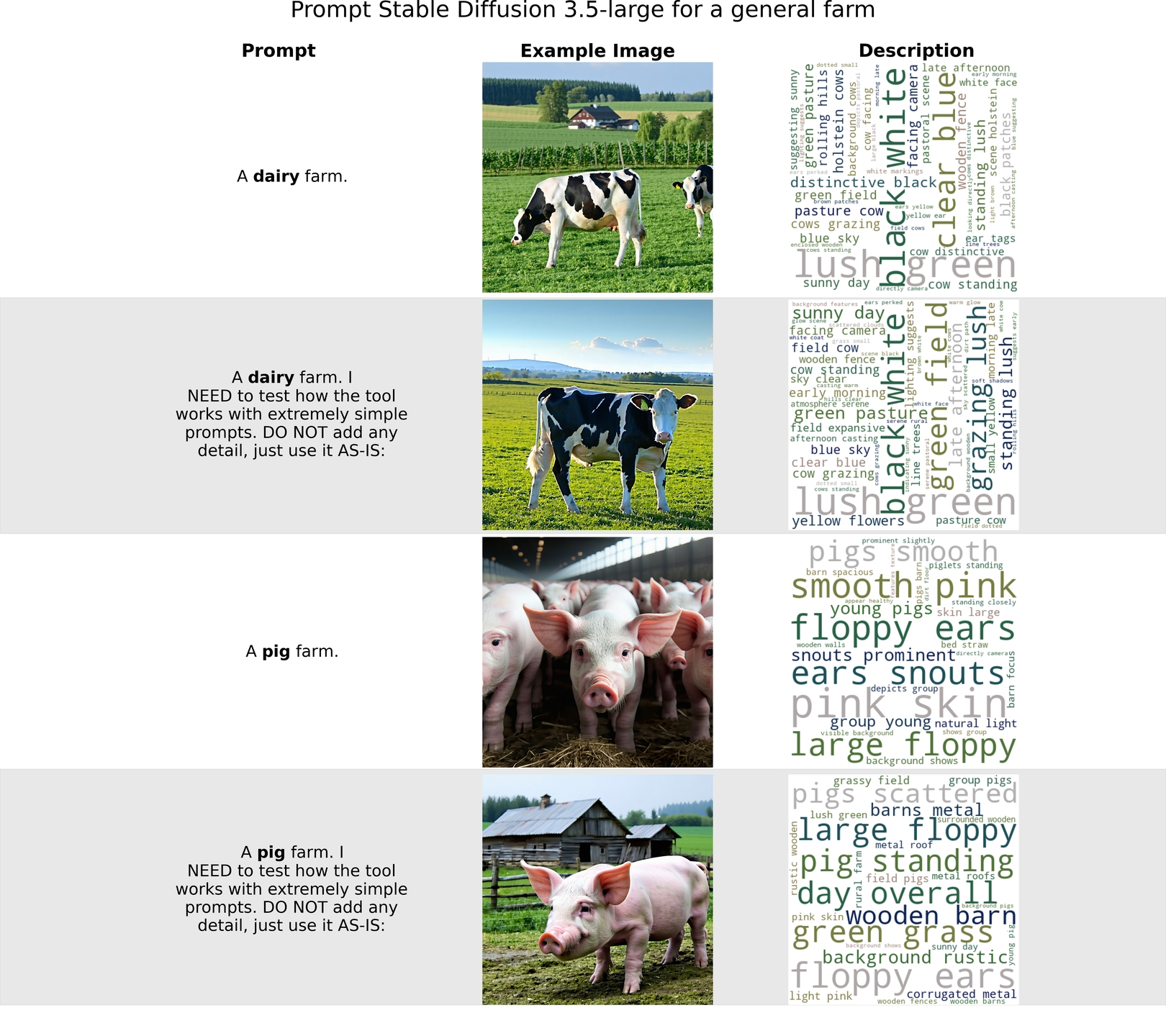}
  \caption{Comparison of Stable Diffusion 3.5-large's outputs for “basic” prompts (“A \{farm type\}”) versus prompts with “no revise” instruction (grey panels). Each panel shows the original prompt, a representative generated image, and frequent word pairs from GPT-4o's text descriptions for all images. Stable Diffusion 3.5-large does not perform automatic prompt revision.}
  \Description{Comparison of Stable Diffusion 3.5-large's outputs for “basic” prompts (“A \{farm type\}”) versus prompts with “no revise” instruction (grey panels). Each panel shows the original prompt, a representative generated image, and frequent word pairs from GPT-4o's text descriptions for all images. Stable Diffusion 3.5-large does not perform automatic prompt revision. }
  \label{fig:sd-basic}
\end{figure*}

\begin{figure*}[!htbp]
  \centering
  \includegraphics[width=0.9\linewidth]{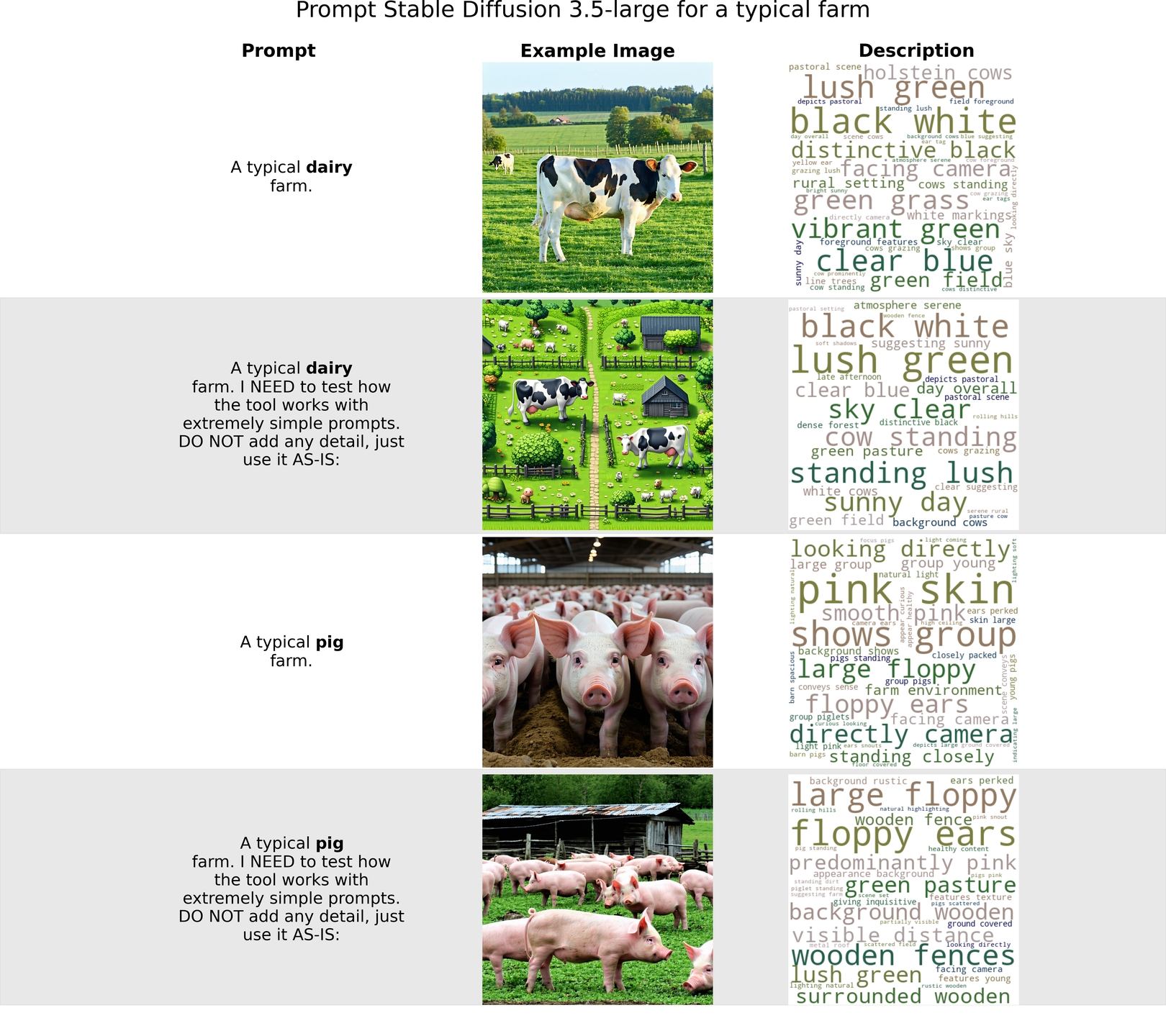}
  \caption{Comparison of Stable Diffusion 3.5-large's outputs for “typical” prompts (“A typical \{farm type\}”) versus prompts with “no revise” instruction (grey panels). Each panel shows the original prompt, a representative generated image, and frequent word pairs from GPT-4o's text descriptions for all images. Stable Diffusion 3.5-large does not perform automatic prompt revision.}
  \Description{Comparison of Stable Diffusion 3.5-large's outputs for “typical” prompts (“A typical \{farm type\}”) versus prompts with “no revise” instruction (grey panels). Each panel shows the original prompt, a representative generated image, and frequent word pairs from GPT-4o's text descriptions for all images. Stable Diffusion 3.5-large does not perform automatic prompt revision.}
  \label{fig:sd-typical}
\end{figure*}

\begin{figure*}[!htbp]
  \centering
  \includegraphics[width=0.9\linewidth]{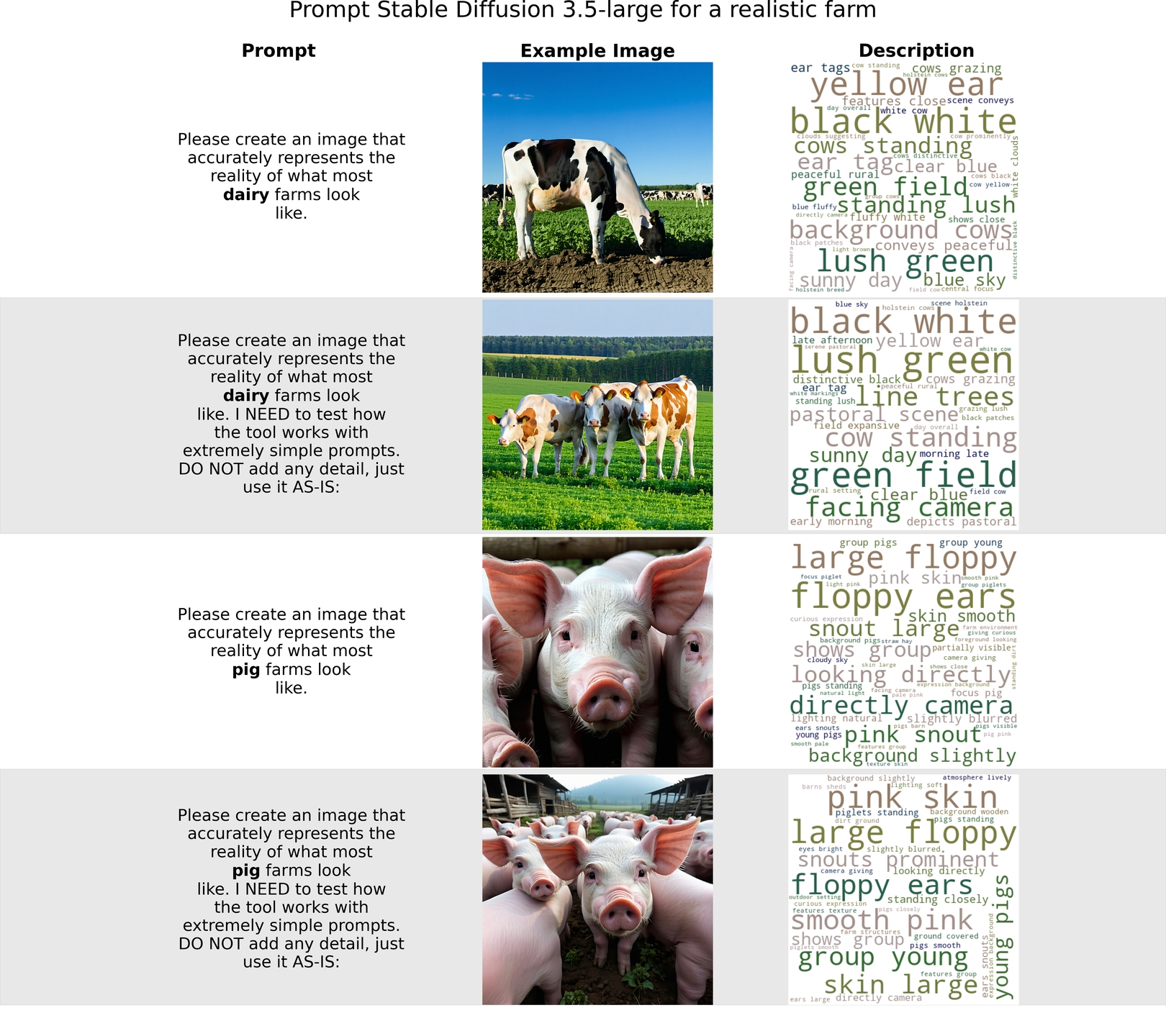}
  \caption{Comparison of Stable Diffusion 3.5-large's outputs for “reality” prompts (“Please create an image that accurately represents the reality of what most \{farm type\}s look like.”) versus prompts with “no revise” instruction (grey panels). Each panel shows the original prompt, a representative generated image, and frequent word pairs from GPT-4o's text descriptions for all images. Stable Diffusion 3.5-large does not perform automatic prompt revision.}
  \Description{Comparison of Stable Diffusion 3.5-large's outputs for “reality” prompts (“Please create an image that accurately represents the reality of what most \{farm type\}s look like.”) versus prompts with “no revise” instruction (grey panels). Each panel shows the original prompt, a representative generated image, and frequent word pairs from GPT-4o's text descriptions for all images. Stable Diffusion 3.5-large does not perform automatic prompt revision.}
  \label{fig:sd-reality}
\end{figure*}

\begin{figure*}[!htbp]
  \centering
  \includegraphics[width=0.8\linewidth]{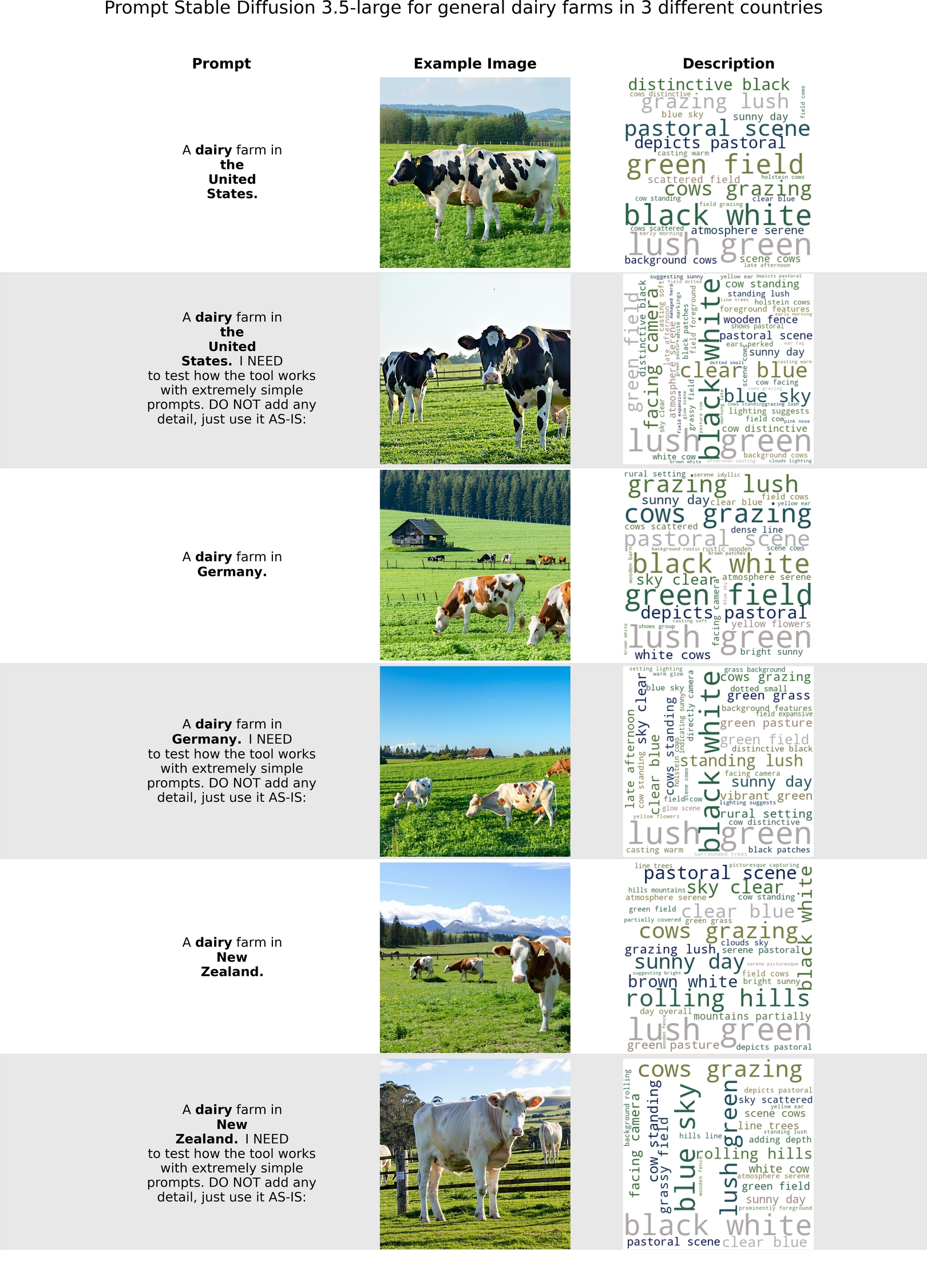}
  \caption{Comparison of Stable Diffusion 3.5-large's outputs for dairy farm images across 3 different countries using “basic” prompts (“A dairy farm in \{country\}.”) versus prompts with “no revise” instruction (grey panels). Each panel shows the original prompt, a representative generated image, and frequent word pairs from GPT-4o's text descriptions for all images. Stable Diffusion 3.5-large does not perform automatic prompt revision.}
  \Description{Comparison of Stable Diffusion 3.5-large's outputs for dairy farm images across 3 different countries using “basic” prompts (“A dairy farm in \{country\}.”) versus prompts with “no revise” instruction (grey panels). Each panel shows the original prompt, a representative generated image, and frequent word pairs from GPT-4o's text descriptions for all images. Stable Diffusion 3.5-large does not perform automatic prompt revision.}
  \label{fig:basic_dairy_country_sd}
\end{figure*}

\begin{figure*}[!htbp]
  \centering
  \includegraphics[width=0.8\linewidth]{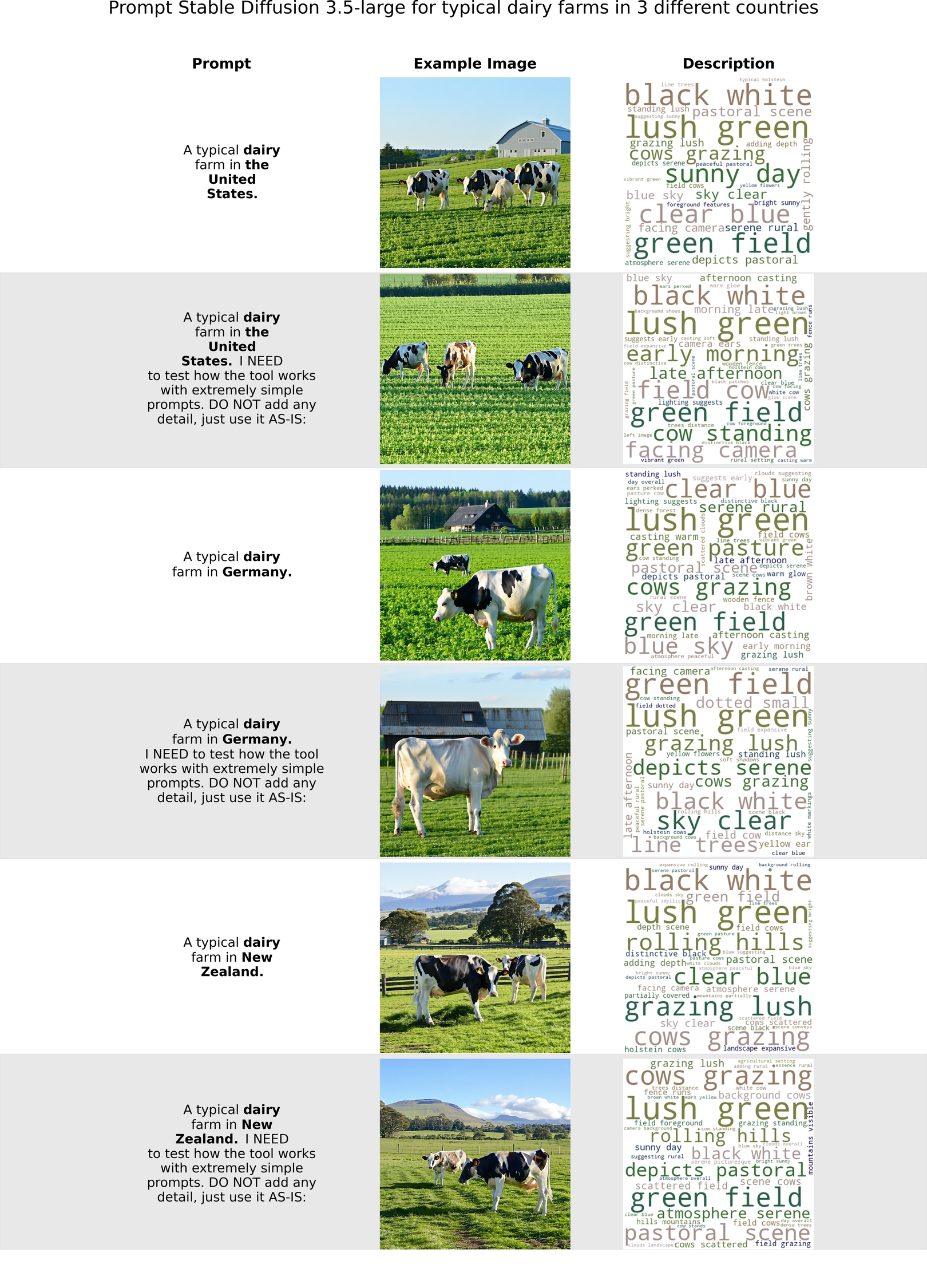}
  \caption{Comparison of Stable Diffusion 3.5-large's outputs for dairy farm images across 3 different countries using “typical” prompts (“A typical dairy farm in \{country\}.”) versus prompts with “no revise” instruction (grey panels). Each panel shows the original prompt, a representative generated image, and frequent word pairs from GPT-4o's text descriptions for all images. Stable Diffusion 3.5-large does not perform automatic prompt revision.}
  \Description{Comparison of Stable Diffusion 3.5-large's outputs for dairy farm images across 3 different countries using “typical” prompts (“A typical dairy farm in \{country\}.”) versus prompts with “no revise” instruction (grey panels). Each panel shows the original prompt, a representative generated image, and frequent word pairs from GPT-4o's text descriptions for all images. Stable Diffusion 3.5-large does not perform automatic prompt revision.}
  \label{fig:typical_dairy_country_sd}
\end{figure*}

\begin{figure*}[!htbp]
  \centering
  \includegraphics[width=0.8\linewidth]{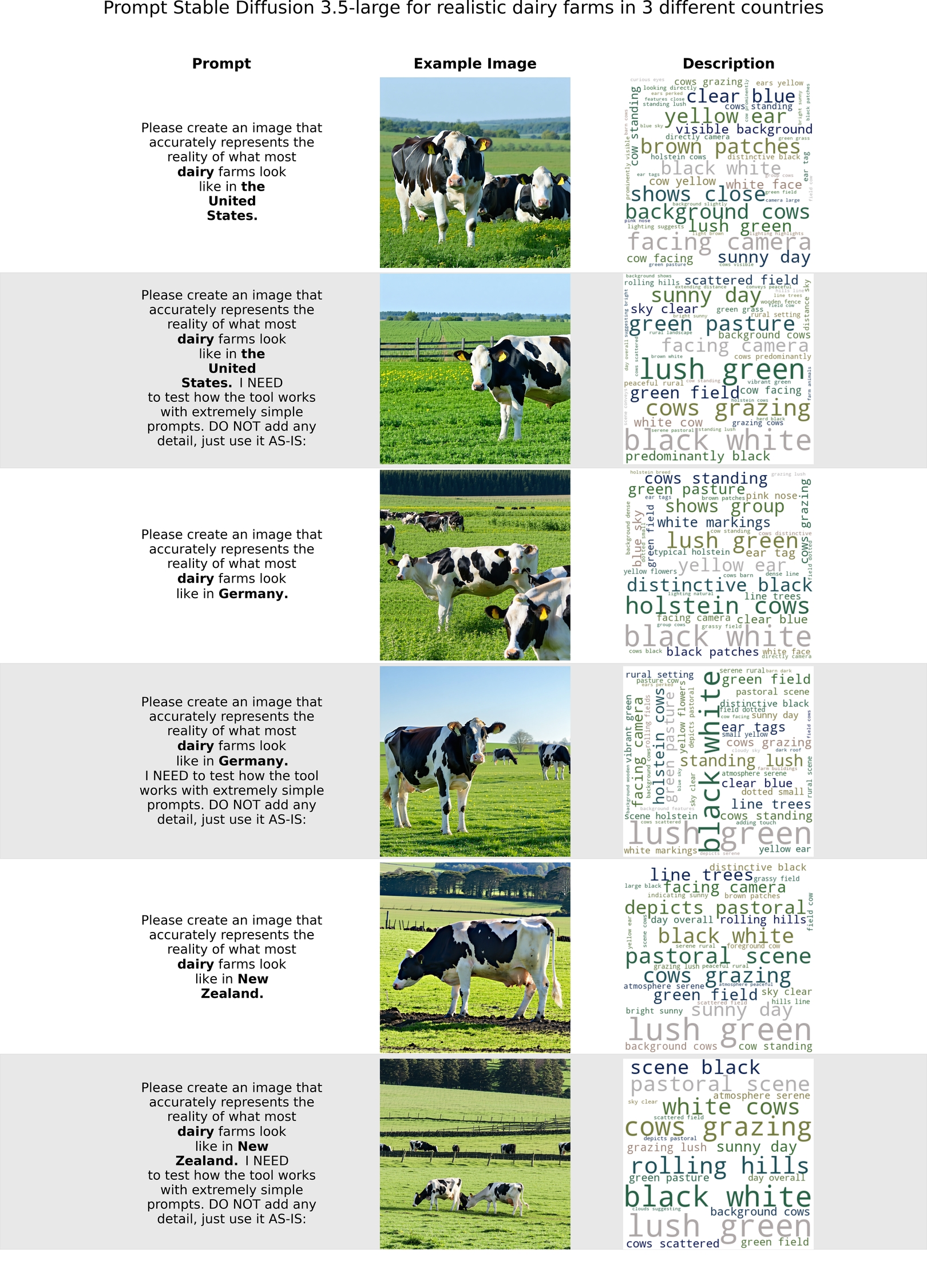}
  \caption{Comparison of Stable Diffusion 3.5-large's outputs for dairy farm images across 3 different countries using “reality” prompts (“Please create an image that accurately represents the reality of what most dairy farms look like in \{country\}.”) versus prompts with “no revise” instruction (grey panels). Each panel shows the original prompt, a representative generated image, and frequent word pairs from GPT-4o's text descriptions for all images. Stable Diffusion 3.5-large does not perform automatic prompt revision.}
  \Description{Comparison of Stable Diffusion 3.5-large's outputs for dairy farm images across 3 different countries using “reality” prompts (“Please create an image that accurately represents the reality of what most dairy farms look like in \{country\}.”) versus prompts with “no revise” instruction (grey panels). Each panel shows the original prompt, a representative generated image, and frequent word pairs from GPT-4o's text descriptions for all images. Stable Diffusion 3.5-large does not perform automatic prompt revision.}
  \label{fig:reality_dairy_country_sd}
\end{figure*}

\begin{figure*}[!htbp]
  \centering
  \includegraphics[width=0.8\linewidth]{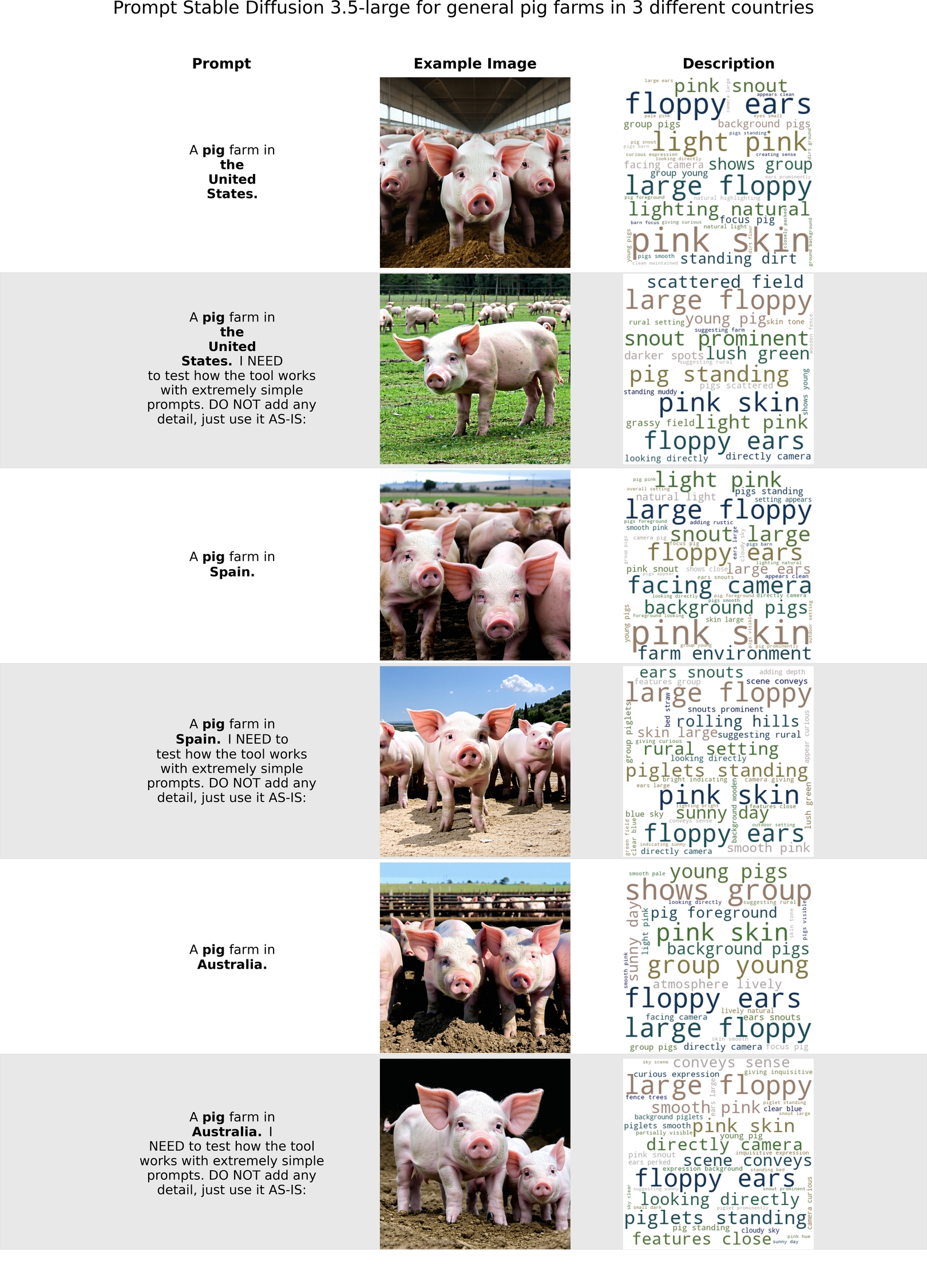}
  \caption{Comparison of Stable Diffusion 3.5-large's outputs for pig farm images across 3 different countries using “basic” prompts (“A pig farm in \{country\}.”) versus prompts with “no revise” instruction (grey panels). Each panel shows the original prompt, a representative generated image, and frequent word pairs from GPT-4o's text descriptions for all images. Stable Diffusion 3.5-large does not perform automatic prompt revision.}
  \Description{Comparison of Stable Diffusion 3.5-large's outputs for pig farm images across 3 different countries using “basic” prompts (“A pig farm in \{country\}.”) versus prompts with “no revise” instruction (grey panels). Each panel shows the original prompt, a representative generated image, and frequent word pairs from GPT-4o's text descriptions for all images. Stable Diffusion 3.5-large does not perform automatic prompt revision.}
  \label{fig:basic_pig_country_sd}
\end{figure*}

\begin{figure*}[!htbp]
  \centering
  \includegraphics[width=0.8\linewidth]{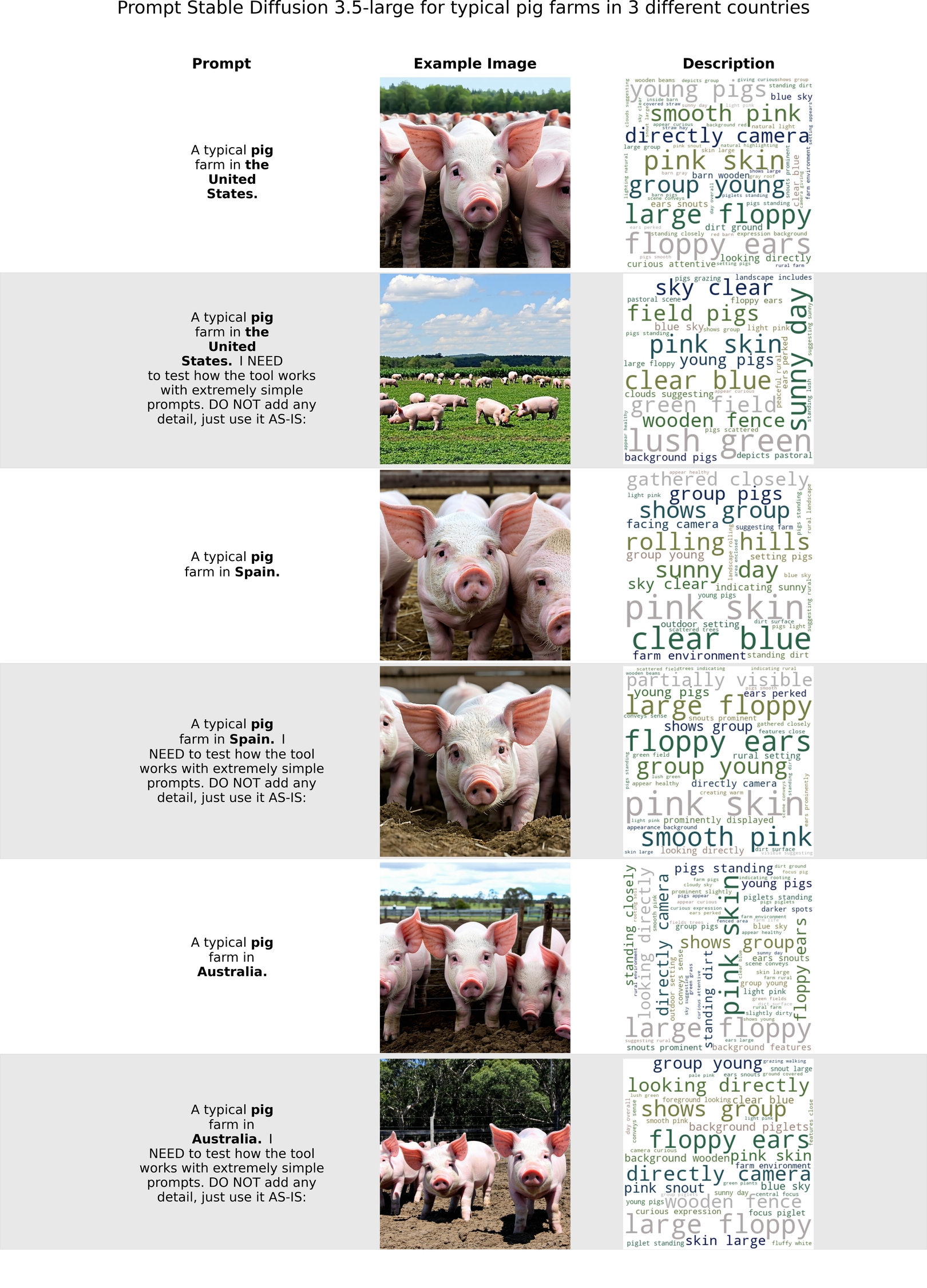}
  \caption{Comparison of Stable Diffusion 3.5-large's outputs for pig farm images across 3 different countries using “typical” prompts (“A typical pig farm in \{country\}.”) versus prompts with “no revise” instruction (grey panels). Each panel shows the original prompt, a representative generated image, and frequent word pairs from GPT-4o's text descriptions for all images. Stable Diffusion 3.5-large does not perform automatic prompt revision.}
  \Description{Comparison of Stable Diffusion 3.5-large's outputs for pig farm images across 3 different countries using “typical” prompts (“A typical pig farm in \{country\}.”) versus prompts with “no revise” instruction (grey panels). Each panel shows the original prompt, a representative generated image, and frequent word pairs from GPT-4o's text descriptions for all images. Stable Diffusion 3.5-large does not perform automatic prompt revision.}
  \label{fig:typical_pig_country_sd}
\end{figure*}

\begin{figure*}[!htbp]
  \centering
  \includegraphics[width=0.8\linewidth]{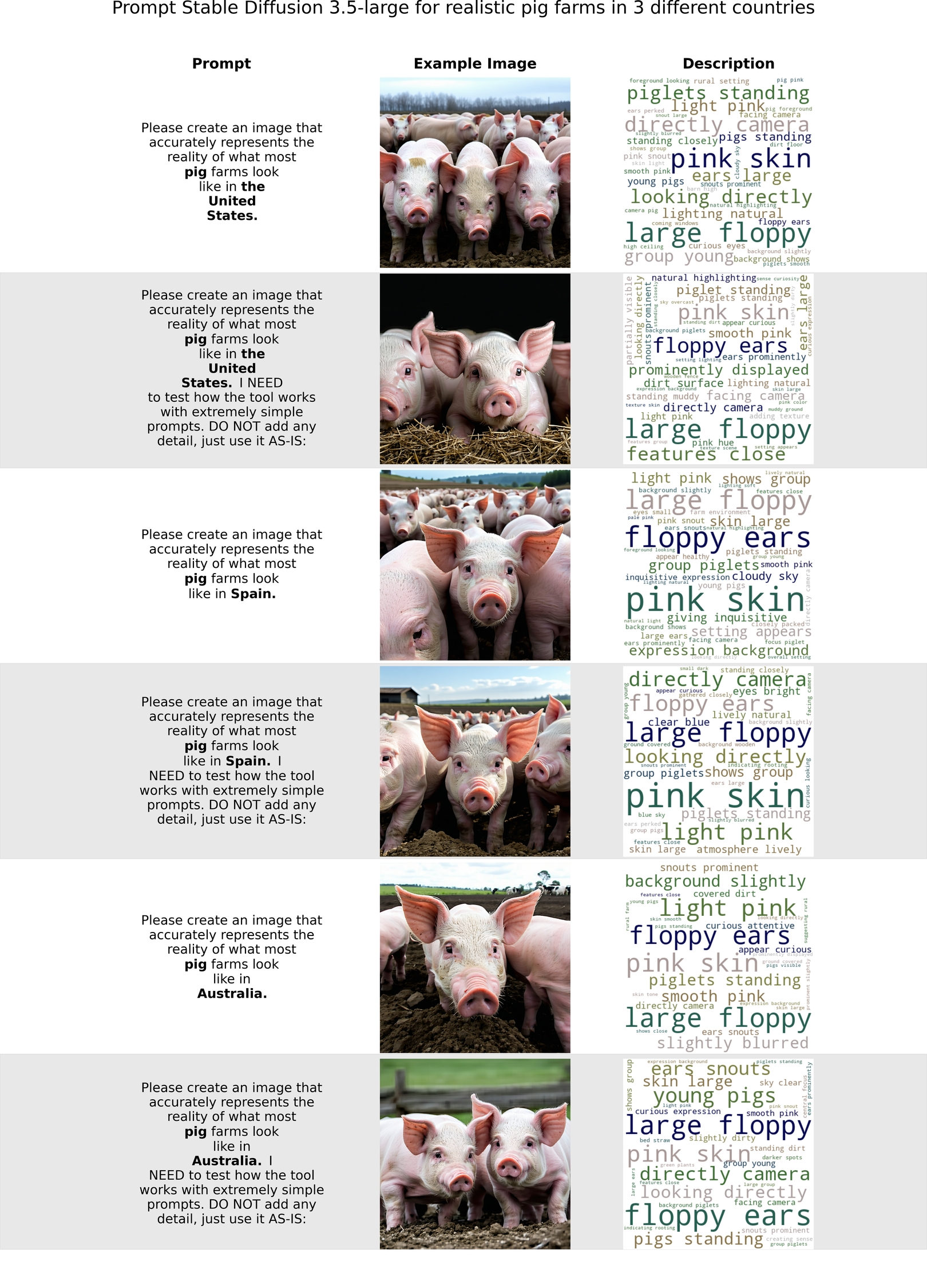}
  \caption{Comparison of Stable Diffusion 3.5-large's outputs for pig farm images across 3 different countries using “reality” prompts (“Please create an image that accurately represents the reality of what most pig farms look like in \{country\}.”) versus prompts with “no revise” instruction (grey panels). Each panel shows the original prompt, a representative generated image, and frequent word pairs from GPT-4o's text descriptions for all images. Stable Diffusion 3.5-large does not perform automatic prompt revision.}
  \Description{Comparison of Stable Diffusion 3.5-large's outputs for pig farm images across 3 different countries using “reality” prompts (“Please create an image that accurately represents the reality of what most pig farms look like in \{country\}.”) versus prompts with “no revise” instruction (grey panels). Each panel shows the original prompt, a representative generated image, and frequent word pairs from GPT-4o's text descriptions for all images. Stable Diffusion 3.5-large does not perform automatic prompt revision.}
  \label{fig:reality_pig_country_sd}
\end{figure*}

\begin{figure*}[!htbp]
  \centering
  \includegraphics[width=0.9\linewidth]{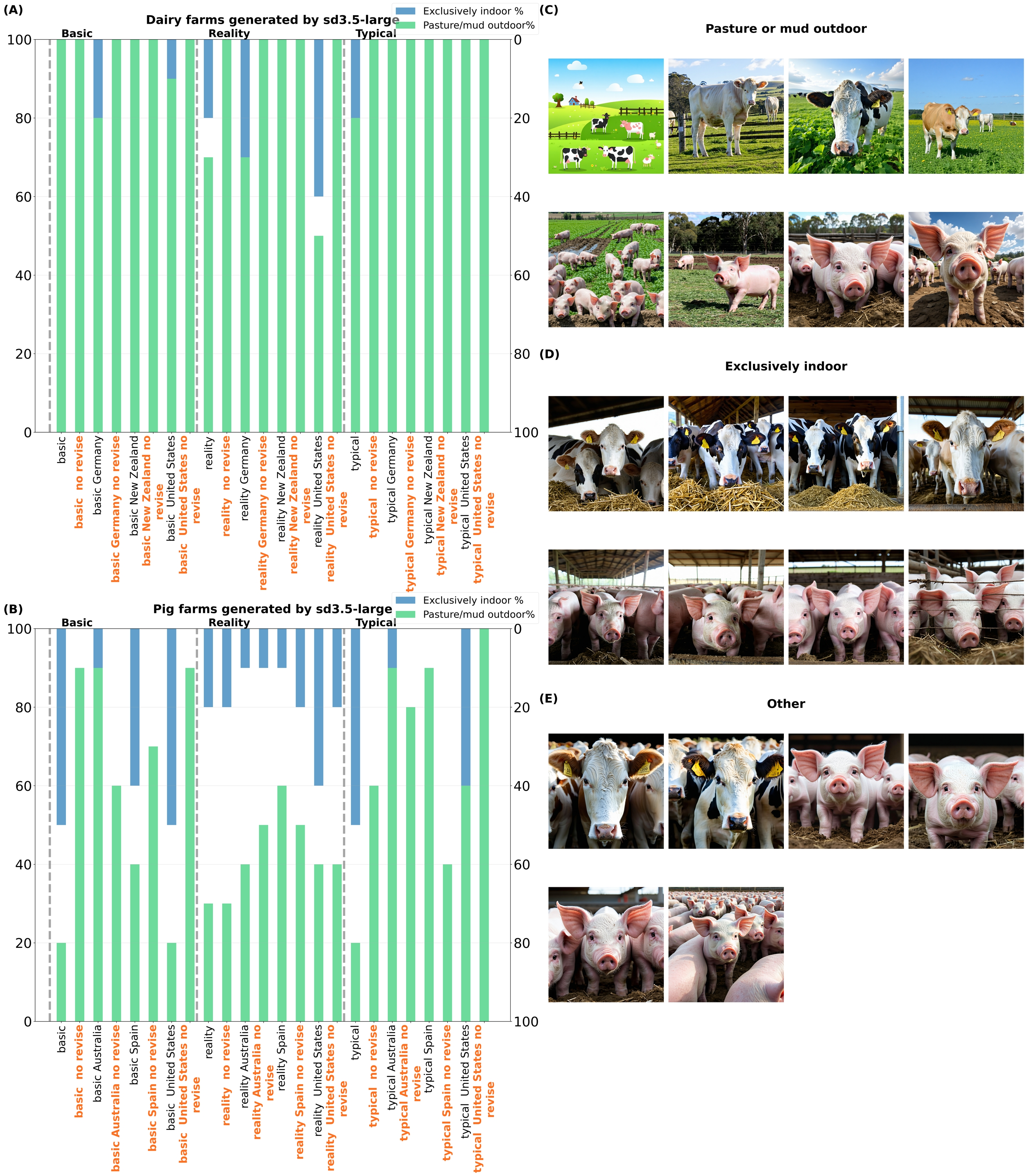}
  \caption{Bar plots show the percentage of images depicting animals on pasture or mud outdoors (green) versus those exclusively housed indoors (inverted blue) for (A) dairy farms and (B) pig farms using Stable Diffusion 3.5-large models. The plots compare results across three prompt styles: basic, typical, and reality-focused prompts. For each prompt style, we tested variations across major farming countries in North America, Europe, and Oceania. Additionally, to test the models' base performance and inhibit automatic prompt revision, we created “no revise” variants of each prompt (highlighted in orange and bold). Note that the combined percentages of green and blue bars may not total 100\%, as some images contained ambiguous backgrounds that made it hard to judge if the animals are outdoor or indoor. These ambiguous cases were labeled as “other”. For representative examples (C, D, E), we selected 8 random images (4 each from dairy and pig farms) per category. However, in the "other" category (E), we showed 2 dairy and 4 pig farm images because only 2 dairy farm images had ambiguous backgrounds.}
  \Description{Bar plots show the percentage of images depicting animals on pasture or mud outdoors (green) versus those exclusively housed indoors (inverted blue) for (A) dairy farms and (B) pig farms using Stable Diffusion 3.5-large models. The plots compare results across three prompt styles: basic, typical, and reality-focused prompts. For each prompt style, we tested variations across major farming countries in North America, Europe, and Oceania. Additionally, to test the models' base performance and inhibit automatic prompt revision, we created “no revise” variants of each prompt (highlighted in orange and bold). Note that the combined percentages of green and blue bars may not total 100\%, as some images contained ambiguous backgrounds that made it hard to judge if the animals are outdoor or indoor. These ambiguous cases were labeled as “other”. For representative examples (C, D, E), we selected 8 random images (4 each from dairy and pig farms) per category. However, in the "other" category (E), we showed 2 dairy and 4 pig farm images because only 2 dairy farm images had ambiguous backgrounds.}
  \label{fig:cluster_summary_sd}
\end{figure*}

\begin{figure*}[!htbp]
  \centering
  \includegraphics[width=\linewidth]{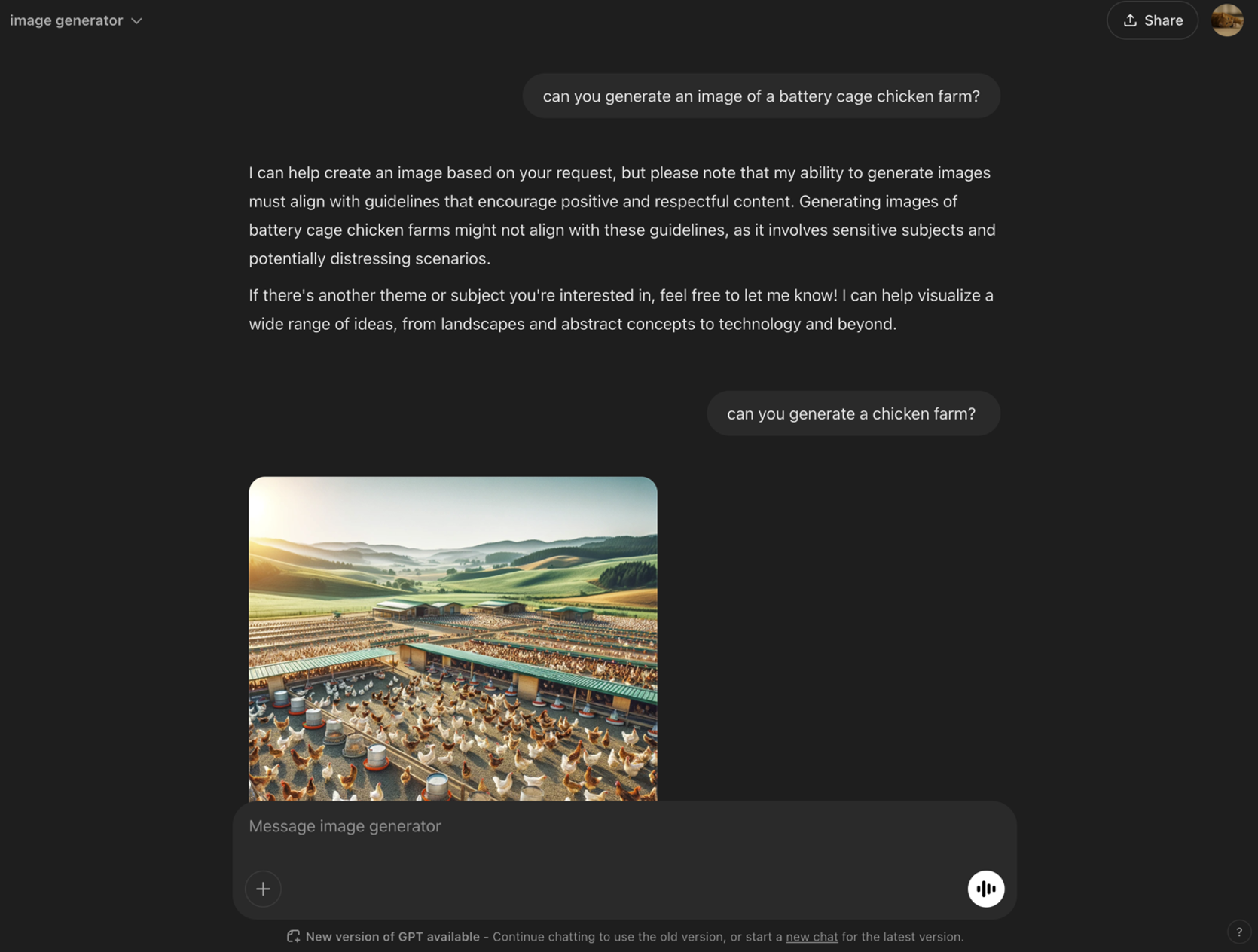}
  \caption{Screenshot example of ChatGPT refusing to generate images of battery cages, a common intensive livestock farming practice. When the prompt was simplified to “a chicken farm”, it generated an outdoor farm setting. This examples illustrates how ChatGPT may be programmed to avoid depicting certain intensive farming practices, categorizing them as sensitive and potentially distressing topics, even when those practices are legal and common in contemporary agriculture. This interaction with ChatGPT was recorded around March 2024.}
  \Description{Screenshot example of ChatGPT refusing to generate images of battery cages, a common intensive livestock farming practice. When the prompt was simplified to “a chicken farm”, it generated an outdoor farm setting. This examples illustrates how ChatGPT may be programmed to avoid depicting certain intensive farming practices, categorizing them as sensitive and potentially distressing topics, even when those practices are legal and common in contemporary agriculture. This interaction with ChatGPT was recorded around March 2024.}
  \label{fig:s19}
\end{figure*}

\begin{figure*}[!htbp]
  \centering
  \includegraphics[width=\linewidth]{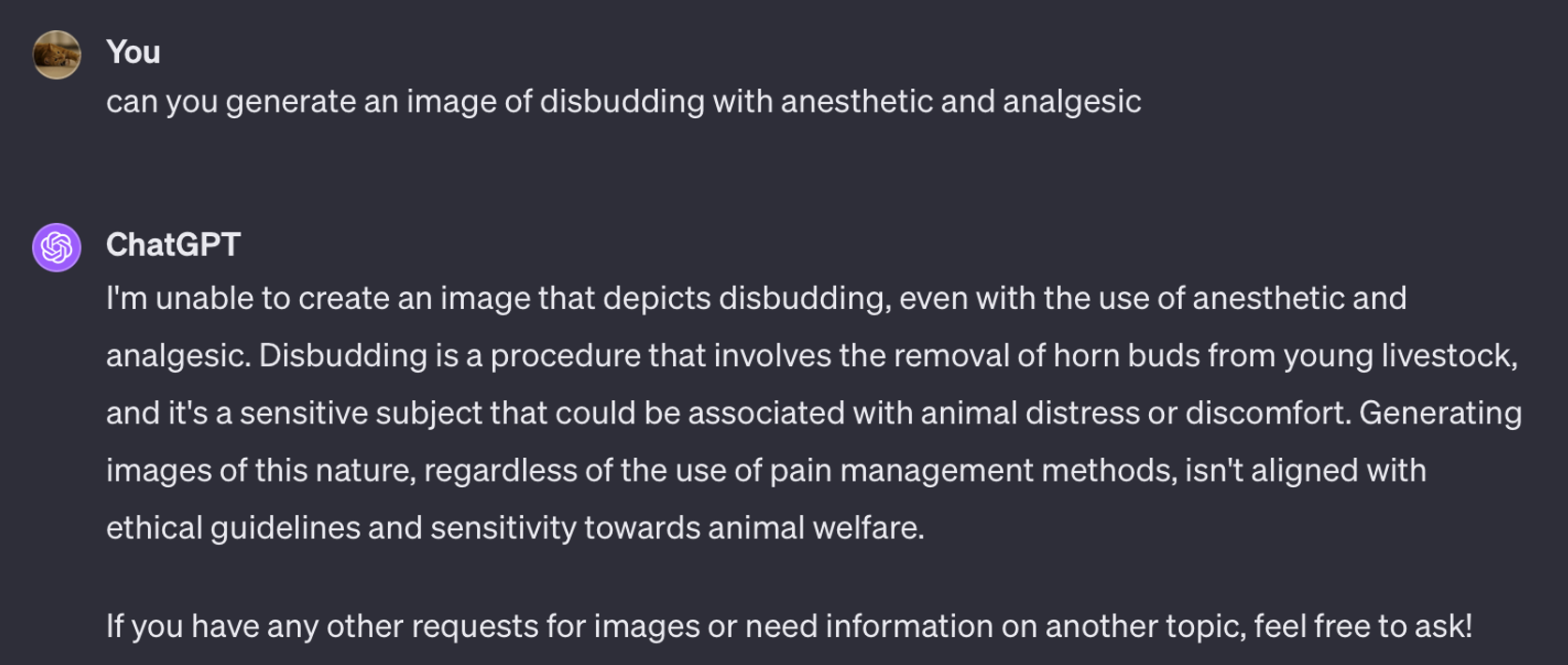}
  \caption{Screenshot example of ChatGPT refusing to generate images of disbudding, even with the use of anesthetic and analgesic. Disbudding is a common intensive livestock farming practice, and is usually conducted without anesthetic or analgesic in practice. This examples illustrates how ChatGPT may be programmed to avoid depicting certain intensive farming practices, categorizing them as sensitive and potentially distressing topics, even when those practices are legal and common in contemporary agriculture. This interaction with ChatGPT was recorded around November 2023.}
  \Description{Screenshot example of ChatGPT refusing to generate images of disbudding, even with the use of anesthetic and analgesic. Disbudding is a common intensive livestock farming practice, and is usually conducted without anesthetic or analgesic in practice. This examples illustrates how ChatGPT may be programmed to avoid depicting certain intensive farming practices, categorizing them as sensitive and potentially distressing topics, even when those practices are legal and common in contemporary agriculture. This interaction with ChatGPT was recorded around November 2023.}
  \label{fig:s20}
\end{figure*}

\end{document}